\documentclass[11pt,letter]{article}
\usepackage{jheppub}
\usepackage{amsmath,amssymb}
\usepackage[dvipsnames]{xcolor}
\usepackage{xspace}
\usepackage{ifdraft}
\usepackage{epstopdf}
\usepackage{slashed}
\usepackage{diagbox}
\usepackage{colortbl}
\usepackage{shuffle}
\usepackage{bbold}
\usepackage{enumitem}
\usepackage{tabularx}
\usepackage{graphicx}
\usepackage{stackengine}

\usepackage{tikz-feynman} 

\immediate\write18{mkdir -p images}  %% Create `pgf-img` directory
\usetikzlibrary{external}  
\immediate\write18{mkdir -p images}  %% Create `pgf-img` directory
\tikzset{graviton/.style={decorate, decoration={snake, amplitude=.4mm, segment length=1.5mm, pre length=.5mm, post length=.5mm}, double}}

\usepackage[framemethod=default]{mdframed}
\newmdenv[skipabove=7pt,
skipbelow=7pt,
rightline=false,
leftline=false,
topline=false,
bottomline=false,
backgroundcolor=gray!15,
linecolor=gray,
innerleftmargin=5pt,
innerrightmargin=5pt,
innertopmargin=5pt,
innerbottommargin=5pt,
leftmargin=0cm,
rightmargin=0cm,
linewidth=4pt]{eBox}

\usepackage{xcolor}
\hypersetup{
    colorlinks,
    linkcolor={nhpRed},
    citecolor={nhpBlue},
    urlcolor={nhpBlue}
}

\definecolor{jaxoblue}{HTML}{0086FF}

\usepackage{xspace}

\newcommand{\cHL}[0]{C^{\text{H.L.}}}

\definecolor{nhpRed}{RGB}{161,0,0}
\definecolor{nhp4}{RGB}{203, 4, 31}
\definecolor{nhp3}{RGB}{244,99,30}
\definecolor{nhp2}{RGB}{255,159,0}
\definecolor{nhp1}{RGB}{48,152,152}
\definecolor{nhpBlue}{RGB}{0,100,144}
\definecolor{cutred}{RGB}{219,56,49}
\definecolor{hgreen}{RGB}{25,176,146}
\definecolor{hgreen1}{RGB}{175,230,175}
\definecolor{hblue}{RGB}{52,152,219}
\definecolor{hbluedark}{RGB}{36, 106, 160}
\definecolor{hblue1}{RGB}{255,255,166}
\definecolor{hred}{RGB}{139,1,0}
\definecolor{hreddark}{RGB}{151, 58, 81}
\definecolor{hred1}{RGB}{255,155,155}
\definecolor{cutred}{RGB}{219,56,49}
\definecolor{hgrey4}{RGB}{75,75,75}
\definecolor{hgrey5}{RGB}{50,50,50}
\definecolor{hgrey3}{RGB}{100,100,100}
\definecolor{hgrey}{RGB}{125,125,125}
\definecolor{hgrey2}{RGB}{125,125,125}
\definecolor{hgrey1}{RGB}{150,150,150}
\definecolor{hgrey0}{RGB}{175,175,175}
\definecolor{hgreyLight}{RGB}{242,242,242}
\definecolor{darkgreen}{RGB}{59,126,108}

\newcommand{\scalelessIntAscalar}{ {
\begin{tikzpicture}[baseline=(current  bounding  box.center)]
\begin{feynman}
\vertex (a1) at (-.8, -0.5) {};
\vertex (a2) at (-.8, 0.5) {};
\vertex [dot, scale=2](mid1) at (0,0){};
\vertex [dot, scale=2](mid2) at (0,0){};
\vertex [dot, scale=1.5,hgrey0](mid3) at (0,0){};
\vertex (a3) at (.8, 0.5) {};
\vertex (a4) at (.8, -0.5) {};
\diagram{
(mid1) --[ultra thick,](a1),
(mid1) --[ultra thick,](a2),
(mid1) --[ultra thick,](a3),
(mid1) --[ultra thick,](a4),
(mid1) -- [ ultra thick,out=120,in=60,min distance=1cm](mid2),
(mid1) --[ultra thick,out=-120,in=-60,min distance=1cm](mid2),
};
\end{feynman}
\end{tikzpicture}
}
}

\newcommand{\simpleBox}{ {
\begin{tikzpicture}[baseline=(current  bounding  box.center)]
\begin{feynman}
\vertex (a1) at (-1,1){2};
\vertex (a2) at (1,-1){4};
\vertex (a3) at (1,1){3};
\vertex (a4) at (-1,-1){1};
\vertex (mid3) at (.5,.5);
\vertex (mid4) at (.5,-.5);
\vertex (mid5) at (-.5,.5);
\vertex (mid6) at (-.5,-.5);
\diagram{
(a4) --[ultra thick,](mid6),
(a3) --[ultra thick,](mid3),
(a2) --[ultra thick,](mid4),
(a1) --[ultra thick,](mid5),
(mid3) --[ultra thick,](mid4),
(mid5) --[ultra thick,](mid6),
(mid5) --[ultra thick,](mid3),
(mid4) --[ultra thick,](mid6),
};
\end{feynman}
\end{tikzpicture}
}
}

\newcommand{\xBox}{ {
\begin{tikzpicture}[baseline=(current  bounding  box.center)]
\begin{feynman}
\vertex (a1) at (-1.5,1){2};
\vertex (a2) at (1.5,-1){4};
\vertex (a3) at (1.5,1){3};
\vertex (a4) at (-1.5,-1){1};
\vertex (mid1) at (0,-.5);
\vertex (mid2) at (0,.5);
\vertex (mid3) at (1,.5);
\vertex (mid4) at (1,-.5);
\vertex (mid5) at (-1,.5);
\vertex (mid6) at (-1,-.5);
\vertex (mid7) at (-.5,0) {};
\diagram{
(a4) --[ultra thick,](mid6),
(a3) --[ultra thick,](mid3),
(a2) --[ultra thick,](mid4),
(a1) --[ultra thick,](mid5),
(mid1) --[ultra thick,](mid6),
(mid3) --[ultra thick,](mid2),
(mid1) --[ultra thick,](mid4),
(mid5) --[ultra thick,](mid7),
(mid7) --[ultra thick,](mid1),
(mid2) --[ultra thick,](mid6),
(mid5) --[ultra thick,](mid2),
(mid3) --[ultra thick,](mid4),
};
\end{feynman}
\end{tikzpicture}
}
}

\newcommand{\dBox}{ {
\begin{tikzpicture}[baseline=(current  bounding  box.center)]
\begin{feynman}
\vertex (a1) at (-1.5,1){2};
\vertex (a2) at (1.5,-1){4};
\vertex (a3) at (1.5,1){3};
\vertex (a4) at (-1.5,-1){1};
\vertex (mid1) at (0,-.5);
\vertex (mid2) at (0,.5);
\vertex (mid3) at (1,.5) ;
\vertex (mid4) at (1,-.5) ;
\vertex (mid5) at (-1,.5);
\vertex (mid6) at (-1,-.5) ;
\diagram{
(a4) --[ultra thick,](mid6),
(a3) --[ultra thick,](mid3),
(a2) --[ultra thick,](mid4),
(a1) --[ultra thick,](mid5),
(mid1) --[ultra thick,](mid2),
(mid3) --[ultra thick,](mid2),
(mid1) --[ultra thick,](mid6),
(mid1) --[ultra thick,](mid4),
(mid5) --[ultra thick,](mid2),
(mid5) --[ultra thick,](mid6),
(mid3) --[ultra thick,](mid4),
};
\end{feynman}
\end{tikzpicture}
}
}

\newcommand{\scalelessIntBscalar}{ {
\begin{tikzpicture}[baseline=(current  bounding  box.center)]
\begin{feynman}
\vertex (a1) at (-.8, -0.6) {};
\vertex (a2) at (-.8, 0.6) {};
\vertex (a3) at (-1, 0) {};
\vertex [dot, scale=2](mid1) at (0,0){};
\vertex [dot, scale=1.5,hgrey0](mid2) at (0,0){};
\vertex [dot, scale=2](mid3) at (1,0){};
\vertex [dot, scale=1.5,hgrey0](mid4) at (1,0){};
\vertex (a4) at (1.8, 0) {};
\diagram{
(mid1) --[ultra thick,](a1),
(mid1) --[ultra thick,](a2),
(mid1) --[ultra thick,](a3),
(mid1) --[ultra thick,](mid3),
(mid1) --[ultra thick,out=60,in=120,min distance=0.4cm](mid3),
(mid1) --[ultra thick,out=-60,in=-120,min distance=0.4cm](mid3),
(mid3) --[ultra thick](a4),
};
\end{feynman}
\end{tikzpicture}
}
}

\newcommand{\scaleIntAscalarsmall}[4]{ {
\begin{tikzpicture}[baseline=2]
\begin{feynman}
\vertex (a1) at (-1.4*.7, .5*.7) {#1};
\vertex (a3) at (-1.3*.7, 1.3*.7) {#3};
\vertex (a2) at (1.3*.7, 1.3*.7) {#2};
\vertex (a4) at (1.4*.7, .5*.7) {#4};
\vertex [dot,scale=2*.7](mid1) at (0.5*.7,0.5*.7){};
\vertex [dot,scale=1.5*.7,hgrey0](mid2) at (0.5*.7,0.5*.7){};
\vertex [dot,scale=2*.7](mid3) at (-0.5*.7,0.5*.7){};
\vertex [dot,scale=1.5*.7,hgrey0](mid4) at (-0.5*.7,0.5*.7){};
\vertex [dot,scale=2*.7](mid5) at (0,-.1){};
\vertex [dot,scale=2*.7](mid7) at (0,-.1){};
\vertex [dot,scale=1.5*.7,hgrey0](mid6) at (0,-.1){};
\diagram{
(mid3) --[ ultra thick](a1),
(mid3) --[ ultra thick](a3),
(mid1) --[ ultra thick](a2),
(mid1) --[ ultra thick](a4),
(mid1) --[ ultra thick](mid3),
(mid1) --[ ultra thick](mid5),
(mid3) --[ ultra thick](mid5),
(mid5) --[ ultra thick,out=-135,in=-45,min distance=.5cm](mid7),
};
\end{feynman}
\end{tikzpicture}
}
}

\newcommand{\scaleIntBscalarsmall}[4]{ {
\begin{tikzpicture}[baseline=2]
\begin{feynman}
\vertex (a1) at (-1.4*.7, .5*.7) {#1};
\vertex (a2) at (1.3*.7, 1.3*.7) {#2};
\vertex [dot,scale=2*.7](mid1) at (0.5*.7,0.5*.7){};
\vertex [dot,scale=1.5*.7,hgrey0](mid2) at (0.5*.7,0.5*.7){};
\vertex [dot,scale=2*.7](mid3) at (-0.5*.7,0.5*.7){};
\vertex [dot,scale=1.5*.7,hgrey0](mid4) at (-0.5*.7,0.5*.7){};
\vertex [dot,scale=2*.7](mid5) at (0,0){};
\vertex [dot,scale=1.5*.7,hgrey0](mid6) at (0,0){};
\vertex (a3) at (-.7*.7, -.7*.7) {#3};
\vertex (a4) at (.7*.7, -.7*.7) {#4};
\diagram{
(mid3) --[ ultra thick](a1),
(mid1) --[ ultra thick](a2),
(mid1) --[ ultra thick,out=120,in=60,min distance=0.1cm](mid3),
(mid1) --[ ultra thick](mid3),

(mid1) --[ ultra thick](mid5),
(mid3) --[ ultra thick](mid5),

(mid5) --[ ultra thick](a4),
(mid5) --[ ultra thick,](a3)
};
\end{feynman}
\end{tikzpicture}
}
}

\newcommand{\scaleIntBscalar}[4]{ {
\begin{tikzpicture}[baseline=6]
\begin{feynman}
\vertex (a1) at (-1.1, .5) {#1};
\vertex (a2) at (1, 1) {#2};
\vertex [dot,scale=2](mid1) at (0.5,0.5){};
\vertex [dot,scale=1.5,hgrey0](mid2) at (0.5,0.5){};
\vertex [dot,scale=2](mid3) at (-0.5,0.5){};
\vertex [dot,scale=1.5,hgrey0](mid4) at (-0.5,0.5){};
\vertex [dot,scale=2](mid5) at (0,0){};
\vertex [dot,scale=1.5,hgrey0](mid6) at (0,0){};
\vertex (a3) at (-.5, -.5) {#3};
\vertex (a4) at (.5, -.5) {#4};
\diagram{
(mid3) --[ ultra thick](a1),
(mid1) --[ ultra thick](a2),
(mid1) --[ ultra thick,out=120,in=60,min distance=0.1cm](mid3),
(mid1) --[ ultra thick](mid3),

(mid1) --[ ultra thick](mid5),
(mid3) --[ ultra thick](mid5),

(mid5) --[ ultra thick](a4),
(mid5) --[ ultra thick,](a3)
};
\end{feynman}
\end{tikzpicture}
}
}

\newcommand{\Hspin}{ {
\begin{tikzpicture}[baseline=-3]
\begin{feynman}
\vertex (a1) at (-.7,-.7) {};
\vertex (a2) at (-.7,.7) {};
\vertex (a3) at (.7,.7) {};
\vertex (a4) at (.7,-.7) {};
\vertex (mid1) at (-.35,0);
\vertex (mid2) at (.35,0);
\vertex (mid1A) at (-.35,.15);
\vertex (mid2a) at (.35,.15);
\vertex (mid1B) at (-.35,-.15);
\vertex (mid2b) at (.35,-.15);
\diagram{
(mid1) -- [photon,ultra thick] (a1);
(mid1) -- [photon,ultra thick] (a2);
(mid2) -- [photon,ultra thick] (a3);
(mid2) -- [photon,ultra thick] (a4);
(mid1) -- [photon,ultra thick,hred] (mid2);
(mid1A) -- [ photon,ultra thick,hred](mid2a);
(mid1B) -- [ photon,ultra thick,hred] (mid2b);

};
\end{feynman}
\end{tikzpicture}
}
}

\newcommand{\vectorCon}{ {
\begin{tikzpicture}[baseline=-3]
\begin{feynman}
\vertex (a1) at (-.7,-.7) {};
\vertex (a2) at (-.7,.7) {};
\vertex (a3) at (.7,.7) {};
\vertex (a4) at (.7,-.7) {};
\vertex (mid1) at (0,0);
\diagram{
(mid1) -- [photon,ultra thick] (a1);
(mid1) -- [photon,ultra thick] (a2);
(mid1) -- [photon,ultra thick] (a3);
(mid1) -- [photon,ultra thick] (a4);
};
\end{feynman}
\end{tikzpicture}
}
}

\newcommand{\pionCon}{ {
\begin{tikzpicture}[baseline=-3]
\begin{feynman}
\vertex (a1) at (-.7,-.7) {$\pi$};
\vertex (a2) at (-.7,.7) {$\pi$};
\vertex (a3) at (.7,.7) {$\pi$};
\vertex (a4) at (.7,-.7) {$\pi$};
\vertex (mid1) at (0,0);
\diagram{
(mid1) -- [ultra thick] (a1);
(mid1) -- [ultra thick] (a2);
(mid1) -- [ultra thick] (a3);
(mid1) -- [ultra thick] (a4);
};
\end{feynman}
\end{tikzpicture}
}
}

\newcommand{\scaleIntAsmallNoNum}{ {
\begin{tikzpicture}[baseline=-3]
\begin{feynman}
\vertex (a1) at (1*0.7, -0.6*0.7) ;
\vertex (a2) at (1*0.7, 0.6*0.7) ;
\vertex [dot,scale=2*0.7](mid3) at (1*0.7,0){};
\vertex [dot,scale=2*0.7](mid4) at (1*0.7,0){};
\vertex [dot,scale=1.5*0.7,hgrey0](mid4) at (1*0.7,0){};
\vertex [dot,scale=2*0.7](mid5) at (2*0.7,0){};
\vertex [dot,scale=1.5*0.7,hgrey0](mid6) at (2*0.7,0){};
\vertex (a3) at (2.8*0.7, .6*0.7) {};
\vertex (a4) at (2.8*0.7, -.6*0.7) {};
\diagram{
(mid3) --[ ultra thick,](a1),
(mid3) --[ ultra thick,](a2),
(mid3) --[ ultra thick,out=60,in=120,min distance=0.5*0.7cm](mid5),
(mid3) --[ ultra thick,out=-140,in=140,min distance=.8cm](mid4),
(mid3) --[ ultra thick,out=-60,in=-120,min distance=0.5*0.7cm](mid5),
(mid5) --[ ultra thick](a4),
(mid5) --[ ultra thick,](a3)
};
\end{feynman}
\end{tikzpicture}
}
}

\newcommand{\scaleIntBsmallNoNum}{ {
\begin{tikzpicture}[baseline=-3]
\begin{feynman}
\vertex (a1) at (1.95, 0) ;
\vertex (a2) at (.25,0) ;
\vertex [dot,scale=2*0.7](mid3) at (1*0.7,0){};
\vertex [dot,scale=2*0.7](mid4) at (1*0.7,0){};
\vertex [dot,scale=1.5*0.7,hgrey0](mid4) at (1*0.7,0){};
\vertex [dot,scale=2*0.7](mid5) at (2*0.7,0){};
\vertex [dot,scale=1.5*0.7,hgrey0](mid6) at (2*0.7,0){};
\vertex (a3) at (2.8*0.7, .6*0.7) {};
\vertex (a4) at (2.8*0.7, -.6*0.7) {};
\diagram{
(mid5) --[ ultra thick,](a1),
(mid3) --[ ultra thick,](a2),
(mid3) --[ ultra thick,out=60,in=120,min distance=0.5*0.7cm](mid5),
(mid3) --[ ultra thick,out=-60,in=-120,min distance=0.5*0.7cm](mid5),
(mid3) --[ ultra thick](mid5),
(mid5) --[ ultra thick](a4),
(mid5) --[ ultra thick,](a3)
};
\end{feynman}
\end{tikzpicture}
}
}

\newcommand{\scaleIntCsmallNoNum}{ {
\begin{tikzpicture}[baseline=-3]
\begin{feynman}
\vertex (a1) at (-.8*0.7, -0.6*0.7) {};
\vertex (a2) at (-.8*0.7, 0.6*0.7) {};
\vertex [dot,scale=2*0.7](mid1) at (0,0){};
\vertex [dot,scale=1.5*0.7,hgrey0](mid2) at (0,0){};
\vertex [dot,scale=2*0.7](mid3) at (1*0.7,0){};
\vertex [dot,scale=1.5*0.7,hgrey0](mid4) at (1*0.7,0){};
\vertex [dot,scale=2*0.7](mid5) at (2*0.7,0){};
\vertex [dot,scale=1.5*0.7,hgrey0](mid6) at (2*0.7,0){};
\vertex (a3) at (2.8*0.7, .6*0.7) {};
\vertex (a4) at (2.8*0.7, -.6*0.7) {};
\diagram{
(mid1) --[ ultra thick,](a1),
(mid1) --[ ultra thick,](a2),
(mid1) --[ ultra thick,out=60,in=120,min distance=0.5*0.7cm](mid3),
(mid1) --[ ultra thick,out=-60,in=-120,min distance=0.5*0.7cm](mid3),
(mid3) --[ ultra thick,out=60,in=120,min distance=0.5*0.7cm](mid5),
(mid3) --[ ultra thick,out=-60,in=-120,min distance=0.5*0.7cm](mid5),
(mid5) --[ ultra thick](a4),
(mid5) --[ ultra thick,](a3)
};
\end{feynman}
\end{tikzpicture}
}
}

\newcommand{\scaleIntCsmall}{ {
\begin{tikzpicture}[baseline=-3]
\begin{feynman}
\vertex (a1) at (-.8*0.7, -0.6*0.7) {1};
\vertex (a2) at (-.8*0.7, 0.6*0.7) {2};
\vertex [dot,scale=2*0.7](mid1) at (0,0){};
\vertex [dot,scale=1.5*0.7,hgrey0](mid2) at (0,0){};
\vertex [dot,scale=2*0.7](mid3) at (1*0.7,0){};
\vertex [dot,scale=1.5*0.7,hgrey0](mid4) at (1*0.7,0){};
\vertex [dot,scale=2*0.7](mid5) at (2*0.7,0){};
\vertex [dot,scale=1.5*0.7,hgrey0](mid6) at (2*0.7,0){};
\vertex (a3) at (2.8*0.7, .6*0.7) {3};
\vertex (a4) at (2.8*0.7, -.6*0.7) {4};
\diagram{
(mid1) --[ ultra thick,](a1),
(mid1) --[ ultra thick,](a2),
(mid1) --[ ultra thick,out=60,in=120,min distance=0.5*0.7cm](mid3),
(mid1) --[ ultra thick,out=-60,in=-120,min distance=0.5*0.7cm](mid3),
(mid3) --[ ultra thick,out=60,in=120,min distance=0.5*0.7cm](mid5),
(mid3) --[ ultra thick,out=-60,in=-120,min distance=0.5*0.7cm](mid5),
(mid5) --[ ultra thick](a4),
(mid5) --[ ultra thick,](a3)
};
\end{feynman}
\end{tikzpicture}
}
}

\newcommand{\scaleIntBubbleProbeA}[2]{ {
\begin{tikzpicture}[baseline=-2.5]
\begin{feynman}
\vertex (a1) at (-.6,.3) {#1};
\vertex (a2) at (-.6, -.3) {#2};
\vertex (a4) at (.5, .7) {${}^{\alpha_1=1}$};
\vertex (a6) at (.5, -.8) {${}^{\alpha_4=1}$};
\vertex [dot,scale=2](mid1) at (0,0){};
\vertex [dot,scale=1.5,hgrey0](mid2) at (0,0){};
\vertex [dot,scale=2](mid5) at (1,0){};
\vertex [dot,scale=1.5,hgrey0](mid6) at (1,0){};
\vertex (a3) at (1.6, .3) {#2};
\vertex (a4) at (1.6, -.3) {#2};
\diagram{
(a1) --[ultra thick,](mid1),
(a2) --[ultra thick,](mid1),
(mid1) --[ ultra thick,out=60,in=120,min distance=0.5cm](mid5),
(mid1) --[ ultra thick,out=30,in=150,min distance=0.2cm](mid5),
(mid1) --[ ultra thick,out=-30,in=-150,min distance=0.2cm](mid5),
(mid1) --[ ultra thick,out=-60,in=-120,min distance=0.5cm](mid5),
(mid5) --[ultra thick,](a3),
(mid5) --[ultra thick,](a4)
};
\end{feynman}
\end{tikzpicture}
}
}

\newcommand{\scaleIntBubbleProbeB}[2]{ {
\begin{tikzpicture}[baseline=-2.5]
\begin{feynman}
\vertex (a1) at (-.6,.3) {#1};
\vertex (a2) at (-.6, -.3) {#2};
\vertex [dot,scale=2](mid1) at (0,0){};
\vertex (a4) at (.5, .5) {${}^{\alpha_2=D/2-2}$};
\vertex (a6) at (.5, -.6) {${}^{\alpha_4=1}$};
\vertex [dot,scale=1.5,hgrey0](mid2) at (0,0){};
\vertex [dot,scale=2](mid5) at (1,0){};
\vertex [dot,scale=1.5,hgrey0](mid6) at (1,0){};
\vertex (a3) at (1.6, .3) {#2};
\vertex (a4) at (1.6, -.3) {#2};
\diagram{
(a1) --[ultra thick,](mid1),
(a2) --[ultra thick,](mid1),
(mid1) --[ ultra thick,out=45,in=135,min distance=0.3cm](mid5),
(mid1) --[ ultra thick](mid5),
(mid1) --[ ultra thick,out=-45,in=-135,min distance=0.3cm](mid5),
(mid5) --[ultra thick,](a3),
(mid5) --[ultra thick,](a4)
};
\end{feynman}
\end{tikzpicture}
}
}

\newcommand{\scaleIntBubbleProbeC}[2]{ {
\begin{tikzpicture}[baseline=-2.5]
\begin{feynman}
\vertex (a1) at (-.6,.3) {#1};
\vertex (a2) at (-.6, -.3) {#2};
\vertex (a4) at (.5, .4) {${}^{\alpha_3=D/2-2}$};
\vertex (a6) at (.5, -.5) {${}^{\alpha_4=1}$};
\vertex [dot,scale=2](mid1) at (0,0){};
\vertex [dot,scale=1.5,hgrey0](mid2) at (0,0){};
\vertex [dot,scale=2](mid5) at (1,0){};
\vertex [dot,scale=1.5,hgrey0](mid6) at (1,0){};
\vertex (a3) at (1.6, .3) {#2};
\vertex (a4) at (1.6, -.3) {#2};
\diagram{
(a1) --[ultra thick,](mid1),
(a2) --[ultra thick,](mid1),
(mid1) --[ ultra thick,out=30,in=150,min distance=0.3cm](mid5),
(mid1) --[ ultra thick,out=-30,in=-150,min distance=0.3cm](mid5),
(mid5) --[ultra thick,](a3),
(mid5) --[ultra thick,](a4)
};
\end{feynman}
\end{tikzpicture}
}
}

\newcommand{\scaleIntBubbleProbe}[2]{ {
\begin{tikzpicture}[baseline=-2]
\begin{feynman}
\vertex (b1) at (-.8,.8) {};
\vertex (b2) at (1.8,.8) {};
\vertex (b3) at (1.8,-.8) {};
\vertex (b4) at (-.8,-.8) {};
\vertex (a1a) at (-.8,.6) {};
\vertex (a1b) at (-.6,.1) {$\vdots$};
\vertex (a1c) at (-.8,-.6) {};
\vertex (a2) at (.6,.7) {$\alpha_1$};
\vertex (a4) at (.6,-.7) {$\alpha_2$};
\vertex [dot,scale=2](mid1) at (0,0){};
\vertex [dot,scale=1.5,hgrey0](mid2) at (0,0){};
\vertex [dot,scale=2](mid5) at (1,0){};
\vertex [dot,scale=1.5,hgrey0](mid6) at (1,0){};
\vertex (a3a) at (1.8, .6) {};
\vertex (a3b) at (1.6, .1) {$\vdots$};
\vertex (a3c) at (1.8,-.6) {};
\diagram{
(b1) --[scalar, hblue,ultra thick,](b4),
(b3) --[scalar, hblue,ultra thick,](b2),
(a1a) --[fermion, ultra thick,](mid1),
(a1c) --[fermion, ultra thick,](mid1),
(mid1) --[ ultra thick,out=60,in=120,min distance=0.4cm](mid5),
(mid1) --[ ultra thick,out=-60,in=-120,min distance=0.4cm](mid5),
(mid5) --[fermion, ultra thick,](a3a),
(mid5) --[fermion, ultra thick,](a3c)
};
\end{feynman}
\end{tikzpicture}
}
}

\newcommand{\scaleIntCscalar}[4]{ {
\begin{tikzpicture}[baseline=(current  bounding  box.center)]
\begin{feynman}
\vertex (a1) at (-.6, -0.6) {#1};
\vertex (a2) at (-.6, 0.6) {#2};
\vertex [dot,scale=2](mid1) at (0,0){};
\vertex [dot,scale=1.5,hgrey0](mid2) at (0,0){};
\vertex [dot,scale=2](mid3) at (.8,0){};
\vertex [dot,scale=1.5,hgrey0](mid4) at (.8,0){};
\vertex [dot,scale=2](mid5) at (1.6,0){};
\vertex [dot,scale=1.5,hgrey0](mid6) at (1.6,0){};
\vertex (a3) at (2.2, .6) {#3};
\vertex (a4) at (2.2, -.6) {#4};
\diagram{
(mid1) --[ ultra thick,](a1),
(mid1) --[ ultra thick,](a2),
(mid1) --[ ultra thick,out=60,in=120,min distance=0.4cm](mid3),
(mid1) --[ ultra thick,out=-60,in=-120,min distance=0.4cm](mid3),
(mid3) --[ ultra thick,out=60,in=120,min distance=0.4cm](mid5),
(mid3) --[ ultra thick,out=-60,in=-120,min distance=0.4cm](mid5),
(mid5) --[ ultra thick](a4),
(mid5) --[ ultra thick,](a3)
};
\end{feynman}
\end{tikzpicture}
}
}

\newcommand{\scaleIntAfermion}[4]{ {
\begin{tikzpicture}[baseline=-3]
\begin{feynman}
\vertex (a1) at (-.8, -0.6) {#1};
\vertex (a2) at (-.8, 0.6) {#2};
\vertex [dot,scale=3](mid1) at (0,0){};
\vertex [dot,scale=2.5,hgrey0](mid2) at (0,0){};
\vertex [dot,scale=3](mid3) at (1,0){};
\vertex [dot,scale=2.5,hgrey0](mid4) at (1,0){};
\vertex (a3) at (1.8, .6) {#3};
\vertex (a4) at (1.8, -.6) {#4};
\diagram{
(mid1) --[photon, ultra thick,](a1),
(mid1) --[photon, ultra thick,](a2),
(mid1) --[fermion, ultra thick,out=60,in=120,min distance=0.4cm,hred](mid3),
(mid3) --[fermion, ultra thick,out=-120,in=-60,min distance=0.4cm,hred](mid1),
(mid3) --[photon, ultra thick](a4),
(mid3) --[photon, ultra thick,](a3)
};
\end{feynman}
\end{tikzpicture}
}
}

\newcommand{\scaleIntAScalar}[4]{ {
\begin{tikzpicture}[baseline=-3]
\begin{feynman}
\vertex (a1) at (-.8, -0.6) {#1};
\vertex (a2) at (-.8, 0.6) {#2};
\vertex [dot,scale=3](mid1) at (0,0){};
\vertex [dot,scale=2.5,hgrey0](mid2) at (0,0){};
\vertex [dot,scale=3](mid3) at (1,0){};
\vertex [dot,scale=2.5,hgrey0](mid4) at (1,0){};
\vertex (a3) at (1.8, .6) {#3};
\vertex (a4) at (1.8, -.6) {#4};
\diagram{
(mid1) --[photon, ultra thick,](a1),
(mid1) --[photon, ultra thick,](a2),
(mid1) --[ ultra thick,out=60,in=120,min distance=0.4cm,hblue](mid3),
(mid1) --[ultra thick,out=-60,in=-120,min distance=0.4cm,hblue](mid3),
(mid3) --[photon, ultra thick](a4),
(mid3) --[photon, ultra thick,](a3)
};
\end{feynman}
\end{tikzpicture}
}
}

\newcommand{\scaleIntAvectorODD}[6]{ {
\begin{tikzpicture}[baseline=0]
\begin{feynman}
\vertex (a5) at (1.3,.7) {#6};
\vertex (a6) at (0,.7) {#5};
\vertex (a1) at (-.8, -0.6) {#1};
\vertex (a2) at (-.8, 0.6) {#2};
\vertex [dot,scale=3](mid1) at (0,0){};
\vertex [dot,scale=2.5,hgrey4](mid2) at (0,0){};
\vertex [dot,scale=3](mid3) at (1,0){};
\vertex [dot,scale=2.5,hgrey0](mid4) at (1,0){};
\vertex (a3) at (1.8, .6) {#3};
\vertex (a4) at (1.8, -.6) {#4};
\diagram{
(mid1) --[photon, ultra thick,](a1),
(mid1) --[photon, ultra thick,](a2),
(mid1) --[photon, ultra thick,out=60,in=120,min distance=0.4cm](mid3),
(mid1) --[photon, ultra thick,out=-60,in=-120,min distance=0.4cm](mid3),
(mid3) --[photon, ultra thick](a4),
(mid3) --[photon, ultra thick,](a3)
};
\end{feynman}
\end{tikzpicture}
}
}

\newcommand{\bubCutStates}[4]{ {
\begin{tikzpicture}[baseline=-4]
\begin{feynman}
\vertex (a1) [dot,scale=1.6]at (-.6,0){};
\vertex (a2) [dot,scale=1.6]at (.6,0){};
\vertex (a1) [dot,hgrey0,scale=1.3]at (-.6,0){};
\vertex (a2) [dot,hgrey0,scale=1.3]at (.6,0){};
\vertex (a1a) at (-1.3,.6){#2};
\vertex (a1b) at (-1.3,-.6){#1};
\vertex (a1c) at (-1.1,0){};
\vertex (a2a) at (1.3,.6){#3};
\vertex (a2b) at (1.3,-.6){#4};
\vertex (a2c) at (1.1,0){};
\vertex (a5) at (0,.4){};
\vertex (a7) at (0,-.4){};
\vertex (a8) at (-.5,0){};
\vertex (b1) at (0,.7);
\vertex (b2) at (0,-.7);
\vertex (b3) at (.7,0);
\vertex (b4) at (-.7,0);
\vertex (b5) at (0,0){};
\diagram{
(b1) --[scalar,hblue,ultra thick,](b5),
(b2) --[scalar,hblue,ultra thick,](b5),
(a5) --[ ultra thick,in=60,out=180,min distance=0.1cm](a1),
(a5) --[ ultra thick,out=0,in=120,min distance=0.1cm](a2),
(a2) --[ ultra thick,out=-120,in=0,min distance=0.1cm](a7),
(a1) --[ ultra thick,out=-60,in=180,min distance=0.1cm](a7),
(a1a) --[ ultra thick,](a1),
(a1b) --[ ultra thick,](a1),
(a2a) --[ ultra thick,](a2),
(a2b) --[ ultra thick,](a2),
};
\end{feynman}
\end{tikzpicture}
}
}

\newcommand{\bubCut}{ {
\begin{tikzpicture}[baseline=-4]
\begin{feynman}
\vertex (a1) [dot,scale=1.6]at (-.6,0){};
\vertex (a2) [dot,scale=1.6]at (.6,0){};
\vertex (a1) [dot,hgrey0,scale=1.3]at (-.6,0){};
\vertex (a2) [dot,hgrey0,scale=1.3]at (.6,0){};
\vertex (a1a) [dot,scale=1.6]at (-.9,.25);
\vertex (a1b) [dot,scale=1.6]at (-.9,-.25);
\vertex (a1c) at (-1.1,-0.05){};
\vertex (a2a) [dot,scale=1.6]at (.9,.25);
\vertex (a2b) [dot,scale=1.6]at (.9,-.25);
\vertex (a2c) at (1.1,-0.05){${}^{P_i}$};
\vertex (a5) at (0,.4){};
\vertex (a7) at (0,-.4){};
\vertex (a8) at (-.5,0){};
\vertex (b1) at (0,.7);
\vertex (b2) at (0,-.7);
\vertex (b3) at (.7,0);
\vertex (b4) at (-.7,0);
\vertex (b5) at (0,0){};
\diagram{
(b1) --[scalar,hblue,ultra thick,](b5),
(b2) --[scalar,hblue,ultra thick,](b5),
(a5) --[ ultra thick,in=60,out=180,min distance=0.1cm](a1),
(a5) --[ ultra thick,out=0,in=120,min distance=0.1cm](a2),
(a2) --[ ultra thick,out=-120,in=0,min distance=0.1cm](a7),
(a1) --[ ultra thick,out=-60,in=180,min distance=0.1cm](a7),
(a1a) --[ ultra thick,](a1),
(a1b) --[ ultra thick,](a1),
(a2a) --[ ultra thick,](a2),
(a2b) --[ ultra thick,](a2),
};
\end{feynman}
\end{tikzpicture}
}
}

\newcommand{\triCut}{ {
\begin{tikzpicture}[baseline=-4]
\begin{feynman}
\vertex (a1) [dot,scale=1.6] at (-.5,.5){};
\vertex (a2) [dot,scale=1.6]at (.6,0){};
\vertex (a4) [dot,scale=1.6]at (-.5,-.5){};
\vertex (a1) [dot,hgrey0,scale=1.3] at (-.5,.5){};
\vertex (a2) [dot,hgrey0,scale=1.3]at (.6,0){};
\vertex (a4) [dot,hgrey0,scale=1.3]at (-.5,-.5){};
\vertex (a1a) [dot,scale=1.6] at (-.5,.8);
\vertex (a1b) [dot,scale=1.6] at (-.8,.5);
\vertex (a1c) at (-.8,.8){${}^{P_i}$};
\vertex (a2a) [dot,scale=1.6]at (.9,.25);
\vertex (a2b) [dot,scale=1.6]at (.9,-.25);
\vertex (a2c) at (1.1,-0.1){${}^{P_j}$};
\vertex (a4a) [dot,scale=1.6]at (-.5,-.8);
\vertex (a4b) [dot,scale=1.6]at (-.8,-.5);
\vertex (a4c) at (-.8,-.9){};
\vertex (a5) at (0,.3){};
\vertex (a7) at (0,-.3){};
\vertex (a8) at (-.5,0){};
\vertex (b1) at (0,.7);
\vertex (b2) at (0,-.7);
\vertex (b3) at (.7,0);
\vertex (b4) at (-.7,0);
\vertex (b5) at (0,0){};
\diagram{
(b1) --[scalar,hblue,ultra thick,](b5),
(b2) --[scalar,hblue,ultra thick,](b5),
(b4) --[scalar,hblue,ultra thick,](b5),
(a5) --[ ultra thick,](a1),
(a5) --[ ultra thick,](a2),
(a2) --[ ultra thick,](a7),
(a4) --[ ultra thick,](a7),
(a8) --[ ultra thick,](a4),
(a1) --[ ultra thick,](a8),
(a1a) --[ ultra thick,](a1),
(a1b) --[ ultra thick,](a1),
(a2a) --[ ultra thick,](a2),
(a2b) --[ ultra thick,](a2),
(a4a) --[ ultra thick,](a4),
(a4b) --[ ultra thick,](a4),
};
\end{feynman}
\end{tikzpicture}
}
}

\newcommand{\boxCut}{ {
\begin{tikzpicture}[baseline=-4]
\begin{feynman}
\vertex (a1) [dot,scale=1.6] at (-.5,.5){};
\vertex (a2) [dot,scale=1.6]at (.5,.5){};
\vertex (a3) [dot,scale=1.6]at (.5,-.5){};
\vertex (a4) [dot,scale=1.6]at (-.5,-.5){};
\vertex (a1) [dot,hgrey0,scale=1.3] at (-.5,.5){};
\vertex (a2) [dot,hgrey0,scale=1.3]at (.5,.5){};
\vertex (a3) [dot,hgrey0,scale=1.3]at (.5,-.5){};
\vertex (a4) [dot,hgrey0,scale=1.3]at (-.5,-.5){};
\vertex (a1a) [dot,scale=1.6] at (-.5,.8);
\vertex (a1b) [dot,scale=1.6] at (-.8,.5);
\vertex (a1c) at (-.8,.8){${}^{P_i}$};
\vertex (a2a) [dot,scale=1.6]at (.5,.8);
\vertex (a2b) [dot,scale=1.6]at (.8,.5);
\vertex (a2c) at (.9,.8){${}^{P_j}$};
\vertex (a3a) [dot,scale=1.6]at (.5,-.8);
\vertex (a3b) [dot,scale=1.6]at (.8,-.5);
\vertex (a3c) at (.9,-.9){${}^{P_k}$};
\vertex (a4a) [dot,scale=1.6]at (-.5,-.8);
\vertex (a4b) [dot,scale=1.6]at (-.8,-.5);
\vertex (a3c) at (-.8,-.9){};
\vertex (a5) at (0,.5){};
\vertex (a6) at (.5,0){};
\vertex (a7) at (0,-.5){};
\vertex (a8) at (-.5,0){};
\vertex (b1) at (0,.7);
\vertex (b2) at (0,-.7);
\vertex (b3) at (.7,0);
\vertex (b4) at (-.7,0);
\vertex (b5) at (0,0){};
\diagram{
(b1) --[scalar,hblue,ultra thick,](b5),
(b2) --[scalar,hblue,ultra thick,](b5),
(b3) --[scalar,hblue,ultra thick,](b5),
(b4) --[scalar,hblue,ultra thick,](b5),
(a5) --[ ultra thick,](a1),
(a5) --[ ultra thick,](a2),
(a2) --[ ultra thick,](a6),
(a6) --[ ultra thick,](a3),
(a4) --[ ultra thick,](a7),
(a7) --[ ultra thick,](a3),
(a8) --[ ultra thick,](a4),
(a1) --[ ultra thick,](a8),
(a1a) --[ ultra thick,](a1),
(a1b) --[ ultra thick,](a1),
(a2a) --[ ultra thick,](a2),
(a2b) --[ ultra thick,](a2),
(a3a) --[ ultra thick,](a3),
(a3b) --[ ultra thick,](a3),
(a4a) --[ ultra thick,](a4),
(a4b) --[ ultra thick,](a4),
};
\end{feynman}
\end{tikzpicture}
}
}

\newcommand{\scaleIntApion}{ {
\begin{tikzpicture}[baseline=(current  bounding  box.center)]
\begin{feynman}
\vertex (a1) at (-.8, -0.6) {1};
\vertex (a2) at (-.8, 0.6) {2};
\vertex [dot,scale=3](mid1) at (0,0){};
\vertex [dot,scale=2.5,hgrey0](mid2) at (0,0){};
\vertex [dot,scale=3](mid3) at (1,0){};
\vertex [dot,scale=2.5,hgrey0](mid4) at (1,0){};
\vertex (a3) at (1.8, .6) {3};
\vertex (a4) at (1.8, -.6) {4};
\diagram{
(mid1) --[ ultra thick,](a1),
(mid1) --[ultra thick,](a2),
(mid1) --[ ultra thick,out=60,in=120,min distance=0.4cm](mid3),
(mid1) --[ ultra thick,out=-60,in=-120,min distance=0.4cm](mid3),
(mid3) --[ultra thick](a4),
(mid3) --[ ultra thick,](a3)
};
\end{feynman}
\end{tikzpicture}
}
}

\newcommand{\scaleIntAvector}[4]{ {
\begin{tikzpicture}[baseline=-3]
\begin{feynman}
\vertex (a1) at (-.8, -0.6) {#1};
\vertex (a2) at (-.8, 0.6) {#2};
\vertex [dot,scale=3](mid1) at (0,0){};
\vertex [dot,scale=2.5,hgrey0](mid2) at (0,0){};
\vertex [dot,scale=3](mid3) at (1,0){};
\vertex [dot,scale=2.5,hgrey0](mid4) at (1,0){};
\vertex (a3) at (1.8, .6) {#3};
\vertex (a4) at (1.8, -.6) {#4};
\diagram{
(mid1) --[photon, ultra thick,](a1),
(mid1) --[photon, ultra thick,](a2),
(mid1) --[photon, ultra thick,out=60,in=120,min distance=0.4cm](mid3),
(mid1) --[photon, ultra thick,out=-60,in=-120,min distance=0.4cm](mid3),
(mid3) --[photon, ultra thick](a4),
(mid3) --[photon, ultra thick,](a3)
};
\end{feynman}
\end{tikzpicture}
}
}

\newcommand{\intScaleless}{ {
\begin{tikzpicture}[baseline=(current  bounding  box.center)]
\begin{feynman}
\vertex [dot,scale=3](mid1) at (0,0){};
\vertex [dot,scale=3](mid3) at (0,0){};
\vertex [dot,scale=2.5,hgrey0](mid2) at (0,0){};
\vertex (a1) at (-.9, -.75) ;
\vertex (a2) at (-.55,-.75) ;
\vertex (a3) at (.55,-.75) ;
\vertex (a4) at (.9,-.75) ;
\vertex (a5) at (0,-.55) {$\,\cdots$};
\vertex (b1) at (-1.2,-.8);
\vertex (b6) at (1.2,-.8);
\diagram{
(mid1) --[ ultra thick,](a1),
(b1) --[ scalar,ultra thick,hblue](b6),
(mid1) --[ultra thick,](a2),
(mid1) --[ultra thick,](a3),
(mid1) --[ultra thick,](a4),
(mid1) --[ultra thick,out=60,in=120,min distance=1.2cm](mid3),
};
\end{feynman}
\end{tikzpicture}
}
}

\newcommand{\extScaleless}{ {
\begin{tikzpicture}[baseline=(current  bounding  box.center)]
\begin{feynman}
\vertex [dot,scale=3](mid1) at (0,0){};
\vertex [dot,scale=3](mid3) at (0,0){};
\vertex [dot,scale=2.5,hgrey0](mid2) at (0,0){};
\vertex (a1) at (-.9, -.75) ;
\vertex (a2) at (-.55,-.75) ;
\vertex (a3) at (.55,-.75) ;
\vertex (a4) at (.9,-.75) ;
\vertex (c1) at (-1.1, 1) {1};
\vertex (c2) at (-.45,1) {2};
\vertex (c3) at (.25,1) {$\cdots$};
\vertex (c4) at (1.1,1){$n\!-\!1$} ;
\vertex (a5) at (0,-.55) {$\,\cdots$};
\vertex (b1) at (-1.2,-.8);
\vertex (b6) at (1.2,-.8);
\diagram{
(mid1) --[ ultra thick,](a1),
(b1) --[ scalar,ultra thick,hblue](b6),
(mid1) --[ultra thick,](a2),
(mid1) --[ultra thick,](a3),
(mid1) --[ultra thick,](a4),
(mid1) --[ fermion,ultra thick,](c1),
(mid1) --[fermion,ultra thick,](c2),
(mid1) --[fermion,ultra thick,](c4),
};
\end{feynman}
\end{tikzpicture}
}
}

\newcommand{\scaleTree}[1]{ {
\begin{tikzpicture}[baseline=(current  bounding  box.center)]
\begin{feynman}
\vertex (a1) at (-.8, -0.6) {};
\vertex (a2) at (-.8, 0.6) {};
\vertex [dot,scale=3](mid1) at (0,0){};
\vertex [dot,scale=2.5,#1](mid2) at (0,0){};
\vertex (a3) at (.8, .6) {};
\vertex (a4) at (.8, -.6) {};
\diagram{
(mid1) --[photon, ultra thick,](a1),
(mid1) --[photon, ultra thick,](a2),
(mid1) --[photon, ultra thick](a4),
(mid1) --[photon, ultra thick,](a3)
};
\end{feynman}
\end{tikzpicture}
}
}

\newcommand{\scaleIntBtune}[8]{ {
\begin{tikzpicture}[baseline=4]
\begin{feynman}
\vertex (a1) at (-1.4,.5);
\vertex (a2) at (1.2,1);
\vertex [dot,scale=3](mid1) at (0.7,0.5){};
\vertex [dot,scale=2.5,hgrey0](mid2) at (0.7,0.5){};
\vertex [dot,scale=3](mid3) at (-0.7,0.5){};
\vertex [dot,scale=2.5,hgrey0](mid4) at (-0.7,0.5){};
\vertex [dot,scale=3](mid5) at (0,0){};
\vertex [dot,scale=2.5,hgrey0](mid6) at (0,0){};
\vertex (a3) at (-.5, -.5);
\vertex (a4) at (.5, -.5);
\diagram{
(mid3) --[photon, ultra thick](a1),
(mid1) --[photon, ultra thick](a2),
(mid1) --[#1, ultra thick,out=120,in=60,min distance=0.1cm,#5](mid3),
(mid1) --[#2, ultra thick,#6](mid3),

(mid1) --[#3,ultra thick,#7](mid5),
(mid3) --[#4,ultra thick,#8](mid5),

(mid5) --[photon, ultra thick](a4),
(mid5) --[photon, ultra thick,](a3)
};
\end{feynman}
\end{tikzpicture}
}
}

\newcommand{\scaleIntCtune}[4]{ {
\begin{tikzpicture}[baseline=-3]
\begin{feynman}
\vertex (a1) at (-.8, -0.6) {};
\vertex (a2) at (-.8, 0.6) {};
\vertex [dot,scale=3](mid1) at (0,0){};
\vertex [dot,scale=2.5,hgrey0](mid2) at (0,0){};
\vertex [dot,scale=3](mid3) at (1,0){};
\vertex [dot,scale=2.5,hgrey0](mid4) at (1,0){};
\vertex [dot,scale=3](mid5) at (2,0){};
\vertex [dot,scale=2.5,hgrey0](mid6) at (2,0){};
\vertex (a3) at (2.8, .6) {};
\vertex (a4) at (2.8, -.6) {};
\diagram{
(mid1) --[photon, ultra thick,](a1),
(mid1) --[photon, ultra thick,](a2),
(mid1) --[#1, ultra thick,out=60,in=120,min distance=0.4cm,#2](mid3),
(mid3) --[#1, ultra thick,in=-60,out=-120,min distance=0.4cm,#2](mid1),
(mid3) --[#3, ultra thick,out=60,in=120,min distance=0.4cm,#4](mid5),
(mid5) --[#3, ultra thick,in=-60,out=-120,min distance=0.4cm,#4](mid3),
(mid5) --[photon, ultra thick](a4),
(mid5) --[photon, ultra thick,](a3)
};
\end{feynman}
\end{tikzpicture}
}
}

 \def\draftnote#1{{\color{red}\it #1}} 
\def\sect#1{section~\ref{#1}}
\def\Fig#1{fig.~{\ref{#1}}}
\def\Fig#1{Fig.~{\ref{#1}}}

\def\spa#1.#2{\left\langle#1\,#2\right\rangle}
\def\spb#1.#2{\left[#1\,#2\right]}
\def\spash#1.#2{\spa{\smash{#1}}.{\smash{#2}}}
\def\spbsh#1.#2{\spb{\smash{#1}}.{\smash{#2}}}
\def\sand#1.#2.#3{%
\left\langle\smash{#1}{\vphantom1}^{-}\right|{#2}%
\left|\smash{#3}{\vphantom1}^{-}\right\rangle}
\def\sandpp#1.#2.#3{%
\left\langle\smash{#1}{\vphantom1}^{+}\right|{#2}%
\left|\smash{#3}{\vphantom1}^{+}\right\rangle}
\def\sandpm#1.#2.#3{%
\left\langle\smash{#1}{\vphantom1}^{+}\right|{#2}%
\left|\smash{#3}{\vphantom1}^{-}\right\rangle}
\def\sandmp#1.#2.#3{%
\left\langle\smash{#1}{\vphantom1}^{-}\right|{#2}%
\left|\smash{#3}{\vphantom1}^{+}\right\rangle}

\def\sec#1{section~\ref{#1}}
\def\eqn#1{eq.~(\ref{#1})}

\definecolor{NUpurple}{RGB}{078,042,132}

\author[1]{John Joseph M. Carrasco,}
\author[1]{Nicolas H. Pavao}

\affiliation[1]{Department of Physics and Astronomy, Northwestern
  University, Evanston, Illinois 60208, USA}

\title{Even-point Multi-loop Unitarity and its Applications: Exponentiation, Anomalies and Evanescence}

\abstract{
We identify novel  structure in newly computed multi-loop amplitudes and quantum actions for even-point effective field theories, including both the nonlinear sigma model (NLSM) and double-copy gauge theories such as Born-Infeld and its supersymmetric generalizations.  
We exploit special properties of all even-point theories towards efficient unitarity based amplitude construction. 
In doing so, we find evidence that the leading IR divergence of NLSM amplitudes exponentiates when the symmetry group is $\mathbb{CP}^1\cong SU(2)/U(1)$. 
We then systematically compute the two-loop anomalous behavior of Born-Infeld, and find that the counterterms needed to restore $U(1)$ invariant behavior at loop-level can be constructed via a symmetric-structure double-copy.  
We also demonstrate that the divergent part of the one-minus $(-+++)$ two-loop anomaly vanishes upon introducing an evanescent operator. In addition to these pure photon counterterms, we verify through explicit calculation that the anomalous matrix elements that violate $U(1)$ duality invariance can be alternatively cancelled by summing over internal $\mathcal{N}=4$ DBIVA superfields. 
Finally we find that $\mathcal{N}=4$ Dirac-Born-Infeld-Volkov-Akulov (DBIVA) amplitudes permit double-copy construction through two-loop order by reproducing our unitarity based result with a double copy between color-dual $\mathcal{N}=4$ super-Yang-Mills  and our two-loop NLSM amplitudes. 
This result supports the possibility of  color-dual representations for NLSM beyond one-loop. 
We conclude with an overview of how $D$-dimensional four-photon counterterms can be constructed in generality with the symmetric-structure double-copy, and outline a convenient way of counting evanescent operators using Hilbert series as generating functions. }

\newpage

\begin{document}
\maketitle
\flushbottom
 
\setstackgap{S}{6pt}
\setstackgap{L}{7pt}

\section{Introduction}\label{sec:intro}

Recent decades have seen  significant advances in our ability to compute scattering amplitudes to higher orders in perturbation theory. At the epicenter of this explosion in the generation of sharp $S$-matrix data is the unitarity method \cite{UnitarityMethod, Fusing, BDKUniarityReview}, which bypasses the standard Feynman diagram approach by constructing higher order amplitudes directly from lower order on-shell information. At one-loop, the unitarity method allows one to extract amplitudes by directly computing a series of unitarity cuts \cite{Forde:2007mi}. 
On-shell methods have furthermore unveiled novel amplitude-level structure, like the  duality between color and kinematics \cite{BCJ} and associated double-copy construction \cite{BCJLoop}, which are frequently obscured by the traditional Lagrangian description of the theories. 
While much of the research in perturbative calculations thus far has focused on gauge theory and gravity, recent literature has investigated to what extent the $S$-matrix of effective field theory is constrained by on-shell data \cite{Adams:2006sv,Cheung:2014dqa,Cheung:2015ota,Cheung:2016drk,Cheung:2018oki,Low:2019ynd,Carrasco:2019yyn,Arkani-Hamed:2020blm,Carrasco:2021ptp,Chi:2021mio,Bonnefoy:2021qgu,Carrasco:2022lbm,Carrasco:2022sck,Green:2022slj,Pavao:2022kog,Chen:2022shl,Chen:2023dcx,Brown:2023srz,Li:2023wdm}. 

In this paper we make the quantum leap to the multi-loop sector of effective field theory.  Namely we study even-point (EP) effective field theory (EFT) at the multi-loop level using a combination of generalized unitarity \cite{UnitarityMethod, Fusing, BDKUniarityReview} and the double-copy construction \cite{BCJ, BCJLoop}. The EFTs that we consider are the nonlinear sigma model (NLSM) and a related family of gauge theories involving Born-Infeld theory and supersymmetric generalizations known as Dirac-Born-Infeld-Volkov-Akulov (DBIVA) theories. These families are known at tree-level to be double-copies between NLSM and Yang-Mills theories with varying amounts of supersymmetry. While we focus on these specific examples, the methods we develop can be used for perturbative calculations in any even-point effective field theory. The motivation for this work is three-fold.

1. First, perturbative calculations in NLSM and DBIVA effective field theories are in many ways significantly simpler than the typical multi-loop calculations of more phenomenological theories, like quantum-chromodynamics (QCD). As we will show, the multi-loop amplitudes for these theories allow us to recycle many of the $D$-dimensional integration tools that are so powerful at one-loop order. To approach this problem, we expand on the now standard procedure of Forde \cite{Forde:2007mi}, that computes one-loop amplitudes directly from 4D unitarity cuts. In \sect{sec:EMU} we'll show that there are only two basis integrals needed at two-loop four-point for even-point effective field theory amplitudes. In the spirit of one-loop unitarity, the integral coefficients can be directly extracted from Even-point Multi-loop Unitarity (EMU) and a tensor reduction algorithm. Our method of EMU is a $D$-dimensional approach to integrand construction. This allows us to compute a large catalog of $D$-dimensional two-loop amplitudes, which provides fertile ground for cultivating insights about the multi-loop structure of effective field theories. The perturbative depth of our calculations sheds light on the exponential structure of IR divergences for $\mathbb{CP}^1$ NLSM amplitudes through two-loops, the anomalous behavior $U(1)$ duality invariance through two-loop order in DBIVA theories, as well as the emergence of evanescent operators relevant for anomaly cancellation in pure Born-Infeld theory. We believe these surprisingly rich physical structures could serve as a theoretical laboratory for future studies of the interplay between effective field theory operators and perturbative calculations in quantum field theory. 

2. The second motivation is that very little is known about the duality between color and kinematics at multi-loop level for generic models in the web of theories \cite{BCJreview}. For a comprehensive review of color-dual representations, see Refs. \cite{BCJreview, Bern:2022wqg, Adamo:2022dcm} and references therein. While there has been tremendous success of applying the color-kinematics duality to $\mathcal{N}=4$ super-Yang-Mills, for which color-dual integrands are known through four-point four-loop \cite{Bern:2012uf, Neq44np, GravityFour}, five-point through three-loops, and to seven-point at one-loop \cite{Bjerrum-Bohr:2013iza,Edison:2020uzf,Edison:2022jln}, there are many obstructions for generic gauge/gravity theories. Presently, the state-of-the-art for nonlinear sigma model (NLSM) color-dual numerators come from the $XY\!Z$ model of Cheung and Shen \cite{Cheung2016prv}, while $D$-dimensional integrands for pure Yang-Mills have only been identified through five-point one-loop \cite{He:2017spx}. At present, there are no known $D$-dimensional representations for either of these theories at two-loop that globally manifest the duality between color and kinematics, despite attempts in the literature \cite{OneTwoLoopPureYMBCJ, Mogull:2015adi, Bern:2015ooa, Geyer:2019hnn}. Furthermore, similar bottlenecks exists for less-than-maximal $\mathcal{N}<2$ sYM \cite{Johansson:2017bfl}. While there have been a number of recent developments in constructing manifestly color-dual Feynman rules \cite{Chen:2019ywi,Chen:2021chy,Brandhuber:2021bsf,Cheung:2021zvb,Ben-Shahar:2021zww,Cheung:2022mix, Ben-Shahar:2022ixa}, a precise definition the kinematic algebra off-shell remains elusive. Considering the known tree-level double-copy relationships between NLSM and DBIVA and higher-derivative corrections \cite{Cachazo:2014xea,Carrasco:2016ldy}, all of the amplitudes we compute should in principle participate in a double-copy formulation of our results. At the very least, the amplitudes we compute will serve as important checks on future breakthroughs in multi-loop studies of color-dual representations. In addition to traditional double-copy construction, we find that the quantum effective actions generated by loop effects permit a rather compact construction in terms of {symmetric-structure double-copy}, which was introduced in recent work by the authors~\cite{Carrasco:2022jxn}. 

3.  Finally, DBIVA effective field theories touch a wide range research areas at the forefront of high energy physics. Of more formal interest, is the presence of quantum anomalies that violate the $U(1)$ duality invariance, which the theory enjoys at tree-level \cite{Elvang:2020kuj}. The presence of these $U(1)$ anomalies in gravity is closely linked to ultraviolet divergences computed at loop-level \cite{Bern:2007xj,Carrasco:2013ypa,Craig:2019zkf,Monteiro:2022nqt}. It has been argued that cancelling these anomalous matrix elements with $R^n$ counterterms can lead to enhanced UV cancellations \cite{Bern:2017tuc,Bern:2017rjw,Bern:2019isl}. This relationship has been demonstrated both for pure Einstein-Hilbert gravity \cite{Goroff:1985sz,Goroff:1985th,vandeVen:1991gw}, and also for less than maximal $\mathcal{N}\leq 4$ supergravity \cite{Bern:2013uka}. While DBIVA theories are themselves are ultraviolet divergent in $D=4$, due to their simplicity they serve as an essential laboratory for probing anomaly cancellation at high loop order. Moreover, in addition to their formal theory relevance, DBI theories have garnered wide phenomenological interest in cosmology, both for sourcing non-gaussianities in CMB bispectrum \cite{Alishahiha:2004eh,Creminelli:2005hu,Fergusson:2008ra}, and for their compatibility with the observed CMB tensor mode suppression \cite{Carrasco:2015pla,Carrasco:2015rva,Carrasco:2015uma,BICEP:2021xfz,Kallosh:2021mnu}. Thus, with inflationary data on the horizon, understanding the perturbative structure of DBI could be particularly relevant for modeling early universe quantum fluctuations. 

The outline of the paper is as follows: in \sect{sec:Review}, we will provide a review of the on-shell methods and integration techniques needed to probe the multi-loop physics studied in this paper. Then in \sect{sec:EMU}, we introduce the method of Even-point Multi-loop Unitarity (EMU) and compute the requisite two-loop tensor integrals. In \sect{sec:Loops}, we present our results, beginning with a warm-up calculation in \sect{sec:NLSMU} where we construct integrands for NLSM through two-loop order, and compute the fully integrated amplitudes. We show that in $D=2-2\epsilon$, the leading IR divergences of the $\mathbb{CP}^1$ model exponentiate. With the scaffolding of $D$-dimensional integration in hand, we then compute two-loop amplitudes for $\mathcal{N}=4$ DBIVA theory and pure-photon Born-Infeld theory in \sect{sec:DBIU}. There we compute the anomalies present beyond one-loop order.  Then, using the NLSM integrands of \sect{sec:NLSMU}, we perform a multi-loop double-copy to $\mathcal{N}=4$ DBIVA observables with the color-dual basis numerators available in the literature in \sect{sec:DBIvDC}. After computing the catalog of two-loop amplitudes, we study the construction of quantum effective actions in \sect{sec:Actions} as they relate to the anomalous matrix elements present in two-loop Born-Infeld amplitudes. We demonstrate in \sect{sec:Anomalies} that the one-minus anomaly at two-loop requires the introduction of an evanescent operator at $\mathcal{O}(\alpha'^4)$. Then in \sect{sec:ActionDC} we take the opportunity to discuss the application of symmetric-structure double copy to cancel these anomalies, along with their relationship to higher-spin modes as studied in a recent work by the authors. Finally, in \sect{sec:EOpsCounting} we lay out a Hilbert series framework for counting evanescent operators at general orders in mass-dimension. To conclude, in \sect{sec:Discussion} we discuss many directions of future work and summarize the insights gained from this study.

\section{Review} 
\label{sec:Review}
\subsection{Color-dressed and ordered amplitudes}
Here we provide an overview of the amplitudes nomenclature and organizational principles we use throughout the work. When working with scattering amplitudes, $\mathcal{A}$, in an on-shell framework it is convenient to introduce the following graphical description at general multiplicity, $n$, and loop order, $L$,
\begin{equation}
\label{fullAmp}
\mathcal{A}_{n,L} = \int \prod_{i=1}^L\frac{d^D l_i}{(2\pi)^D} \sum_{g} \frac{1}{S_g}\frac{\mathcal{N}_g}{D_g}\,.
\end{equation}
As written above, $S_g$ are the internal symmetry factors for a Feynman graph, $g$, the propagator structure is captured by the denominator function, $D_g$, and the theory-dependent interaction vertices determine numerator functions, $\mathcal{N}_g$. As written, the numerator functions carry all non-propagator kinematic information and any color-data that would be encoded in the interaction Feynman rules of the theory. By color-data we mean the typical weighting of feynman diagrams by the representation of the particles described.  Most of the theories we describe are defined in the adjoint, so the color-data involves the usual dressing of vertices with  antisymmetric structure constants, $f^{abc}$.  Although, as we will see, it will be useful to describe certain counterterms in terms of symmetric $d^{abc}$ color-weights,

For  gauge theories, it is often useful to further decompose the full, or color-dressed, amplitudes of \eqn{fullAmp} into purely kinematic building blocks where the color information is stripped away. For example, at tree level, gauge theory $n$-point amplitudes can be re-expressed in terms of a trace basis decomposition as follows:
\begin{equation}
\mathcal{A}_{n,\text{tree}} = \sum_{\sigma \in S^{n-1}} \text{Tr}(T^1 T^{\sigma(2)}\cdots T^{\sigma(n)}) A_n(1,\sigma(2),...,\sigma(n))\,,
\end{equation}
where the sum is taken over the inequivalent orderings of group theory generator traces. A similar decomposition applies to one-loop gauge theory amplitudes, which we will describe in \sect{sec:Loops}. The color-ordered functions, $A_n$, are called {partial amplitudes}, as they only contain on-shell information for a particular color ordering. However, like full amplitudes, they factor on poles to lower-point partial amplitudes. For vector theories partial amplitudes are independently gauge invariant since they weight linearly independent color traces. 

If we consider the kinematic gauge-invariant ordered amplitudes themselves, many gauge theories reveal hidden redundancy. The $(n-1)!$ partial amplitudes for a large catalog \cite{BCJreview} of adjoint gauge theories, including Yang-Mills and NLSM, are linearly related to a smaller set of basis amplitudes. The first such set of partial amplitude relations identified by Kleiss and Kluijf \cite{Kleiss:1988ne}, known formally as KK relations, relates amplitudes different color orderings and signature to a basis of $(n-2)!$ partial amplitudes:
\begin{equation}
\label{KK}
A(1,\alpha,n,\beta)=(-1)^{|\alpha|}\sum_{\sigma\in \alpha \shuffle \beta^T} A(1,\sigma,n)\,,
\end{equation}
where $\beta^T$ denotes the inverse ordering of $\beta$, and $\shuffle$ is the shuffle product that runs over ordered permutations of $\alpha$ and $\beta^T$. More recently, Bern, Johansson and one of the authors, (BCJ) further identified a set of kinematic relations that reduced the size of $n$-point partial amplitude basis down to $(n-3)!$ via BCJ relations \cite{BCJ}:
\begin{equation}\label{BCJ}
\sum_{i=2}^{n-1}k_1\cdot(k_2+\cdots +k_i) A(2,...,i,1,...,n)=0\,.
\end{equation}
Underlying both sets of amplitude relations are hidden graphical  principles that manifest the amplitude level redundancy of \eqn{KK} and \eqn{BCJ}, which we  describe in the next section. Before moving on, we emphasize that everything described thus far applies in arbitrary spacetime dimensions. As such, these building-blocks are well suited for constructing loop integrands compatible with  dimensional regularization.
\label{subsec:AmpReview}

\subsection{Color-Kinematics Duality and the Double-Copy}
\label{subsec:DCReview}
To apprehend this graph-based structure for color-dressed theories, we first rewrite the full amplitude of \eqn{fullAmp} as a sum over {cubic graphs}, as follows:
\begin{equation}\label{fullAmpCubic}
\mathcal{A}_{n,L} = \int \prod_{i=1}^L\frac{d^D l_i}{(2\pi)^D} \sum_{g\in \Gamma^{(3)}_{n,L}} \frac{1}{S_g}\frac{C_gN_g}{D_g}\,,
\end{equation}
where $N_g$ and $C_g$ are the kinematic numerators and color factors, respectively, and $ \Gamma^{(3)}_{n,L}$ denotes the set of cubic diagrams at $n$-point $L$-loop. In this formulation of the amplitude, contact diagrams are absorbed into cubic graphs by multiplying by inverse propagators. The procedure of assigning contact diagrams to cubic graphs is referred to as {generalized gauge freedom} \cite{BCJ}. In practice, there are many different ways to assign contact diagrams to cubic graphs. 

The gauge theories that we study in this paper have sufficient generalized gauge freedom to assign contact diagrams to kinematic numerators such that at every multiplicity at tree level one can write the amplitude in terms of $N_g$ that obey the same algebraic relations as the color factors, $C_g$. For color factors that are composed of adjoint structure constants, $f^{abc}$ they must be anti-symmetric around vertices, and be related by Jacobi relations about each edge.   For $SU(N_c)$ we can take $f^{abc} \propto \text{Tr}(T^a[T^b,T^c])$ for arbitrary representation $T$ of the gauge group.

We will also consider generalization beyond the adjoint to color-weights that can include $d^{abc}$ and associated algebraic relations.   For $SU(N_c)$ we can take $d^{abc} \propto \text{Tr}[T^a\{T^b,T^c\}]$.  Symmetric-structure color factors obey symmetric algebraic relations rather than anti-symmetric ones.

When a gauge theory can be expressed in a form such that the color and kinematic factors obey the same algebraic relations, we call such theories \textit{color-dual}.   For ordered amplitudes, the ability to find kinematic numerators that obey antisymmetry and Jacobi relations is one-to-one with the ordered amplitudes satisfying KK and BCJ relations, respectively.  

An important consequence of the duality between color and kinematics is the  ability to double-copy color-dual numerators with each other~\cite{BCJ}, while simultaneously preserving factorization and gauge invariance of scattering amplitudes.  Because cubic-graph color-weights are not linearly independent, gauge-invariance of the full amplitude is encoded in algebraic relation between the color factors. Thus, equipped with a set of color-dual numerators, $\tilde{N}_g$, that obey the same algebraic relations as $C_g$, we can make the simple replacement, $C_g \rightarrow \tilde{N}_g$, giving a new gauge invariant amplitude, $\mathcal{M}$, as follows:
\begin{equation}\label{fullAmpCubicDouble}
\mathcal{M}_{n,L} = \int \prod_{i=1}^L\frac{d^D l_i}{(2\pi)^D} \sum_{g\in \Gamma^{(3)}_{n,L}} \frac{1}{S_g}\frac{\tilde{N}_gN_g}{D_g}\,.
\end{equation}
Note that $\mathcal{M}$ is colorless. It satisfies manifest gauge invariance on the $N_g$ side of the double copy due to the algebraic properties of $\tilde{N}_g$. In the event that $\tilde{N}_g$ also belongs to a vector theory, then $\mathcal{M}$ describes amplitudes in a gravitational theory. In such a case, the gauge invariance of constituent $\tilde{\mathcal{A}}$ and $\mathcal{A}$ in \eqn{fullAmpCubic} conspire to generate linearized diffeomorphism invariance in \eqn{fullAmpCubicDouble}.  As long as both kinematic factors obey the same algebraic constraints as the color algebra, this double copy construction works for integrands at the multi-loop level since it globally manifests color-kinematics on all possible unitarity cuts~\cite{BCJ}.

Throughout the text we will reserve $\mathcal{M}$ to denote double-copy amplitudes.  As shorthand, when the kinematic weights of two theories, $X$ and $Y$, participate in the double copy construction of $\mathcal{M}^{XY}$ we will use the outer-product to mean double copy construction as in,
\begin{equation}\label{doubleCopyProd}
\mathcal{M}^{\text{XY}} = X \otimes Y \equiv \int \prod_{i=1}^L\frac{d^D l_i}{(2\pi)^D} \sum_{g\in \Gamma^{(3)}_{n,L}} \frac{1}{S_g}\frac{N^X_gN^Y_g}{D_g}\,.
\end{equation}
At tree-level, such double-copy construction in the adjoint case can be understood equivalently \cite{Bern:1998sv,Vaman:2010ez,Bjerrum-Bohr:2010diw,Bjerrum-Bohr:2010kyi,Bjerrum-Bohr:2010pnr,Bjerrum-Bohr:2012kaa} in terms of a KLT or momentum kernel matrix, $S(a|b)$, and a BCJ spanning set of ordered amplitudes for each of the $X$ and $Y$ theories,
\begin{equation}
 X \otimes Y = \sum_{a,b \in S^{n-3}_{(2,\ldots,n-2)}} A^X(1,\{a\},n,n-1) S(a|b) A^Y(1,\{b\},n-1,n)\,.
\end{equation}
The sum runs over both sets, $a$ and $b$, of $(n-3)!$ distinct ordered amplitudes. No such momentum kernel has yet been constructed for the symmetric case~\cite{Carrasco:2022jxn} .

As we will see, double-copy construction between symmetric color-dual kinematic numerators~\cite{Carrasco:2022jxn} is quite natural for capturing Born-Infeld counterterms needed for anomaly cancellation beyond one loop. Indeed, many of the higher derivative counterterms captured by double-copying symmetric kinematic factors are inaccessible\footnote{As identified in Ref.~\cite{Carrasco:2022jxn} they can technically be described in terms of adjoint double-copy but between presumably unphysical theories that require factorization involving higher-spin exchange.} to the traditional double copy between adjoint kinematics. We will describe this non-adjoint color-dual double-copy in more detail in \sect{sec:ActionDC}, where we will explicate the tension between local symmetric numerators and higher-spin adjoint numerators.

\subsection{4D Spinor Helicity vs. $D$-dimensions}
\label{subsec:4DandDDReview}
All the amplitude methods we have discussed thus far are valid in arbitrary dimension. This has a number of advantages that we will touch upon when reviewing integrand construction towards dimensionally regulated amplitudes in \sect{sec:genU}. While most of our calculations will be carried out completely agnostic to the spacetime dimension, the vector amplitudes we compute will have 4D symmetries that hidden in $D$-dimensional kinematics. It is useful therefore to also consider explicitly four-dimensional kinematics. 

When working with bosonic kinematics in arbitrary dimensions, we will employ formal Lorentz covariant polarizations, $\varepsilon^\mu_a$, and momenta, $k^\mu_a$. The mostly minus signature will be used throughout. The only restrictions we will place on physical momenta and polarizations are the standard on-shell constraints: momentum conservation and the null-momenta condition,
\begin{equation}
\sum_{a} k_a^\mu =0\,, \qquad \qquad k_a^2 =0\,.
\end{equation}
Furthermore, we will take external polarizations $\varepsilon_a$ formal and transverse throughout:
\begin{equation}
\varepsilon_a \cdot k_a = 0\,.
\end{equation}
Dimensions will generically be away from four, in $D=4-2\epsilon$, with $\epsilon$ arbitrary. As noted, it will be convenient to take $D\rightarrow 4$ in certain circumstances after integration.  In such cases we will employ appropriate spinor-helicity variables.  Here we apply the same conventions of Ref.~\cite{jjmcTASI2014}, which we quote now. For massless momenta, $k_a$ and $k_b$, we have 
\begin{equation}
s_{ab} = (k_a+k_b)^2= \langle ab \rangle[ba]\,,
\end{equation}
the component definition of our spinor bracket that are consistent with the conventions above,
\begin{align}
\langle ab \rangle &= \frac{(a_1 + i a_2)(b_0+b_3)-(b_1 + i b_2)(a_0+a_3)}{\sqrt{(a_0+a_3)(b_0+b_3)}}\,,
\\
[ab] &= \frac{(b_1 - i b_2)(a_0+a_3)-(a_1 - i a_2)(b_0+b_3)}{\sqrt{(a_0+a_3)(b_0+b_3)}}\,,
\end{align}
where the $a_i$ are component values of the four-vector, $k^\mu_a = (a_0,a_1,a_2,a_3)$. Four-dimensional polarization dot products with fixed helicity states can be mapped as follows:
\begin{equation}\label{eqn:4DPols}
\begin{aligned}
k_a \cdot \varepsilon_b^{(+)} &= \frac{\langle q a \rangle[ab]}{\sqrt{2}\langle q b\rangle}\,,
\qquad\quad \qquad
k_a \cdot \varepsilon_b^{(-)} = -\frac{[qa]\langle ab\rangle}{\sqrt{2}[qb]}\,,
\\
\varepsilon_a^{(-)}\cdot \varepsilon_b^{(+)} &= - \frac{\langle q a\rangle [qb]}{ [qa]\langle q b\rangle} \,,
\qquad \qquad
\varepsilon_a^{(\pm)}\cdot \varepsilon_b^{(\pm)} = 0 \,,
\end{aligned}
\end{equation}
where $q^\mu$ is some null reference momentum that all polarizations are projected along. While all four-dimensional equivalence relayed in this paper can be achieved analytically, in practice it is often much more  convenient to verify equivalence numerically.

\subsection{One-loop integral basis and tensor reduction}
\label{sec:1loopMethods}

It is well known that one-loop amplitudes are spanned by a small basis of scalar integrals \cite{Forde:2007mi,Badger:2008cm,ElvangHuangReview}. This property can be exposed via a $D$-dimensional integral reduction algorithm due to Passarino and Veltman \cite{Passarino:1978jh}. At one-loop, the basis of irreducible scalar products (ISPs) that include loop momenta grows in lock step with the number of propagators. Explicitly, an $N$-gon integral can have $N-1$ factors of $(k_i\cdot l)$, due to momentum conservation, and exactly one factor of $l^2$; this matches the number of $N$-gon propagators. 

Furthermore, factors of $(\varepsilon_i\cdot l)$ can be mapped to tensor structures of external kinematics with factors of $(\varepsilon_i \cdot k_j)(k_i  \cdot l)/(k_i \cdot k_j)$ using polarization completeness relations~\cite{Bern:2017tuc} . Thus, any $N$-gon tensor integral permits a partial fraction decomposition in terms of inverse propagators, and thus can be mapped to a scalar basis of integrals:
\begin{equation}\label{TenRed1loop}
T_{\mu_1\mu_2...\mu_r} I_N^{\mu_1\mu_2...\mu_r} = \sum_{M=1}^N C_M I_M\,,
\end{equation}
where $T_{\mu_1\mu_2...\mu_r}$ and $C_M$ are functions exclusively of external kinematics. This relationship holds $D$-dimensionally for any $r$-rank $N$-gon one-loop integral. Thus, any one-loop amplitude can be expressed completely in terms of a basis of one-loop scalar integrals. This special property of one-loop integration is {not} the case for generic loop order, where the basis of ISPs grows faster than number of inverse propagators. As such, any universal integral basis beyond one-loop must include integrals with non-unit powers of the propagators.  

The remarkable simplicity of one-loop amplitudes is even more dramatic in a fixed spacetime dimension. In $D=4$, the kinematic projection to a basis of scalar integrals saturates at the box integral due to the appearance of Gramm determinants. Roughly speaking, $N$-gon integral coefficients of \eqn{TenRed1loop} come dressed with a factor of $\mathcal{G}_N$,
\begin{equation}
\mathcal{G}_N = \text{det} (k_i \cdot k_j)\,,
\end{equation}
where $k_i$ are the external momenta flowing into the $N$-gon integral. In fixed spacetime dimension, $D$, the momenta are $D$-component vectors and thus the Lorentz product matrix $(k_i \cdot k_j)$ must have a null space for $N>D$. As a result, $C_M \neq 0$, only when $M\leq D+1$. Furthermore, in $D=4-2\epsilon$ dimensions, any scalar pentagon integral can be rewritten up to $\mathcal{O}(\epsilon)$ in terms of five pinched scalar box integrals \cite{Bern:1993kr}. All the information of one-loop 4D amplitudes can therefore be completely determined by evaluating up to box integrals, regardless of the multiplicity. We will demonstrate in the next section how this can be used in 4D integral construction.  

The last important feature of one-loop basis integrals is that many of them can be evaluated in arbitrary dimension for a very general set of parameters appearing in the exponents of the propagators. Take for example the $D$-dimensional massless triangle and bubble integrals \cite{Smirnov:2004ym} that we will use throughout the text:
\begin{equation}\label{eq:integrals}
{
\begin{aligned}
I_{2,(K)}^{(\alpha_1,\alpha_2)}&=\int \frac{d^{D} l}{(2\pi)^{D}} \frac{1}{[l^2]^{\alpha_1}[(l+K)^2]^{\alpha_2}} 
\\
&=i\left[-\frac{K^2}{4\pi}\right]^{D/2}\frac{\Gamma(D/2-\alpha_1)\Gamma(D/2-\alpha_2)\Gamma(\alpha_{12}-D/2)}{[K^2]^{\alpha_{12}}\Gamma(\alpha_1)\Gamma(\alpha_2)\Gamma(D-\alpha_{12})} \,,
 \\\\
I_{3,(K_{12})}^{(\alpha_1,\alpha_2,\alpha_3)}&= \int \frac{d^{D} l}{(2\pi)^{D}} \frac{1}{[l^2]^{\alpha_1}[(l+K_1)^2]^{\alpha_2}[(l+K_{12})^2]^{\alpha_3}} 
\\
&=i\left[-\frac{K_{12}^2}{4\pi}\right]^{D/2}\frac{\Gamma(D/2-\alpha_{12})\Gamma(D/2-\alpha_{23})\Gamma(\alpha_{123}-D/2)}{[K_{12}^2]^{\alpha_{123}}\Gamma(\alpha_1)\Gamma(\alpha_3)\Gamma(D-\alpha_{123})} \,,
\end{aligned}
}
\end{equation}
where we have introduced the subscript notation for indexed sums $X_{12...n} = X_1+X_2+ \cdots +X_n$. Throughout the text we will suppress the regularization scale that appears in the argument of the logarithms as they can be restored by dimensional analysis. The above integral expressions hold for massless external kinematics, $K_i^2=0$, which applies to all the integrals needed for the physical processes described in the text. 

The existence of closed form expressions, like those in \eqn{eq:integrals}, while common for one-loop integrals, are incredibly rare for higher loop order outside of very specialized kinematic regimes. Luckily, the integral basis at four-point two-loop order for even-point theories can be reconstructed from the one-loop expressions in \eqn{eq:integrals}. This is a special property of what we refer to as \textit{recursively one-loop} amplitudes. We can exemplify this property in the simple context of scalar $\varphi^{2k}$ theory.

Consider the interactions needed to construct two loop amplitudes for the generic arbitrarily weighted $\varphi^{2k}$-scalar theory:
\begin{equation}\label{eq:evenPointL}
\mathcal{L}_{\varphi^{2k}} = \frac{1}{2}(\partial \varphi)^2 + c_4 \varphi^4 + c_6 \varphi^6+ c_8 \varphi^8+\cdots 
\end{equation}
 The two-loop amplitude for $\varphi^{2k}$ theory is  simply generated by evaluating the following set of scalar 1-particle-reducible (1PI) integrals:
\begin{equation}
\label{eq:2loopEven}
\begin{aligned}
\mathcal{M}_{\varphi^{2k}}^{\text{2-loop}}  &= \frac{c_4^3}{4}\scaleIntCscalar{}{}{}{} +  \frac{c_4^3}{2}\scaleIntBscalar{}{}{}{} + \frac{c_8}{4}\scalelessIntAscalar 
\\
&+  \frac{c_4 c_6}{6}\scalelessIntBscalar+\text{perms}(1,2,3,4)
\end{aligned}
\end{equation}
\begin{figure}[t]
    \centering
    \begin{equation*}\Omega_{(2,4)} = \left\{\scaleIntBscalar{}{}{}{},\scaleIntCscalar{}{}{}{} \right\}\end{equation*}
    \caption{Recursive one-loop integrals contributing to the  two-loop four-point calculation in even-point effective field theories, we refer to as the ostrich integral and double-bubble from left to right.}
    \label{fig:contrib2Loop}
\end{figure}
Since the external states are massless, only the first two integrals, those appearing in \Fig{fig:contrib2Loop}, are non-vanishing in dimensional regularization -- the rest are scaleless. 
Therefore the above even-point perturbative predictions of $\varphi^{2k}$ are equivalent to $\varphi^4$ theory at four-point through two-loop order: 
\begin{equation}
\mathcal{M}_{\varphi^{2k}}^{\text{2-loop}} \equiv \mathcal{M}_{\varphi^{4}}^{\text{2-loop}} \,.
\end{equation}
In addition to being perturbatively equivalent to the quartic sector of the theory at two-loop, we can see above that the integrals for the Feynman diagrams in \eqn{eq:2loopEven} are {recursively one-loop}. In other words, to integrate the full amplitude at two-loop, we only need the iterated basis of one-loop integrals. This observation comes with the added advantage of permitting the application of one-loop Passarino-Veltman tensor reduction \cite{Passarino:1978jh} to two loop integrals. This property of recursively one-loop integrals will dramatically simplify the higher derivative even-point theories that we consider.  

\subsection{Even-point Effective Field Theories}
\label{subsec:EPEFTReview}
In this section we provide a catalog of the even-point effective field theories studied throughout the paper. 
Some of the theories we consider, like Volkov-Akulov and supersymmetric extensions of Born-Infeld, have natural formulations in terms of veirbeins that depend on fermionic degrees of freedom. For such theories, it is useful to distinguish between the coordinate independent actions, $S$, and the coordinate-dependent Lagrangian densities, $\mathcal{L}$,
\begin{equation}
S= \int d^D x \,\mathcal{L}\,.
\end{equation}
\paragraph{Nonlinear Sigma Model}
The simplest effective field theory that we will consider is the {nonlinear sigma model} (NLSM). The chiral NLSM Lagrangian density with diagonal symmetry group $G=SU(N_c)$ can be expressed in terms of the chiral current, $j_\mu = U^\dagger \partial_\mu U$, as follows,
\begin{equation}\label{eq:NLSMLag}
\mathcal{L}^{\text{NLSM}}= \frac{1}{2}\text{tr}[(\partial_\mu U)^\dagger (\partial ^\mu U)] = \frac{1}{2}\text{tr}\left[\frac{(\partial_\mu\pi )( \partial^\mu \pi) }{(1-f_\pi^{-2}\pi^2)^2}\right]\,,
\end{equation}
where the trace is over the color indices of the gauge group, $\pi \equiv \pi^a T^a$. On the right hand side we have applied the Cayley parameterization to the $SU(N_c)$ group elements, $U= \frac{1+\pi/f_\pi}{1-\pi/f_\pi}$, where $f_\pi$ is the dimensionful pion decay width. By construction, this model has a left-right global symmetry, $G\times G$, that is nonlinearly realized, and a linearly realized diagonal subgroup, $G$. This theory is the unique two-derivative EFT that is invariant under constant shifts of the pion field in color space:
\begin{equation}\label{eq:shiftSym}
\pi^a \rightarrow \pi^a + c^a\,.
\end{equation}
We could also consider higher derivative generalizations of the NLSM that are invariant under the shift symmetry of \eqn{eq:shiftSym}.  Such  EFTs correspond to the so-called chiral limit of {chiral perturbation theory}~\cite{Weinberg:1978kz,Gasser:1983yg,Kampf:2006yf,Kampf:2013vha}, describing massless pion scattering below the chiral symmetry breaking scale of QCD:
\begin{equation}\label{eq:XPTLag}
\mathcal{L}^{\chi\text{PT}}=\frac{1}{2}\text{tr}[(\partial_\mu U)^\dagger (\partial ^\mu U)] + \frac{\beta_1}{f_\pi^2} \text{tr}[(\partial_\mu U)^\dagger (\partial ^\mu U)]^2 + \frac{\beta_2}{f_\pi^2} \text{tr}[(\partial_\mu U)^\dagger( \partial_\nu U)]\text{tr}[(\partial^\mu U)^\dagger( \partial^\nu U)] +\cdots
\end{equation}
By construction, each of the higher derivative operators appearing a bove are invariant under the shift symmetry of \eqn{eq:shiftSym}. These higher derivative generalizations of NLSM have been studied both in the context of the $S$-matrix bootstrap \cite{Manohar:2008tc,Bellazzini:2016xrt,Guerrieri:2020bto}, and soft theorem bootstraps of Refs. \cite{Cheung:2014dqa,Cheung:2015ota,Cheung:2016drk,Cheung:2018oki,Low:2019ynd}. Given the relevance of \eqn{eq:XPTLag} in Standard Model pion scattering, the amplitudes for $G= SU(2)$ isospin pions have been computed through two-loop order \cite{Bijnens:1995yn,Bijnens:1997vq,Girlanda:1997ed}. In this work, we add to the literature by computing the two-loop amplitudes of \eqn{eq:NLSMLag} for arbitrary $SU(N_c)$ color structure. We then use these amplitudes to compute $\mathcal{N}=4$ DBIVA amplitudes via the double copy in \sect{sec:DBIvDC}.

In addition to the chiral NLSM described above, our methods apply equally well to a target space model describing pion dynamics on a coset manifold $G/H$, with a nonlinearly realized global symmetry $G$ broken down to a linearly realized isometry group $H$. Recent literature that study the universal soft behavior of these models can be found in Refs.~\cite{Low:2014nga,Low:2017mlh,Liu:2018vel,Low:2018acv}. One such model that we will study in the text is the $\mathbb{CP}^N$ model where the global $SU(N+1)$ is spontaneously broken down to a local $U(N)$ symmetry on the target space. The Lagrangian for the theory we provide below:
\begin{equation}\label{eq:CPNLag}
\mathcal{L}^{\text{NLSM}}_{\mathbb{CP}^N} = \frac{1}{2}P(z,\bar{z})^{ij}(\partial_\mu \bar{z}_i)(\partial^\mu z_j)=\frac{1}{2}\frac{(f_\pi^2+\bar{z} z)\delta^{ij}-z^i\bar{z}^j}{(f_\pi^{2}+\bar{z} z)^2}(\partial_\mu \bar{z}_i)(\partial^\mu z_j)\,,
\end{equation}
where $z=(z_1,z_2,...,z_N)$ is a complex vector in the fundamental representation of $U(N)$, and $P(z,\bar{z})$ is the Fubini-Study metric on $\mathbb{CP}^N$. The full $SU(N+1)$ symmetry of the theory is realized due to the imbedding of $\mathbb{CP}^N$ in a complex space one dimension higher, $\mathbb{C}^{N+1}$. In \sect{sec:NLSMU}, we will demonstrate through explicit calculation that the leading IR divergences of the $\mathbb{CP}^1$ model exponentiates in $D=2$, the critical dimension of the theory. 
\paragraph{Born-Infeld Photons and Dirac Scalars}
The second even-point EFT we will study is Born-Infeld theory \cite{Born:1934gh}. This theory describes the dynamics of open string endpoints with $U(1)$ charge attached to a $D$-dimensional spacetime with Dirichlet boundary conditions. Given the stringy dynamics orthogonal to the spacetime, the electric field generated by the charged endpoints has a maximum field strength of order the inverse string tension, $1/\alpha'$. The effective Lagrangian for this theory
\begin{equation}
 \label{biLag}
  \alpha'^{\,2}\mathcal{L}^{\text{BI}} = 1-\sqrt{\text{det}(\eta_{\mu\nu}+\alpha' F_{\mu\nu})} \,.
\end{equation}
A close cousin of Born-Infeld theory is the dimensionally reduction to Dirac scalars, which describes a $D$-dimensional spacetime propagating in a $(D+1)$-dimensional background:
\begin{equation}
 \label{DBI}
  \alpha'^{\,2}\mathcal{L}^{\text{DBI}} =1- \sqrt{\text{det}(\eta_{\mu\nu}+\alpha'^{\,2} \partial_\mu \varphi \partial_\nu \varphi)}\,,
\end{equation}
where $\varphi$ is the spacetime coordinate in the orthogonal space. This theory is uniquely determined by considering the most general polynomial  of $X=(\partial \varphi)^2$, $P(X)$,  constrained to be invariant under the field redefinition \cite{Cheung:2016drk}:
\begin{equation}\label{DBIShift}
\varphi \rightarrow \varphi + c + b^{\mu} (x_\mu + \varphi \partial_\mu \varphi)\,.
\end{equation}
When the transformation law leaves the polynomial $P(X)$ theory invariant, there is an additional set of $D+1$ conserved charges, and the Poincar\'{e} symmetry of $P(X)$ EFT is promoted from $D$ to $(D+1)$ dimensions. One can verify that applying \eqn{DBIShift} to \eqn{DBI} shifts the Lagrangian by a total derivative~\cite{deRham:2010eu}. Similar to the soft theorem constraints on the building blocks of NLSM and $\chi \text{PT}$, the DBI action can be bootstrapped by requiring the amplitudes vanish at $\mathcal{O}(p^2)$ when taking external momentum, $p$, soft~\cite{Cheung:2016drk}.

\paragraph{Volkov-Akulov Fermions and Supersymmetry}
Now we will describe the supersymmetric extensions of the Born-Infeld action given above in \eqn{biLag}. First we start by defining the theory of shift symmetric fermions first written down by Volkov, Akulov (VA)~\cite{Volkov:1973ix}. The easiest way to define this shift symmetric VA theory is by constructing a volume form in the veirbein frame out of manifestly supersymmetric 1-forms. That is, define a 1-form, $ \omega^m = e^m_\mu dx^\mu$, where the veirbeins (frame fields), $e^m_\mu$, are expressed in terms of the fermion fields, $\lambda$, as follows:
 \begin{equation}
 e^m_\mu = \delta^m_\mu + i \bar{\lambda} \Gamma^m\! \stackrel{\leftrightarrow}{\partial_\mu}\!  \lambda = \delta^m_\mu + i (\bar{\lambda} \Gamma^m \partial_\mu  \lambda - \partial_\mu \bar{\lambda} \Gamma^m   \lambda)\,.
 \end{equation}
 This frame field is manifestly invariant under the superspace shift:
\begin{equation}
\delta \lambda \rightarrow \eta \qquad  \delta \bar{\lambda} \rightarrow \bar{\eta} \qquad  \delta x^\mu \rightarrow x^\mu - i (\bar{\lambda} \,\Gamma^\mu \eta- i \bar{\eta} \,\Gamma^\mu \lambda)\,.
\end{equation}
 The Volkov-Akulov action in the veirbein frame is then simply the fermionic world-volume integral over the $D$-form \cite{Kallosh:1997aw},
 \begin{equation}
 S^{\text{VA}} = \int \omega^1 \wedge \omega^2 \wedge \cdots \wedge \omega^D\,.
 \end{equation}
 This is not dissimilar from the construction of the pure scalar DBI action above. Likewise, we can write the Lagrangian in the spacetime coordinate frame in parallel to \eqn{biLag} and \eqn{DBI}. Transforming back into spacetime coordinates yields the following Lagrangian density for self interaction VA fermions upto four-point interactions:
\begin{equation}
 \label{vaLag}
  \alpha'^{\,2}\mathcal{L}^{\text{VA}} = \text{det}(e^m_\mu) =  i\bar{\lambda}\!\stackrel{\leftrightarrow}{\slashed{\partial}}\!\lambda+ \frac{1}{2}(\bar{\lambda}\Gamma^\mu \partial^\nu \lambda)(\bar{\lambda}\Gamma_\mu \partial_\nu \lambda)+\mathcal{O}(\partial^4\lambda^6) 
  \,.
\end{equation}
Furthermore, a convenient choice of field redefinition was proposed by Komargodski and Seiberg (KS) \cite{Komargodski:2009rz,Kuzenko:2010ef}, which defines a new fermion field, $\psi = \psi(\lambda,\bar{\lambda})$, in a way that is perturbatively equivalent to the VA fermions. This construction is equivalent to a theory of nilpotent chiral superfields \cite{Rocek:1978nb,Casalbuoni:1988xh,Ferrara:2014kva}, relevant for the construction of $\alpha$-attractor models of inflation \cite{Kallosh:2013yoa}. Applying the KS nonlinear field redefinition yields the considerably simpler Lagrangian for VA theory:
 \begin{equation}
   \alpha'^{\,2}\mathcal{L}^{\text{VA}} = i\bar{\psi}\slashed{\partial}\psi +\frac{1}{2} \bar{\psi}^2 \partial^2 \psi^2 + \frac{1}{4}\psi^2\bar{\psi}^2 \partial^2 \psi^2\partial^2 \bar{\psi}^2\,.
 \end{equation}
With the world-volume formulation of VA theory in hand, we can also trivially construct the action for maximal $\mathcal{N}=1$ supersymmetric Born-Infeld theory in $D=10$ \cite{Tseytlin:1999dj}. We simply replace the flat space background of \eqn{biLag} with the fermionic background of \eqn{vaLag},
\begin{equation}
 \label{DBIVALag}
S^{\text{DBIVA}}_{\mathcal{N}=1} = \int \omega^1 \wedge\omega^2 \wedge \cdots \wedge \omega^{10} \sqrt{\text{det}(\eta_{mn}+ \alpha F_{mn})}\,.
\end{equation}
It is worth noting that the fermion-vector interaction is introduced via the {non-minimal} coupling between the field strength and the fermionic spacetime metric \cite{Bergshoeff:2013pia}. This differs from the linearly realized supersymmetry of super-Yang-Mills theory, which minimally couples the vector mode to the chiral fermion. This non-minimal coupling is a requirement of supersymmtry due to the even-point nature of abelian vector theories like Born-Infeld.

The action above in \eqn{DBIVALag} can be similarly re-expressed in terms of our canonical spacetime coordinates, giving us the following flat-space Lagrangian for maximally supersymmetric Born-Infeld theory in $D=10$: 
\begin{equation}
 \label{dbivaLag}
    \alpha'^{\,2}\mathcal{L}^{\text{DBIVA}}s = \sqrt{\text{det} \left(\eta_{\mu\nu}- \alpha' F_{\mu\nu} + \alpha'^{\,2} (\bar{\lambda} \Gamma_\mu \partial_\nu \lambda)- \alpha'^4(\bar{\lambda} \Gamma^\rho \partial_\mu \lambda )(\bar{\lambda} \Gamma_\rho \partial_\nu \lambda) \right)}
  \,.
\end{equation}
To recover the $D=4$ theory, we simply dimensionally reduce in a way that preserves the maximum number of supercharges, which is $\mathcal{N}=4$ in $D=4$. Implementing a dimensional reduction that preserves the supercharges of \eqn{dbivaLag} is rather straightforward in an on-shell framework:
\begin{equation}
\mathcal{M}^{\text{DBIVA}} = A^{\text{NLSM}} \otimes A^{\text{sYM}}\,.
\end{equation}
Since the double copy construction holds $D$-dimensionally at tree-level \cite{BCJreview}, double-copying maximal sYM with NLSM will yield the analogous spectrum of maximally supersymmetry DBIVA theory, whether in $D=10$, as described above, and in $D=4$. 
\paragraph{Abelian Open String}
Similar to the amplitudes of DBIVA theories described above, the tree-level amplitudes for the open superstring (OSS)~\cite{Mafra:2011nv,Mafra:2011nw} likewise permit a double copy construction \cite{Broedel:2013tta}. To construct open-superstring amplitudes we simply double copy maximally supersymmetric super-Yang-Mills (sYM) and Chan-Paton dressed $Z$-theory amplitudes~\cite{Carrasco:2016ldy,Carrasco:2016ygv,Mafra:2016mcc}, as follows,
\begin{equation}
A^{\text{OSS}}= Z \otimes A^{ \text{sYM}}\,,
\end{equation}
where $\otimes$ implements the field theory double copy \cite{KLT, BCJ} defined in \eqn{doubleCopyProd}. Moreover, the amplitudes of DBIVA emerge from the above construction as the field theory limit of the {abelian open superstring} \cite{Green:1982sw}. To recover the abelian sector, one simply sums over all Chan-Paton color orderings on the side of the bi-colored $Z$-theory amplitudes that obey string monodromy relations. In the field theory limit, the observables for so-called {abelian} $Z$-theory are simply the amplitudes generated by the NLSM Lagrangian \cite{Carrasco:2016ldy}. That is, given a $n$-point amplitude in $Z$-theory, with field theory ordering $a=(a_1,a_2,...,a_n)$ and string theoretic color ordering $A=(A_1,A_2,...,A_n)$, one finds the following relation:
\begin{equation}
A^{\text{NLSM}}{(a_1,a_2,...,a_n)} \equiv \lim_{\alpha' \to 0} (\alpha')^{2-n}  Z_{\times}{(a_1,a_2,...,a_n)}\,,
\end{equation}
where $Z_{\times}$ are abelian $Z$-theory amplitudes of \cite{Carrasco:2016ldy}, defined by summing all possible orderings of Chan-Paton factors:
\begin{equation}
Z_{\times}(a_1,a_2,...,a_n)\equiv \sum_{A \in S^{n-1}} Z_{(A_1,...,A_{n-1},n)}{(a_1,a_2,...,a_n)}\,.
\end{equation}
Since the field theory limit of abelian $Z$-theory produces NLSM amplitudes, the field theory limit of the abelian OSS, gives rise to DBIVA observables at leading order in $\alpha'$:
\begin{equation}
\begin{aligned} 
\lim_{\alpha' \to 0 }Z_{\times}&= A^{\text{NLSM}} \otimes A^{\text{BAS}}\equiv A^{\text{NLSM}}\,,
\\
\Rightarrow \quad\lim_{\alpha' \to 0 }A^{\text{OSS}}_{\times}&= A^{\text{NLSM}} \otimes A^{ \text{sYM}}\equiv A^{\text{DBIVA}}\,.
\end{aligned}
\end{equation}
Additional details on the bi-colored $Z$-theory amplitudes these string-theoretic double-copy constructions can be found in \cite{Broedel:2013tta,Carrasco:2016ldy,Carrasco:2016ygv,Mafra:2016mcc,Azevedo:2018dgo} and references therein. We emphasize that while NLSM and DBIVA are the two theories that we will focus on, the methods that we develop in the next section apply much more broadly to the full catalog of even-point effective field theories.  

\paragraph{Duality Invariance and Higher Derivatives}
The four-dimensional amplitudes of DBIVA theories have an additional special property that we will study in his work. At tree-level, the amplitudes generated by \eqn{biLag} and \eqn{DBIVALag} exhibit $U(1)$ \textit{duality invariance} \cite{Bossard:2012xs,Novotny:2018iph}. A theory is said to exhibit duality invariance when the effect of rotating field strengths, $F^{\mu\nu}$, into the dual-fields, $G^{\mu\nu}$, defined below,
\begin{equation}
G^{\mu\nu} = \epsilon^{\mu\nu\rho\sigma}\frac{\partial \mathcal{L}}{\partial F^{\mu\nu}}\,,
\end{equation}
leaves the equations of motion invariant \cite{Gibbons:1995ap,Babaei-Aghbolagh:2013hia}. This dynamical symmetry gives rise to a 4D helicity selection rule \cite{Novotny:2018iph} whereby tree-level matrix elements vanish on-shell outside the aligned-helicity sector,
\begin{equation}
\mathcal{M}^{\text{DBIVA}}_{(n_{-}, n_{+})} =0 \qquad \Leftrightarrow \qquad n_{-}\neq n_{+}\,,
\end{equation}
where $n_{(+)}$ and $n_{(-)}$ is the number of external positive and negative helicity photons, respectively. For $\mathcal{N}=0$ pure Born-Infeld theory \cite{Born:1934gh,Schrodinger:1935oqa}, this symmetry is violated at four-point one-loop in the all-plus $(++++)$ helicity sector \cite{Elvang:2019twd}. In the text, we will show how this anomaly is propagated to two-loop, and demonstrate that $U(1)$ duality invariance is further broken in the $(-+++)$ configuration beyond one-loop. 

Before proceeding, we comment briefly on the higher derivative corrections one might expect to appear in duality invariant effective field theories. In the case of supersymmetric DBIVA at four-points, $U(1)$ symmetry is promoted to an $R$-symmetry, and is thus protected perturbatively to all orders by supersymmetric Ward Identities \cite{Heydeman:2017yww}. We show this explicitly in the case of $\mathcal{N}=4$ DBIVA through two-loop in \sect{sec:DBIU}. However, this supersymmetric enhancement of $U(1)$ duality does not in general apply to higher multiplicity amplitudes. 

While the leading order in $\alpha'$ must be duality invariant at all multiplicity for DBIVA \cite{Pavao:2022kog}, beyond the leading order, $R$-symmetry permits non-vanishing matrix elements outside the split helicity sector. The low energy effective action of the OSS is one such exemplar of an EFT that respects $U(1)$ duality at leading order in $\alpha'$, but produces duality violating matrix elements at higher-orders above four-point~\cite{Carrasco:2016ldy}. 

\subsection{On-shell Unitarity methods}\label{sec:genU}
Above we have reviewed the known behavior of one-loop integral bases in $D=4$ spacetime dimensions. In this section we will briefly demonstrate why traditional 4D unitarity methods are so powerful -- and why we must go beyond them to accommodate the multi-loop calculations of interest here. 
\paragraph{Integrands from 4D Cuts}
Applying one-loop integral reduction in 4D tells us that any one-loop massless amplitude, for which tadpoles are scaleless, can be expressed in terms of box, triangle, and bubble diagrams, with each integral weighted by dimensionally-dependent functions of kinematics:
\begin{equation}\label{eq:4DUn}
\mathcal{A}^{\text{1-loop}}= \sum_{N=2}^4 C_N(D)I_N^D = \sum_{N=2}^4 C_N(4)I_N^D+ \mathcal{R}\,,
\end{equation}
where the scalar basis-integrals, $I_N^D$, are evaluated in $D=4-2\epsilon$, and $\mathcal{R}$ is a rational term that emerges from series expanding around $\epsilon=(4-D)/2$. To construct integral coefficients, $C_N$, we make use of the Optical Theorem for the $S$-matrix, which states that,
\begin{equation}
S^\dagger S = \mathbb{1} \quad \Rightarrow \quad 2\,\text{Im}(T) = T^\dagger T\,,
\end{equation}
where $S=\mathbb{1}+iT$. Applying this relation to \eqn{eq:4DUn} allows us to identify the imaginary part of $\mathcal{A}^{\text{1-loop}}$ with the imaginary parts of each basis integral weighted by the respective 4D coefficient, $C_N(4)$:
\begin{equation}\label{eq:UnMeth1loop}
\text{Im}(\mathcal{A}^{\text{1-loop}}) = \sum_{N=2}^4 C_N(4)\, \text{Im}(I_N^D)\,.
\end{equation}
Using the optical theorem to constrain scattering amplitudes is known as the \textit{unitarity method} \cite{UnitarityMethod, BDKUniarityReview}. In order to isolate each $C_N(4)$ appearing in \eqn{eq:4DUn}, we must place particular internal legs on-shell. That is, for each internal loop propagator that we wish to take on-shell, $1/P_i^2$, we make the following replacement inside the integral:
\begin{equation}
\frac{i}{P_i^2 + i\varepsilon} \rightarrow 2\pi \delta^{(+)}(P_i^2+i\varepsilon)\,.
\end{equation}
By replacing internal propagators with positive-root delta functions, one can isolate the imaginary parts needed to determine the integral coefficients. This approach of taking individual legs on-shell is known as the method of \textit{generalized unitarity} -- and it can be systematically applied to construct all one-loop amplitudes directly from 4D unitarity cuts \cite{Forde:2007mi}. Using generalized unitarity, the integral coefficients correspond to the following set of iterated cuts:
\begin{align}
C^{(P_iP_jP_k)}_4 &\sim \boxCut\,,
\\
C^{(P_iP_j)}_3 &\sim \triCut - \sum_{k} C^{(P_iP_jP_k)}_4\,,
\\
C^{(P_i)}_2 &\sim \bubCut- \sum_{j} C^{(P_iP_j)} - \sum_{j,k} C^{(P_iP_jP_k)}_4\,,
\end{align}
where exposed legs are summed over all internal states crossing the cut, and the labels $P_i$ are external momentum scales flowing into the specified cuts. The subtractions account for the overlapping information in the respective $N$-gon cuts. By momentum conservation, the momentum flow in the unlabeled vertex is completely determined by the other $N-1$ vertices. As a concrete example, consider the 4D cut construction of the bubble coefficient for pure Born-Infeld theory:
\begin{align}\label{eq:4DBICut}
\bubCutStates{$-$}{$-$}{$+$}{$+$} &= \sum_{h_i\in \text{states}} \mathcal{M}^{\text{BI}}(1^-, 2^-, l_2^{h_2},l_1^{h_1})\mathcal{M}^{\text{BI}}(-l_1^{\bar{h}_1},-l_2^{\bar{h}_2},3^+,4^+)
\\
&= \langle12\rangle^2[l_2 l_1 ]^2 \langle l_1 l_2\rangle^2 [34]^2
\\
&= [(l_1+l_2)^2]^2 \langle12\rangle^2 [34]^2
\\
&= s_{12}^2\langle12\rangle^2 [34]^2\,,
\end{align}
where in the last step we have made the replacement $(l_1+l_2) = -(k_1+k_2)$ by momentum conservation. Determining the leading behavior of the 1-loop Born-Infled amplitude thus amounts to specifying all possible 4D cuts. 

While this method is incredible powerful, as we can see from \eqn{eq:UnMeth1loop} it misses a key piece of the amplitude -- namely, the rational terms, $\mathcal{R}$. For some theories like $\mathcal{N}=4$ super-Yang-Mills in $D=4$ \cite{Fusing}, rational terms are absent because the loop power counting for an $N$-gon cut is bounded to be less than $N-2$. However, for generic theories of interest in this paper, like Born-Infeld, rational terms are present and necessary for extracting anomalous matrix elements and higher-loop sub-divergences. While there are methods developed to extract these rational terms at one-loop using $\mu$-integrals \cite{Badger:2008cm}, at generic loop order evaluating these $\mu$-integrals can become  fairly complicated. For this reason, to perform the multi-loop calculations in this paper, we will employ cut construction in general dimensions \cite{Bern:1995db,Bern:1996ja}. 
\paragraph{Integrands from $D$-dimensional Cuts}
The presence of 4D rational terms in the amplitude is due to the appearance of $(D-4)$ factors. By constructing integrands $D$-dimensionally, we can guarantee that our integral coefficients are sensitive to these overall factors. Rather than summing over internal 4D states, we will instead compute $D$-dimensional cuts using the formal polarizations described in \sect{subsec:4DandDDReview}. The $D$-dimensional analog of \eqn{eq:4DBICut} asserts that the one-loop integrand $\mathcal{I}_{\text{1-loop}}$ of pure Born-Infeld theory should satisfy the following constraint:
\begin{equation}
\label{GUCut}
\text{Cut}( \mathcal{I}_{\text{1-loop}}) = \sum_{h_i\in \text{states}} \mathcal{M}^{\text{BI}}_4(l_1^{h_1},-l_2^{\bar{h}_2})\mathcal{M}^{\text{BI}}_4(l_2^{h_2},-l_1^{\bar{h}_1})\,,
\end{equation}  
where $\text{Cut}( \mathcal{I}_{\text{1-loop}}) $ extracts the kinematic numerator by imposing a maximal cut on the one-loop integrand 
\begin{equation}
\text{Cut}( \mathcal{I}_{\text{1-loop}})  \equiv l_1^2l_2^2(\mathcal{I}_{\text{1-loop}})\Big|_{l_1^2,l_2^2\rightarrow 0}\,.
\end{equation}
To carry out the sum over formal polarization states in \eqn{GUCut} we apply the following $D$-dimensional completeness relation for on-shell polarizations,
\begin{equation}\label{stateSum}
\sum_{\text{states}}\varepsilon^\mu_{(l)}\varepsilon^\nu_{(-l)} = \eta^{\mu\nu} - \frac{l^\mu q^\nu+l^\nu q^\mu}{l\cdot q},
\end{equation}
with null reference momentum, $q^2=0$. The dependence of $q$ is a gauge choice that disappears when on-shell kinematic constraints are imposed. At loop level, the state-sum of \eqn{stateSum} can potentially lead to dimension dependence in the $D$-dimensional integrand. This is due to terms that contain products of internal polarizations:
\begin{equation}
\begin{aligned}
 \mathcal{I}_{\text{1-loop}} &\supset \sum_{ \text{states}} \varepsilon^\mu_{(l)}\varepsilon^\nu_{(-l)} \eta_{\mu\nu}
 \\
 &=\left(\eta^{\mu\nu} - \frac{l^\mu q^\nu+l^\nu q^\mu}{l\cdot q}\right) \eta_{\mu\nu}
 \\
 &=D_s-2\,,
 \end{aligned}
\end{equation}
 where $D_s$ is the spin-dimension of the internal photons, and thus $D_s-2$ just counts the number of photon states. To calculate our loop-level amplitudes, we use dimensional regularization to regulate the integrals. We will adopt the conventions of  \cite{collins_1984,Bern:2002zk} and take the spin-dimension, $D_s$, to be
the same as the dimension appearing in dimensional regularization, $D=4-2\epsilon$. We use them interchangeably throughout. The appearance of these $D$-dependent factors in the state sum are the source of rational terms in the Born-Infeld $S$-matrix. Throughout the text, we will use the generalized unitarity cut condition of \eqn{GUCut} to fix our multiloop integrands. With this in hand, in the spirit of Forde's one-loop cut construction, we will now systematize the procedure for capturing all the cut information needed for even-point theories at the multiloop order.

\section{Even-point Multi-loop Unitarity}
\label{sec:EMU}
In this section we will describe the methods we have developed to compute the multi-loop results of this paper. We begin by introducing the notion of {Even-point Multi-loop Unitarity} (EMU), which is an organizational principle at the foundation of our unitarity-based integrand construction. EMU is an extension of the method of maximal cuts~\cite{Bern:2007ct}, which is a hierarchical approach to perturbative calculations~\cite{Carrasco:2021bmu}.  EMU is a {constructive} algorithm aimed at capturing all the perturbative information needed at general loop order and multiplicity. We describe the algorithm below, and provide a 2-loop 4-point example that we choose to study in this paper.

\paragraph{\textbf{Even-point Multi-loop Unitarity (EMU) }} 
\begin{itemize}
\item \textbf{Step 0} -- At $n$-point $l$-loop, enumerate all 1-particle-irreducible (1PI) graphs constructed from $(n+2l-2)/2$ four-point vertices. We call these diagrams the maximal cut (MC) diagrams. Graphs with higher-point blobs are grouped into the $\text{N}^k \text{MC}$ category, where $k$ is the number of collapsed internal propagators relative to the MC graphs. 
\item \textbf{Step 1} -- Culling from $\text{N}^k\text{MC}\to \partial \text{N}^k\text{MC}$:
\begin{itemize}
\item \textbf{Step 1A} -- Discard all diagrams at the given order $\text{N}^k \text{MC}$ that capture scaleless behavior from \textbf{internal} kinematics. For massless theories, this amounts to throwing out diagrams that contain one of the following internal nodes:
\begin{equation}
\intScaleless
\end{equation}
\item \textbf{Step 1B} -- Discard all diagrams at the given order $\text{N}^k \text{MC}$ that capture scaleless behavior from \textbf{external} kinematics. At $n$-point, this amounts to rejecting diagrams that contain $n-1$ external edges attached to a single blob:
\begin{equation}
\extScaleless
\end{equation}
\end{itemize}
\item \textbf{Step 2} -- Collapsing from $\partial \text{N}^k\text{MC}\to \text{N}^{k+1}\text{MC}$:  If the culling procedure of Step 1 produces an empty set of graphs, $\partial \text{N}^k \text{MC} = \varnothing$, then the routine terminates.  If not, collapse one of the internal propagators for all diagrams in all topologically distinct ways. Collapsing a propagator simply means merging the two nodes connected by that propagator's edge and removing that edge all together.  So the resulting graph will have one less internal edge and one less internal node. This collapsing step will take the $\partial \text{N}^k \text{MC}$ set of diagrams to $\text{N}^{k+1} \text{MC}$. Once there, repeat Step 1A and 1B until the routine terminates.
\end{itemize}
This procedure for collecting all the cut information using EMU can be represented diagrammatically as a sequence of culling and collapsing as follows:
\begin{equation}
\text{MC}_{(l,n)} \stackrel{\text{S1}}{\rightarrow} \partial \text{MC}_{(l,n)} \stackrel{\text{S2}}{\rightarrow}\, \cdots\, \stackrel{\text{S2}}{\rightarrow}\text{N}^k\text{MC}_{(l,n)} \stackrel{\text{S1}}{\rightarrow}\partial\text{N}^k\text{MC}_{(l,n)}\equiv \varnothing
\end{equation}
After applying EMU and collecting all the diagrams up to some order $\text{N}^k\text{MC}_{(l,n)}$, the set of cuts we use to fully constrain the $n$-point $l$-loop integrand, $\Omega_{(l,n)}$, is the intersection of these sequence of unitarity cuts:
\begin{equation}
\Omega_{(l,n)}=\bigcap_{i=0}^k \partial\text{N}^i\text{MC}_{(l,n)}
\end{equation}
Taking the intersection accounts for the overlapping information stored in the cut diagrams\footnote{Strictly speaking only the $\partial \text{N}^k\text{MC}_{(l,n)}$ cuts are required as they represent a spanning set -- namely, they contain all the contacts and residues to fully specify the integrand.}  at different orders in $\text{N}^k\text{MC}$. As a concrete example, let's consider the case of two-loop four-point of interest in this paper. First, we obtain the following set of diagrams at the MC level built from three four-point vertices:
\begin{equation}
\text{MC}_{(2,4)} = \left\{\scaleIntBscalarsmall{}{}{}{},\scaleIntCsmallNoNum,\scaleIntAscalarsmall{}{}{}{}\right\}
\end{equation}
Each blob corresponds to an on-shell tree-level amplitude from the even-point theory of interest. The third diagram is scaleless on support of dimensional regularization, and thus we discard it in Step 1A. This gives the following restricted set of MC diagrams:
\begin{equation}\label{eq:2loopBasis}
\partial \text{MC}_{(2,4)} = \left\{\scaleIntBscalarsmall{}{}{}{},\scaleIntCsmallNoNum \right\}
\end{equation}
Since $\partial \text{MC}_{(2,4)}$ is not empty, we proceed to Step 2. After throwing out the scaleless diagram in $\text{MC}_{(2,4)}$, this leads to the following set of $\text{N}^1\text{MC}$ diagrams for two-loop four-point:
\begin{equation}
\text{N}^1\text{MC}_{(2,4)} = \left\{\scaleIntAsmallNoNum,\scaleIntBsmallNoNum \right\}
\end{equation}
Now the first diagram is discarded in Step 1A, and the second diagram is discarded in Step 1B. Thus, we need not consider any next-to-maximal cuts for two-loop four-point amplitudes in even-point theories. Since $\partial \text{N}^1\text{MC}_{(2,4)} =\varnothing$, this concludes the EMU cut construction. 

As we can see from the diagrams that contribute at two-loops four-point in \eqn{eq:2loopBasis}, even-point theories can be fully constructed from convolutions of one-loop integrals. As described earlier, we call such integrals {recursively one-loop} since we can iteratively apply one-loop unitarity and tensor reduction methods in order to recover the full multi-loop structure. In the next two sections, we first illustrate the recursive behavior of convolution integrals, and then use their properties to develop multi-loop tensor reduction methods that we use throughout the paper. 

\subsection{Multi-loop recursive integrals}
\label{sec:recInt}

As we found above in our example of EMU, the only diagrams that contribute at two-loop four-point are those given in \Fig{fig:contrib2Loop}. Due to the bubble integral insertions, both of these integrals can be evaluated in terms of the one-loop basis integrals of \eqn{eq:integrals}. This is a direct consequence of the single-scale nature of bubble insertions. Since internal bubbles are single-scale, they will only contribute additional powers of inverse propagators. This has the effect of shifting an $L$-loop convolution integral to $(L-1)$-loop order, with a propagator power determined by the mass-dimension of the evaluated bubble integral. Explicitly, we can evaluate a $D$-dimensional nested bubble with internal momentum, $q^\mu$, flowing into the diagram,
\begin{equation}\label{eq:intBubbleRule}
\stackrel{q^\mu}{\longrightarrow}\! \!\scaleIntBubbleProbe{}{} \sim [q^2]^{D/2-\alpha_1-\alpha_2}\,,
\end{equation}
where $\alpha_1$ and $\alpha_2$ are the degree of the denominators appearing in the bubble integral. For example, consider a banana integral diagram constructed from six-point contacts that would appear in a three-loop application of EMU with external momentum scale $q^\mu = (k_{1}+k_2)^\mu$. Evaluating the simplest case where all $\alpha_i=1$, one finds that recursively applying \eqn{eq:intBubbleRule} yields the following sequence:
\begin{equation}
\stackrel{k_{12}^\mu}{\longrightarrow} \!\!\!\scaleIntBubbleProbeA{}{} \sim \scaleIntBubbleProbeB{}{} \sim \scaleIntBubbleProbeC{}{} \sim s_{12}^{3D/2-4}\,.
\end{equation}
For simplicity, we have dropped the $\Gamma$-function dependent factors at each step in the evaluation. This procedure of evaluating bubble integrals at successive orders also works for tensor bubble-integral insertions as well. As we'll show in this next section, such tensor bubble insertions can be reduced to scalars integrals via Passarino-Veltman \cite{Passarino:1978jh}. While there are a number of tensor reduction algorithms available in the literature \cite{Anastasiou:2004vj,vonManteuffel:2012np,Smirnov:2014hma,vonManteuffel:2014ixa,Smirnov:2019qkx,Smirnov:2020quc,Usovitsch:2020jrk,Maierhofer:2018gpa}, our ability to use Passarino-Veltman on the recursive integrals of \Fig{fig:contrib2Loop} will prove useful for evaluating terms with factors of $(\varepsilon_i \cdot l_j)$, which include formal polarizations. This procedure will aid in our pursuit of extracting the rational part of the two-loop Born-Infeld amplitudes that we study in \sect{sec:2loopBIU}.
 
\subsection{Two-loop tensor reduction}\label{sec:tenRed}
First we will study the canonical one-loop case needed for the nested bubble integrals described above, and then show that it naturally generalizes to the two-loop for the recursive integrals of interest. We begin with a general expression for the rank-$n$ tensor bubble:
\begin{align}
     I^{\mu_1\dots \mu_n}_2(K)&= \int \frac{d^D l}{(2\pi)^D} \frac{l^{\mu_1}l^{\mu_2}\cdots l^{\mu_n} }{l^2(l+K)^2}
     \\
     &= \sum_{m+2k=n}a_{(m,k)} \mathcal{T}^{(m,k)}_{\text{bub}} \,,\label{tensorReductionBubble}
\end{align}
where we have introduced the following shorthand for the symmeterized rank-$(m+2k)$ tensor:
\begin{equation}\label{eq:symTen1}
\mathcal{T}^{(m,k)}_{\text{bub}}  \equiv K^{(\mu_1}\cdots K^{\mu_m}\eta^{\mu_1\mu_2}\cdots \eta^{\mu_{2k-1}\mu_{2k})}\,.
\end{equation}
By contracting the tensor integral with the internal scale, $K^\mu$, and the metric, $\eta^{\mu\nu}$, we obtain the following two constraint equations:
\begin{align}
K_{\mu_1}I^{\mu_1\dots \mu_n}_2(K) &= \int \frac{d^D l}{(2\pi)^D} \frac{ (K\cdot l) l^{\mu_2}\cdots l^{\mu_n}}{l^2(l+K)^2}= -\frac{K^2}{2}I^{\mu_2\dots \mu_n}_2(K) \,,
\\
\eta_{\mu_1 \mu_2}I^{\mu_1\dots \mu_n}_2(K)&= \int \frac{d^D l}{(2\pi)^D} \frac{  l^{\mu_2}\cdots l^{\mu_n}}{(l+K)^2}=0\,.
\end{align}
Performing the same contractions on the symmeterized tensor defined above in \eqn{eq:symTen1} yields the following:
\begin{align}
K\cdot K^{\otimes_{(m)}}_{\text{sym}}\eta_{\text{sym}}^{\otimes_{(k)}} &= K^2 \mathcal{T}^{(m-1,k)}_{\text{bub}} + (m+1)\mathcal{T}^{(m+1,k-1)}_{\text{bub}}\,,
\\
\eta \cdot K^{\otimes_{(m)}}_{\text{sym}}\eta_{\text{sym}}^{\otimes_{(k)}} &= K^2 \mathcal{T}^{(m-2,k)}_{\text{bub}} + \left[D+2(m+k-1)\right]\mathcal{T}^{(m,k-1)}_{\text{bub}}\,.
\end{align}
These linear relations can be used to line up the coefficients with distinct tensor structures, which gives the following:
\begin{align}
0&= \sum a_{(m,k)}\left[K^2 \mathcal{T}^{(m-1,k)}_{\text{bub}}+ (m+1)\mathcal{T}^{(m+1,k-1)}_{\text{bub}} +\frac{K^2}{2}\mathcal{T}^{(m,k)}_{\text{bub}}\right]
\\
&= \sum \left[a_{(m+2,k)} K^2 + a_{(m,k+1)} (m+1) +a_{(m+1,k)}\frac{K^2}{2}\right]\mathcal{T}^{(m+1,k)}_{\text{bub}}\,,\end{align}
and similarly so for the metric contraction:
\begin{align}
0 &= \sum a_{(m,k)}\left[K^2 \mathcal{T}^{(m-2,k)}_{\text{bub}} + \left[D+2(m+k-1)\right]\mathcal{T}^{(m,k-1)}_{\text{bub}}\right] 
\\
&= \sum \left[a_{(m+2,k)}K^2  + a_{(m,k+1)}\left[D+2(m+k)\right]\right] \mathcal{T}^{(m,k)}_{\text{bub}}\,.
\end{align}
Treating the distinct tensor structures as basis elements, we thus conclude the following set of linear relations between the coefficients for the rank-$n$ tensor bubble:
\begin{align}
0&=K^2 a_{(m+2,k)}+[D+2(m+k)]a_{(m,k+1)}\,,
\\
0&=K^2a_{(m+2,k)}+(m+1)a_{(m,k+1)}+\frac{1}{2}K^2 a_{(m+1,k)}\,,
\end{align}
where $D= \eta_{\mu\nu}\eta^{\mu\nu}$. These constraints can be rearranged to give the following recursive definition for the coefficients $a_{(m,k)}$:
\begin{eBox}
\begin{equation}\label{eq:bubRed}
\begin{aligned}
a_{(m,k)}&=- \left[\frac{K^2}{D+2(m+k-1)}\right]a_{(m+2,k-1)}
\\
 a_{(m,0)}&=-\left[\frac{D+2(m-2)}{2(D+m-3)}\right]a_{(m-1,0)}
 \\
 a_{(0,0)}&=I_2(K)
\end{aligned}
\end{equation}
\end{eBox}
The base step is simply the scalar bubble integral, $a_{(0,0)}=I_2(K)$. Constructing $a_{(m,k)}$ from this recursive definition, the one-loop integrand of \eqn{GUCut} that contains factors of $(\varepsilon_i\!\cdot \!l )$ and $(k_i \!\cdot \! l)$ can be expressed completely in terms of the bubble integral, $I_2(K)$, weighted by dimension dependent numerical factors and external vector kinematics.

\begin{figure}[t]
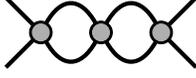

    \centering
    \begin{equation*}\scaleIntCscalar{}{}{}{}\end{equation*}
    \caption{Graphical depiction of the the double-bubble integral. Every exposed internal edge represents a propagator in the integrand.}
    \label{fig:dubBub}
\end{figure}

We note that these coefficients are also sufficient to evaluate the contribution from the double-bubble integral whose propagator structure is sketched in \Fig{fig:dubBub}. This is best demonstrated with an explicit example. Consider the following integral, $I^{\text{ex}.}_{\text{2bub}}$, that functionally captures terms that could appear in the cut construction of the double-bubble integral at two-loop:

\begin{equation}
\label{intg2bub}
I^{\text{ex}.}_{\text{2bub}} \equiv \int \frac{d^D l_{1}d^D l_{2}}{(2\pi)^{2D}} \frac{(l_1\!\cdot\! l_2)^2(l_1\!\cdot\! v_1)(l_2\!\cdot\! v_2)}{l_1^2(l_1+k_{12})^2 l_2^2(l_2+k_{12})^2}\,,
\end{equation}
where $v_i$ is a stand-in for external kinematics, $\varepsilon_i$ or $k_i$. While the numerator mixes factors of $l_1$ and $l_2$, the denominator can be separated. This allows us to re-express the above integral completely in terms of iterated tensor bubbles of \eqn{eq:integrals}:
\begin{equation}
I^{\text{ex}.}_{\text{2bub}} = I_{2}^{\alpha\beta\gamma}(s_{12}) I_2^{\mu\nu\rho}(s_{12})\eta_{\alpha\beta}\eta_{\mu\nu}v_{1 \gamma}v_{2 \rho}\,.
\end{equation}
Then, by applying \eqn{tensorReductionBubble}, and plugging in the expressions for $a_{(m,k)}$, the kinematic numerator of $I^{\text{ex}.}_{\text{2bub}}$ no longer mixes loop momenta. Thus, the integral is separable and can be expressed as a product of scalar bubbles integrated over $l_1$ and $l_2$.

\begin{figure}[t]
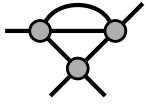

    \centering
    \begin{equation*}\scaleIntBscalar{}{}{}{}\end{equation*}
    \caption{Graphical depiction of the the ostrich integral. Every exposed internal edge represents a propagator in the integrand.}
    \label{fig:ostrich}
\end{figure}
A similar procedure can be applied to the ostrich type diagrams of \Fig{fig:ostrich}. However, the integration procedure is a little more delicate than the $I^{\text{ex}.}_{\text{2bub}}$ of \eqn{intg2bub}.  As $I^{\text{ex}.}_{\text{2bub}}$ could be expressed as an iterated bubble, we only needed to consider {integer powers} of the denominators when constructing the recursion relations. In contrast, ostrich integrals  will lead to {non-integer}, $\epsilon$-dependent powers of loop propagators. We are therefore interested in performing an $x$-dependent tensor reduction on the triangle integral,
\begin{align}\label{triangleTensorInt}
     I^{\mu_1\dots \mu_n}_{3,x}(K_{12})&= \int \frac{d^D l}{(2\pi)^D} \frac{l^{\mu_1}l^{\mu_2}\cdots l^{\mu_n} }{l^2(l+K_{1})^{2x}(l+K_{12})^2}
     \\
     &= \sum_{m+l+2k=n}a^{x}_{(m,l,k)} \mathcal{T}^{(m,l,k)}_{\text{tri}}\,,
\end{align}
with $K_{12}=K_1+K_2$ introduced as shorthand notation. We note that $x$ is a {non-integer} value that will inherit dependence on $\epsilon$ from integrating over the internal bubble, $I^{4-2\epsilon}_2(l+K_1)\sim [(l+K_1)^2]^{-\epsilon}$. The symmetrized triangle tensor takes the following definition:
\begin{equation}
\mathcal{T}^{(m,l,k)}_{\text{tri}} \equiv   K_1^{(\mu_1}\cdots K_1^{\mu_m}K_2^{\mu_1}\cdots K_2^{\mu_l}\eta^{\mu_1\mu_2}\cdots \eta^{\mu_{2k-1}\mu_{2k})}\,.
\end{equation}
Given this definition, when we perform the tensor contractions over $K_1$ and $K_2$, the degree of $x$ will get shifted:
\begin{align}
{K_1}_{\mu_1}I^{\mu_1\dots \mu_n}_{3,x}&= \int \frac{d^D l}{(2\pi)^D} \frac{(K_1\cdot l)l^{\mu_2}\cdots l^{\mu_n} }{l^2(l+K_{1})^{2x}(l+K_{12})^2} = \frac{1}{2}I^{\mu_2\dots \mu_n}_{3,x-1}\,,
\\
{K_2}_{\mu_1}I^{\mu_1\dots \mu_n}_{3,x}&= \int \frac{d^D l}{(2\pi)^D} \frac{(K_2\cdot l)l^{\mu_2}\cdots l^{\mu_n} }{l^2(l+K_{1})^{2x}(l+K_{12})^2} = -\frac{1}{2}\left[I^{\mu_2\dots \mu_n}_{3,x-1}+K_{12}^2 I^{\mu_2\dots \mu_n}_{3,x}\right]\,.
\end{align}
For our purposes, external momenta are taken to be null, $K_1^2=K_2^2 = 0$. By construction, contracting with the metric will yield a scaleless integral, giving us the following constraint:
\begin{equation}
\eta_{\mu_1\mu_2}I^{\mu_1\dots \mu_n}_{3,x}= \int \frac{d^D l}{(2\pi)^D}\frac{l^{\mu_2}\cdots l^{\mu_n} }{(l+K_{1})^{2x}(l+K_{12})^2} = 0\,.
\end{equation}
Note that in the above contractions with $K_1$ and $K_2$ the degree of the denominator get's shifted from $x\rightarrow x-1$. For normal one-loop triangle integral reductions for which $x=1$, this would lead to a tensor bubble that we have computed in the previous section,
\begin{align}
{K_1}_{\mu_1}I^{\mu_1\dots \mu_n}_{3,x=1}& = \frac{1}{2}I^{\mu_2\dots \mu_n}_{2}\,,
\\
{K_2}_{\mu_1}I^{\mu_1\dots \mu_n}_{3,x=1}& = -\frac{1}{2}\left[I^{\mu_2\dots \mu_n}_{2}+K_{12}^2 I^{\mu_2\dots \mu_n}_{3,x=1}\right]\,.
\end{align}
 However, since the two-loop integration leads to $\epsilon$-dependence, this will now induce additional factors of scalar $I_{3,\epsilon - n}$ final tensor reduction. Proceeding with the computation, these integral constraints can be similarly applied to our symmeterized tensors, yielding the following set of contractions:
\begin{align}
K_1\cdot \mathcal{T}^{(m,l,k)}_{\text{tri}} &=\frac{1}{2}K_{12}^2\mathcal{T}^{(m,l-1,k)}_{\text{tri}} + (m+1)\mathcal{T}^{(m+1,l,k-1)}_{\text{tri}} \,,
\\
K_2\cdot \mathcal{T}^{(m,l,k)}_{\text{tri}} &=\frac{1}{2}K_{12}^2\mathcal{T}^{(m-1,l,k)}_{\text{tri}} +(l+1)\mathcal{T}^{(m,l+1,k-1)}_{\text{tri}} \,,
\\
\eta\cdot  \mathcal{T}^{(m,l,k)}_{\text{tri}} &= K_{12}^2\mathcal{T}^{(m-1,l-1,k)}_{\text{tri}} +\left[D+2(m+k+l-1)\right]\mathcal{T}^{(m,l,k-1)}_{\text{tri}} \,.
\end{align}
Keeping this in mind, we obtain the following linear relations between the coefficients,
\begin{equation}
\begin{aligned}
0&=2(m+1)a_{(m,l,k+1)}^{x}+s_{12}a_{(m+1,l+1,k)}^{x}-a_{(m+1,l,k)}^{x-1}\,,
\\
0&=2(l+1)a_{(m,l,k+1)}^{x}+s_{12}[a_{(m+1,l+1,k)}^{x}+a_{(m,l+1,k)}^{x}]+a_{(m,l+1,k)}^{x-1}\,,
\\
0&=[D+2(m+l+k)]a_{(m,l,k+1)}^{x}+s_{12}a_{(m+1,l+1,k)}^{x}\,.
\end{aligned}
\end{equation}
Just as was done for the scalar bubble, these functional expressions can be rearranged to construct family of recursion relations for the tensor coefficients:
\begin{eBox}
\begin{equation}\label{eq:triRed}
\begin{aligned}
a_{(m,l,k)}^x&=-\left[\frac{s_{12}}{D+2(m+l+k-1)}\right]a_{(m+1,l+1,k-1)}^x
\\
a_{(m,l,0)}^x&=-\left[\frac{D+2(m+l-2)}{D+2(m-2))}\right]\left[\frac{1}{s_{12}} a^{x-1}_{(m-1,l,0)}+a^{x}_{(m-1,l,0)}\right]
\\
a^x_{(0,l,0)}&= \frac{1}{s_{12}} a_{(0,l-1,0)}^{x-1} 
\\
a^x_{(0,0,0)} &= I_{3,x}(K_{12})
\end{aligned}
\end{equation}
\end{eBox}
This system of equations uniquely fixes the rank-$n$ massless triangle tensor integral in \eqn{triangleTensorInt} with non-integer powers in the denominator. Using these integral reduction formulae, both the one-loop and two-loop integrals can be expressed completely in terms of scalar one-loop bubble and scalar triangle integrals, which we are provided in \eqn{eq:integrals} of the previous section. Before proceeding to our results, we will introduce a bit of notation that will be relevant for capturing loop-level contributions to the vector amplitudes. 

\subsection{Gauge-invariant on-shell basis}\label{sec:basisT}
The final bit of machinery we introduce for the multi-loop calculations of this paper is a spanning set of $D$-dimensional four-photon on-shell tensor structures. All of the vector integrands constructed in the following sections will be fixed on $D$-dimensional cuts, and projected to a basis of $D$-dimensional photon tensors. 

Being agnostic to dimension allows for a particularly algorithmic approach to identifying rational terms.
In addition, by projecting to a $D$-dimensional basis of gauge-invariant tensor structures, we can more easily track the one-loop divergences that propagate to two-loop order. This is due to the existence of {evanescent operators}, those which vanish when plugging in 4D states, but are non-vanishing in general dimensions. As we will see in pure Born-Infeld amplitudes of \sec{sec:DBIU}, these so-called evanescent operators are the cause of two-loop divergences that are obscured when looking at the one-loop behavior in $D=4$ exclusively. Tracking divergences introduced by evanescent operators has been an active area of research Standard Model effective field theory (SMEFT) \cite{Hartmann:2016pil,Chala:2021cgt,Aebischer:2022tvz,Fuentes-Martin:2022vvu,Isidori:2023pyp}, the UV behavior of quantum gravity \cite{Bern:2015xsa,Bern:2017tuc,Bern:2017puu}, and more generally as an area of formal theory interest \cite{Dugan:1990df,Herrlich:1994kh,Bell:2009nk,Becher:2004kk,DiPietro:2017vsp}. Below we will give a concrete example of an evanescent operator expressed with notation used in the text to provide some justification for our $D$-dimensional tensor basis. 

First, we will use the pair of $D$-dimensional tensor structure, $f_{ij}f_{kl}$ and $f_{ijkl}$ defined previously in \cite{Carrasco:2022jxn},
\begin{align}
f_{ij}= \frac{1}{2}\text{tr}[F_iF_j]\,,\qquad f_{ijkl} = \text{tr}[F_iF_jF_kF_l]\,,
\end{align}
where $F_i^{\mu\nu} = k_i^\mu \varepsilon_i^\nu - k_i^\nu \varepsilon_i^\mu$ are linearized field strengths, and the trace is taken over spacetime indices. With these vector building blocks, there are exactly four $D$-dimensional four-photon matrix elements one can write down at $\mathcal{O}(k^8)$:
\begin{equation}\label{eq:symVecBlocks}
\begin{aligned}
\mathcal{T}_{(2,0)}^{F^2F^2} &= s_{12}^2 f_{12}f_{34} + \text{cyc}(2,3,4)\,,
\\
\mathcal{T}_{(2,0)}^{F^4} &= s_{12}^2 f_{1324} + \text{cyc}(2,3,4)\,,
\\
\mathcal{T}_{(0,1)}^{F^2F^2} &= s_{13}s_{14} f_{12}f_{34} + \text{cyc}(2,3,4)\,,
\\
\mathcal{T}_{(0,1)}^{F^4} &= s_{13}s_{14} f_{1324} + \text{cyc}(2,3,4)\,,
\end{aligned}
\end{equation}
where the Mandelstam invariants are defined as $s_{ij} = (k_i+k_j)^2$.  The index subscripts correspond to powers of Mandelstam invariants that are easly generalized to span arbitrarily high mass-dimension,
\begin{eBox}
\begin{equation}\label{eq:basisTensors}
\begin{aligned}
\mathcal{T}^{F^2F^2}_{(x,y)} &\equiv s_{12}^x (s_{13}s_{14})^y f_{12}f_{34}+\text{cyc}(2,3,4)\,,
\\\\
\mathcal{T}^{F^4}_{(x,y)} &\equiv s_{13}^x (s_{12}s_{14})^y f_{1234}+\text{cyc}(2,3,4)\,,
\\\\
\mathcal{T}^{F^3}_{(x,y)} &\equiv \sigma_3^{x} \sigma_2^y [st A^{F^3}_{(s,t)}]\,,
\end{aligned}
\end{equation}
\end{eBox}
with $\sigma_3 = s_{12}s_{13}s_{14}/8$ and $\sigma_2 = (s_{12}^2+s_{13}^2+s_{14}^2)/8$. The important takeaway from \eqn{eq:symVecBlocks} is that one can write down four independent Lorentz invariant photon tensor structures at $\mathcal{O}(k^8)$ in general dimension. However, this does obscure the 4-dimensional freedom available at this mass dimension. In fact, there are only three distinct helicity structures in $D=4$. One particular basis for these helicity structures, exploiting  the 4D spinor-helicity products reviewed in \sect{subsec:4DandDDReview}, can be written as follows,
\begin{align}
\mathcal{T}^{4\text{D}}_{(++++)} &= (s_{12}^4 + s_{13}^4 + s_{14}^4 ) \frac{[12][34]}{\langle 12\rangle \langle34\rangle }\,,
\\  \label{eq:4D1Tens}
\mathcal{T}^{4\text{D},1}_{(--++)} &= (s_{13}^2+s_{14}^2) \langle 12\rangle^2 [34]^2\,,
\\  \label{eq:4D2Tens}
\mathcal{T}^{4\text{D},2}_{(--++)} &= s_{12}^2 \langle 12\rangle^2 [34]^2\,.
\end{align}
Immediately we can see that the full $D$-dimensional basis must have a non-trivial null space when projected down to $D=4$. Given that the 4D helicity space is overdetermined, we are able to define the following evanescent amplitude $\mathcal{T}^{\text{ev.}}$ that vanishes when constrained to any 4D states:
\begin{equation}
\mathcal{T}^{\text{ev.}} \equiv \mathcal{T}_{(2,0)}^{F^2F^2}-\mathcal{T}_{(0,1)}^{F^2F^2} + \mathcal{T}_{(0,1)}^{F^4}\,.
\end{equation}
Using the helicity projection rules of \eqn{eqn:4DPols}, one can verify that $\mathcal{T}^{\text{ev.}}$ indeed vanishes in $D=4$, while clearly non-vanishing in general dimension. This behavior will be particularly relevant for interpreting the loop level results where $D=4-2\epsilon$. Indeed, it is critically important that we track $D$-dimensional contributions when probing for divergences at multi-loop order.    Consider a two-loop integral for which we need to integrate the following quantity:
\begin{equation}
\scaleIntAvectorODD{}{}{}{}{$\mathcal{T}^{\text{ev.}}$}{$\mathcal{T}^{\text{arb.}}$} =\sum_{\text{states}}\int\frac{d^D l}{(2\pi)^D}\frac{\mathcal{T}^{\text{ev.}}_{(1,2,\bar{l}_1,\bar{l}_2)} \mathcal{T}^{\text{arb.}}_{(l_1,l_2,3,4)}}{l^2(l+k_{12})^2}\,,
\end{equation}
where $\bar{l}_1 =-l_1= l$ and $\bar{l}_2 =-l_2= -(l+k_{12})$. We have dressed the darker vertex with the evanescent contribution, $\mathcal{T}^{\text{ev.}}$, which could be the result of an unspecified loop integral, and take a generic arbitrary vector structure $\mathcal{T}^{\text{arb.}}$ to dress the lighter vertex.   Exposed legs are taken to be on-shell in the numerator of this expression.  

To see what can go wrong with relying only on four-dimensional cut information, we can consider a particularly simple case. Take the vector insertion on the right hand side to be $\mathcal{T}^{\text{arb.}}_{(1234)} \equiv f_{12}f_{34}$. By applying the state sum and tensor reduction formulae from the previous sections, this can be evaluated explicitly. The result is given,
\begin{align}
\scaleIntAvectorODD{}{}{}{}{$\mathcal{T}^{\text{ev.}}$}{$\mathcal{T}^{\text{arb.}}$} &= \frac{(D_s-4)(D_s-3)}{8(D_s-1)}s_{12}^4 I_2(k_{12}) f_{12}f_{34}\,,
\\
& = -\frac{i }{192\pi^2} s_{12}^4 f_{12}f_{34} +\mathcal{O}(\epsilon)\,,
\end{align}
where $I_2(k_{12})$ is the scalar bubble integral in the $s_{12}$-channel, and in the second line we have plugged in $D_s=4-2\epsilon$. Thus far, this is exactly what we should expect. Since the amplitude $\mathcal{T}^{\text{ev.}}$ vanishes in $D=4$, the 4-dimensional cut above must vanish. The Optical Theorem thus disallows imaginary parts from logarithms that would appear post integration. This physical constraint is imposed by the factor of $(D-4)$, which absorbs the divergence of $I_2(k_{12})$ and pushes the logarithm to be subleading at $\mathcal{O}(\epsilon)$. 

However, suppose that $\mathcal{T}^{\text{ev.}}$ came dressed with a $1/\epsilon$-divergence from some nested integral in a full two-loop calculation. That is, suppose the operator insertion came from the leading divergence of a one-loop integral, such that,
\begin{equation}
\mathcal{M}^{1\text{-loop}}\big|_{\text{div.}}\sim \scaleTree{hgrey4}\Bigg|_{\text{div.}} = \frac{1}{\epsilon}\mathcal{T}^{\text{ev.}}_{(1234)}\,.
\end{equation}
Now the two-loop contribution would be {divergent} despite the vanishing 4D cut between the evanescent operator and the example operator, $\mathcal{T}^{f^2}$, chosen above:
\begin{equation}
\scaleIntAvectorODD{}{}{}{}{$\frac{1}{\epsilon}\mathcal{T}^{\text{ev.}}$}{$\mathcal{T}^{\text{arb.}}$}\Bigg|_{\text{div.}}  =- \frac{1}{\epsilon} \frac{i }{192\pi^2}s_{12}^4 f_{12}f_{34}\,.
\end{equation}
We would have missed this if we relied solely on four-dimensional cut information.

It is straightforward to see that the three $D$-dimensional structures given in \eqn{eq:basisTensors} span the predictions of four-photon effective operators at arbitrary mass-dimension.   The first two $\mathcal{T}^{F^2F^2}_{(x,y)}$, and $\mathcal{T}^{F^4}_{(x,y)}$, span the external $(\pm\pm\pm\pm)$ and $(\pm\pm\mp\mp)$ helicity sectors. The third $\mathcal{T}^{F^3}_{(x,y)}$  captures the $(\pm\pm\pm\mp)$ helicity configurations. 

As an aside it is worth noting that the single $F^3$-insertion vector permutation invariant, $st A^{F^3}_{(s,t)}$, can be expressed concisely in terms of $\mathcal{T}^{F^2F^2}_{(x,y)}$ as follows,
\begin{equation}
s t A^{F^3}_{(s,t)} = \frac{\mathcal{T}^{F^2F^2}_{(0,2)} -g_1g_2g_3g_4}{s_{12}s_{13}s_{14}}\,,
\end{equation}
where $g_i \equiv 2 k_{i-1}^\mu F^{\mu\nu}_i k^{\nu}_{i+1}$. While not obvious when expressed in this form, the numerator of the above tensor structure is proportional to the permutation invariant in the denominator, $\mathcal{T}^{F^2F^2}_{(0,2)} -g_1g_2g_3g_4 \propto \sigma_3$, and thus $s t A^{F^3}_{(s,t)}$ is local by construction. 

As \eqn{eq:basisTensors} represents a set of tensor structures which completely spans the space ofpredictions of $D$-dimensional permutation invariant photon effective operators, we can write an on-shell effective amplitude parameterized by numeric Wilson coefficients, $a_{(x,y)}$, as follows,
\begin{eBox}
\begin{equation}
\mathcal{M}^{\text{photon-EFT}}_4 = \sum_{x,y} a_{(x,y)}^{F^2F^2}\mathcal{T}^{F^2F^2}_{(x,y)}+a_{(x,y)}^{F^4}\mathcal{T}^{F^4}_{(x,y)}+a_{(x,y)}^{F^3}\mathcal{T}^{F^3}_{(x,y)}\,.
\end{equation}
\end{eBox}

In Ref. \cite{Carrasco:2022jxn} it was demonstrated that this photonic effective field theory (EFT) amplitude permits a double copy construction by either introducing a set of $d^{abc}$-type symmetric algebraic structures, or more traditional adjoint $f^{abc}$-type kinematics but at the cost of factorizing over higher spin modes. While we will comment on the double-copy properties of this amplitude in \sect{sec:ActionDC}, for now we will just stress that these $D$-dimensional operators will serve as an useful basis for capturing divergences and anomalies in the Born-Infeld $S$-matrix. With that, we are prepared to proceed to the loop level results.

\section{Loop-level results}\label{sec:Loops}
We now apply the EMU approach introduced in \sect{sec:EMU} to construct two-loop amplitudes in NLSM and DBIVA theories and reduce them to an appropriate basis of integrals. We then evaluate the $D$-dimensional scalar integrals in the dimension of interest. We will focus primarily on $D=4-2\epsilon$ for DBIVA amplitudes, but will also consider $D=2-2\epsilon$, which is the dimension for which NLSM is critical. We will then project $D$-dimensional tensor structures along defined 4D spinor helicity states using the conventions described in \sect{subsec:4DandDDReview}. This will clarify where anomalous matrix elements contribute for non-supersymmetric BI, which has a classically conserved $U(1)$ symmetry.

We begin with an instructive example of NLSM pions through the two-loop order. These amplitudes can serve as the scaffolding for constructing $\mathcal{N}=4$ DBIVA integrands using the double copy even without explicitly finding a color-dual representation. An exciting realization of our results is that the loop-level double copy construction of $\mathcal{N}=4$ DBIVA is consistent with the results that we compute directly via unitarity. This is the first application of double-copy construction for multi-loop amplitudes for non-gravitational theories. Furthermore, it suggests that there should exist a color-dual representation of two-loop NLSM integrands, a representation that has yet to be constructed explicitly. 

\subsection{NLSM via EMU}\label{sec:NLSMU}

We begin with the color-dressed NLSM tree amplitudes that will serve as the kinematic building blocks to appear in our unitarity construction. For our purposes, a convenient color basis is the so-called half-ladder\footnote{Called multiperipheral diagrams in \cite{DixonMaltoni}, such half-ladder graphs are also referred to as comb graphs in the literature, see e.g. refs.~\cite{Feng:2020opo,Alday:2023kfm}.} basis of Del Duca, Dixon and Maltoni (DDM) \cite{DixonMaltoni},
\begin{equation}
\mathcal{A}^{\text{NLSM}}_n = \sum_{\sigma \in S^{n-2}}\cHL_{(1\sigma n)} A^{\text{NLSM}}_{(1\sigma n)}\,.
\end{equation}
The half-ladder color factors are defined in terms of the antisymmetric structure constants as, 
\begin{equation}
\cHL_{(1\sigma n)} \equiv f^{1\sigma_2 \beta_2}f^{\beta_2 \sigma_3 \beta_3} \cdots f^{\beta_{n-1}\sigma_{n-1}n}.
\end{equation} 
In the above expression, each half-ladder color factor, associated with a particular channel-graph, dresses a color-ordered amplitude $A^{\text{NLSM}}_{(1\sigma n)}$. At four-point, it is well known that the $(ijkl)$ ordered amplitude for NLSM is simply, 
\begin{equation}
A^{\text{NLSM}}_{(ijkl)} = f_\pi^{-2} \,s_{ik}\,.
\end{equation}

As the $s_{23}$-channel color factor satisfies a Jacobi identity with the $s_{12}$- and $s_{13}$-channel color factors, $\cHL_{(2341)} =\cHL_{(1234)}-\cHL_{(1324)}$, the color-dressed four-point amplitude can be expressed completely in terms of the $s_{12}$-channel and $s_{13}$-channel color factors:
\begin{equation}\label{eq:NLSMamp} 
\mathcal{A}^{\text{NLSM}}_{(1|23|4)} = f_\pi^{-2}\left(\cHL_{(1234)} s_{13}+\cHL_{(1324)}s_{12}\right)\,.
\end{equation}
The color-dressed amplitude is of course entirely Bose-symmetric, but we use the subscript $(1|23|4)$ to emphasize a functional choice of the $(1|\sigma|4)$ half-ladder basis.
 
\subsubsection{One-loop}\label{sec:1loopNLSM}
At one-loop, the unitarity construction of gauge theory amplitudes is particularly simple, since the color factors can be decomposed into a DDM-like ordered color basis. That is, all color factors, $C_g$ associated with the following cubic representation of the integrand:
\begin{equation}
\mathcal{A}^{1\text{-loop}}_{n} = \int \prod_{i=1}^L \frac{d^D l_i}{(2\pi)^D} \sum_{g\in \Gamma^{(3)}} \frac{1}{S_g}\frac{C_g N_g}{D_g}\,,
\end{equation}
can be expressed uniquely as a sum over a canonical ordering of color factors by iteratively applying the color jacobi identity. In doing so, all 1-loop color factors, $C^{1\text{-loop}}_g$, can be expressed as a sum over $n$-gon integrand factors, $C^{n\text{-gon}}_{(a_1a_2...a_n)}$ in the following way,
\begin{equation}
C^{1\text{-loop}}_g = \sum_{\sigma\in S^{n-1}} \beta^{(\sigma)}_g C^{n\text{-gon}}_{(\sigma_{1}\sigma_{2}...\sigma_{n-1}n)} \,,
\end{equation}
where $\beta^{(\sigma)}_g$ are factors of $\{-1,0,1\}$ depending on the color structure of the diagram, $C^{1\text{-loop}}_g$. The $n$-gon basis diagrams take the following definition in terms of adjoint structure constants, $f^{abc}$:
\begin{equation}
C^{n\text{-gon}}_{(a_1a_2...a_n)} \equiv f^{b_1 a_1 b_2}f^{b_2 a_2 b_3}\cdots f^{b_{n} a_{n} b_1}\,.
\end{equation}
This allows us to express the one-loop amplitudes in terms of a sum over the $(n-1)!$ distinct color factors, weighted by the contributions from the color-ordered Feynman diagrams:
\begin{equation}
\mathcal{A}^{1\text{-loop}}_n = \sum_{\sigma \in S^{n-1}} C^{n\text{-gon}}_{(\sigma n)} A^{1\text{-loop}}_{(\sigma n)}\,.
\end{equation}
As this is a minimal basis we will consider cuts that allow a targeted identification of each color-weight's coefficients.

\paragraph{\textbf{Construction}} Using the four-point tree-level NLSM amplitude with the reasoning of \sect{sec:EMU}, it is straightforward to write down an off-shell integrand associated with the one-loop bubble:
\begin{equation}
\scaleIntApion =\sum_{\text{states}}\int \frac{d^D l}{(2\pi)^D} \frac{ \mathcal{A}^{\text{NLSM}}_{(\bar{l}_1|12|\bar{l}_2)}\mathcal{A}^{\text{NLSM}}_{(l_2|34|l_1)}}{l^2(l+k_{12})^2}\,.
\end{equation}
We define the internal loop momenta as, $\bar{l}_1 =-l_1= l$ and $\bar{l}_2 =-l_2= -(l+k_{12})$. The integrand numerator can be easily evaluated, yielding the following expression:
\begin{align}
  \sum_{\text{states}} \mathcal{A}^{\text{NLSM}}_{(\bar{l}_1|12|\bar{l}_2)}\mathcal{A}^{\text{NLSM}}_{(l_2|34|l_1)}&= 4f_\pi^{-4}C^{\text{box}}_{(1234)} \left[(k_2\cdot \bar{l}_1)(k_3\cdot l_1)+(k_1\cdot \bar{l}_1)(k_4\cdot l_1)\right]\\
&+4f_\pi^{-4}C^{\text{box}}_{(1243)} \left[(k_1\cdot \bar{l}_1)(k_3\cdot l_1)+(k_2\cdot \bar{l}_1)(k_4\cdot l_1)\right]\,.
\end{align}
Again, following the reasoning in \sect{sec:EMU} we do not need to worry about any contact terms that may have been missed in this cut -- any such terms do not exist for an even-point theory.

We note that since the internal pions are identical, this integral comes dressed with an internal symmetry factor of $S_{(12|34)}=\frac{1}{2}$. Moreover, to  capture all channels we can sum over all cyclic permutations:
\begin{equation}\label{eq:NLSM1loop}
\mathcal{A}^{\text{NLSM}}_{\text{1-loop}}=\frac{1}{2}\scaleIntApion +\text{cyc}(2,3,4)\,.
\end{equation}
Now, given the appearance of loop momenta in the integrand, we must proceed to the next step in our integration procedure: tensor reduction. 
\paragraph{\textbf{Reduction}} Since there are up to two powers of loop momenta appearing in the integrand, it is clear that there will be dimensional dependence when applying the Passarino-Veltman integral reduction of \eqn{eq:bubRed}. In doing so, we can write the one-loop pion amplitude completely in terms of scalar kinematics and scalar bubble integrals:
\begin{equation}
\mathcal{A}^{\text{NLSM}}_{\text{1-loop}} = f_{\pi}^{-4}C^{\text{box}}_{(1234)}\bigg[\frac{s_{12}I^D_2(k_{12})}{4}\left(s_{12}+\frac{s_{14}-s_{13}}{D-1}\right)+(1\leftrightarrow 3)\bigg] +\text{cyc}(2,3,4)\,.
\end{equation}
At this point, this is a dimensionally agnostic one-loop amplitude for NLSM. The next step is to plug particular values of $D$ into the evaluated analytic expressions of $I^D_2$ provided in \eqn{eq:integrals}. 
\paragraph{\textbf{Integration}}  In the study of NLSM loop-level amplitudes, we provide two examples for the integration dimension, $D=4-2\epsilon$ and $D=2-2\epsilon$, since the critical dimension of NLSM is $D=2$. In an $\overline{\text{MS}}$ renormalization scheme, the $\epsilon$-expanded bubble integrals for these two dimension choices are as follows:
\begin{align}
I_2^{4-2\epsilon}(k_{ij}) = \frac{i}{16\pi^2}\left[\frac{1}{\epsilon} - \text{ln}(-s_{ij})\right]+\mathcal{O}(\epsilon)\,,
\\
I_2^{2-2\epsilon}(k_{ij}) = \frac{i}{2\pi s_{ij}}\left[\frac{1}{\epsilon} - \text{ln}(-s_{ij})\right]+\mathcal{O}(\epsilon)\,.
\end{align}
Plugging these into our $D$-dimensional expressions, and dropping scheme dependent rational terms at subleading order, yields the following color-ordered 1-loop pion amplitudes for each respective dimension:
\begin{align}
A^{4-2\epsilon}_{(1234)} &=\frac{i}{48\pi^2 }f_{\pi}^{-4}\left[\frac{4\sigma_2}{\epsilon}+\left(s_{12}(s_{13}-s_{12})\frac{\text{ln}(-s_{12})}{2}+(1\leftrightarrow 3)\right)+\frac{1}{6}(s_{13}^2 + 2s_{12}s_{23})\right]+\mathcal{O}(\epsilon)
\\
A^{2-2\epsilon}_{(1234)} &= \frac{i}{2\pi }f_{\pi}^{-4}\left[\frac{s_{13}}{\epsilon}-s_{13}\frac{\text{ln}(-s_{12})+\text{ln}(-s_{23})-3}{2}\right] +\mathcal{O}(\epsilon) \,,
\label{eq:1loopPionD2}
\end{align}
where as defined in \sect{sec:basisT}, $\sigma_2=(s_{12}^2+s_{13}^2+s_{14}^2)/8$, and the full amplitude is recovered by summing over cyclic permutations of (2,3,4), 
\begin{equation}
\mathcal{A}^{\text{NLSM}}_{\text{1-loop}} = C^{\text{box}}_{(1234)}A^{D}_{(1234)}+\text{cyc}(2,3,4)\,.
\end{equation} 
It is worth emphasizing  that while the divergence in $D=4-2\epsilon$ is an ultraviolet divergence, the one in $D=2-2\epsilon$ is a logarithmic IR divergence, akin to that of $\mathcal{N}=4$ super-Yang-Mills (sYM) in the critical dimension of $D=4-2\epsilon$. Since all the particle states above are scalars, there are no helicity structures to map into, and thus we will bypass the final integration step of {projection}. Equipped with this one-loop warmup, we are now prepared to take on a two-loop example. 
\subsubsection{Two-loop}\label{sec:2loopNLSM}
The two-loop calculation is exactly the same procedure as one-loop, with some additional details that need to be accounted for when performing the integration step. We begin just as above with integrand construction via $D$-dimensional unitarity. 
\paragraph{\textbf{Construction}} The color-decomposition becomes slightly more complicated at two-loop than the procedure outlined in the previous section at one-loop. Just as before, we will still fix the integrand on color-dressed cuts, however, now we can no longer simply decompose the full color dressed amplitude in terms of integrated color-stripped objects. At two-loop, a convenient basis of adjoint color factors happens to be the {double-box} and the {cross-box}, shown below:
\begin{equation}
C^{\text{2box}}_{(12|34)} = \dBox \qquad C^{\text{Xbox}}_{([12]|34)} = \xBox
\end{equation}
By iteratively applying Jacboi relations on internal edges, any color structure can be expressed completely in terms of functional relabelings of these two graphs. Concretely, these cubic graphs come dressed with the following color factors written in terms of group theory structure constants:
\begin{align}
C^{\text{2box}}_{(12|34)}&\equiv f^{a_1 b_1b_7  }f^{ a_2 b_2 \color{nhpRed}{b_1} } f^{b_2 b_4  \color{nhpRed}{b_3}  } f^{b_3 b_6 b_7 } f^{b_4 a_3 b_5 }f^{b_5 a_4 b_6 } 
\\
C^{\text{Xbox}}_{([12]|34)}& \equiv f^{a_1 b_1b_7  }f^{ a_2 b_2 \color{nhpRed}{b_3} } f^{b_2 b_4  \color{nhpRed}{b_1}  } f^{b_3 b_6 b_7 } f^{b_4 a_3 b_5 }f^{b_5 a_4 b_6 }
\end{align}
The only functional difference between these color basis factors is the swap of the colored indices, $b_1$ and $b_3$, in the second and third vertices. Just as at one-loop, we can carefully choose a spanning set of two-loop unitarity cuts in terms of our color-dressed NLSM operators in order to land directly on this minimal color basis. In doing so, we can focus our attention to integrating the kinematics that weight each of the color basis element. 

As we argued in the introduction, two-loop amplitudes for even-point theories, like NLSM, are completely determined by two recursive integrals, the {double-bubble} and the {ostrich-diagram}, modulo theory-dependent kinematic numerators. For NLSM, the relevant integrals are as follows:

\begin{align}
\scaleIntCsmall &= \sum_{\text{states}}\int \frac{d^D l_{1}d^D l_{2}}{(2\pi)^{2D}}\frac{\mathcal{A}^{\text{NLSM}}_{(p_4|12|p_3)}\mathcal{A}^{\text{NLSM}}_{(\bar{p}_1|\bar{p}_4\bar{p}_3|\bar{p}_2)}\mathcal{A}^{\text{NLSM}}_{(p_2|34|p_1)}}{l_1^2(l_1+k_{12})^2l_2^2(l_2+k_{12})^2}
\\
\scaleIntBscalarsmall{1}{2}{3}{4}&=\sum_{\text{states}}\int \frac{d^D l_{1}d^D l_{2}}{(2\pi)^{2D}}\frac{\mathcal{A}^{\text{NLSM}}_{(2|q_4\bar{q}_3|\bar{q}_1)}\mathcal{A}^{\text{NLSM}}_{(1|\bar{q}_4q_3|\bar{q}_2)}\mathcal{A}^{\text{NLSM}}_{(q_2|34|q_1)}}{l_1^2(l_1+l_2+k_1)^2l_2^2(l_2+k_{12})^2}
\end{align}
where the internal momenta appearing in the NLSM operators, $\bar{p}_i= -p_i$ and $\bar{q}_i = -q_i$,  are defined in terms of the two loop momenta as follows:
\begin{align}
p_1 & = l_2+k_{12} \qquad p_2 =-l_2\qquad p_3 =l_1\qquad p_4 = -(l_1+k_{12})\,,
\\
q_1&= l_2+k_{12}
 \qquad q_2 = -l_2 
\qquad q_3 = l_1
\qquad q_4 =  (l_1+l_2+k_1)\,.
\end{align}
Due to the simplicity of scalar theories like NLSM, the state sum for the double-bubble numerator is quite simple, and can be expressed concisely below:
\begin{equation}\label{eq:NLSM2Bub}
\begin{aligned}
\text{Cut}\left(\! \scaleIntCsmall \! \right)&= \sum_{\text{states}}\mathcal{A}^{\text{NLSM}}_{(p_4|12|p_3)}\mathcal{A}^{\text{NLSM}}_{(\bar{p}_1|\bar{p}_4\bar{p}_3|\bar{p}_2)}\mathcal{A}^{\text{NLSM}}_{(p_2|34|p_1)}
\\
&=C^{\text{2box}}_{(12|34)} \left[\tau^{(1)}_3 \tau_{13}\tau^{(3)}_{1}+\tau^{(1)}_3 \tau_{23}\tau^{(3)}_{2}+\tau^{(2)}_3 \tau_{23}\tau^{(3)}_{1}+\tau^{(2)}_3 \tau_{13}\tau^{(3)}_{2}\right]
\\
&\,+C^{\text{2box}}_{(12|43)}\left[\tau^{(2)}_3 \tau_{23}\tau^{(3)}_{2}+\tau^{(2)}_3 \tau_{13}\tau^{(3)}_{1}+\tau^{(1)}_3 \tau_{23}\tau^{(3)}_{1}+\tau^{(1)}_3 \tau_{13}\tau^{(3)}_{2}\right]
\end{aligned}
\end{equation}
where we have introduced the following notation above that combines internal and external momenta, $\tau^{(i)}_j = (k_i  + p_j)^2$ and $\tau_{ij} = (p_i + p_j)^2$. Plugging in the color-dressed NLSM amplitudes, the same can be done for the ostrich-diagram cut, giving us

\begin{equation}\label{eq:NLSMostrich}
\begin{aligned}
\text{Cut}\left(\! \scaleIntBscalarsmall{1}{2}{3}{4} \! \right)&=\sum_{\text{states}}\mathcal{A}^{\text{NLSM}}_{(2|q_4\bar{q}_3|\bar{q}_1)}\mathcal{A}^{\text{NLSM}}_{(1|\bar{q}_4q_3|\bar{q}_2)}\mathcal{A}^{\text{NLSM}}_{(q_2|34|q_1)}
\\
&=C^{\text{Xbox}}_{([12]|34)}\left[\tau^{(3)}_1+\tau^{(3)}_2\right] \left[ \tau^{(2)}_3\tau^{(1)}_3+ \tau^{(2)}_4\tau^{(1)}_4\right]
\\
&-\left[C^{\text{2box}}_{(12|34)}\tau^{(3)}_1+C^{\text{2box}}_{(12|43)}\tau^{(3)}_2\right] \left[\tau^{(2)}_3\tau^{(1)}_4+ \tau^{(2)}_4\tau^{(1)}_3\right] \,.
\end{aligned}
\end{equation}
The kinematic variables $\tau^{(i)}_j $ and $\tau_{ij}$ are the same, except we have made the replacement $p_i \rightarrow q_i$. The full two-loop NLSM amplitude can thus be computed by summing over the distinct labels of the resulting integrals, each weighted by suitable internal symmetry factors:
\begin{equation}
\label{eq:NLSM2loop}
\mathcal{A}^{\text{NLSM}}_{\text{2-loop}}=\frac{1}{4}\left[ \scaleIntCsmall \right]+ \frac{1}{2}\left[\scaleIntBscalarsmall{1}{2}{3}{4}+\scaleIntBscalarsmall{4}{3}{2}{1}\right]+\text{cyc}(2,3,4)\,.
\end{equation}
We stress that the integrals appearing above are complicated by tensor integrals with many powers of loop momenta. However, as noted in the introduction, since both integrals are recursively one-loop, we can again apply a two-loop generalization of Passarino-Veltman for both the bubble and triangle integrals, stated in \eqn{eq:bubRed} and \eqn{eq:triRed}, respectively. With this, we proceed to the next step in the calculation.

\paragraph{\textbf{Reduction}} Just as with one-loop, we will now reduce the integrals of \eqn{eq:NLSM2loop} to the a basis of $D$-dimensional scalar integrals using the EMU loop reduction. The double-bubble contribution yields the following:
\begin{equation}
\label{eq:NLSM2bub}
\begin{aligned}
\scaleIntCsmall &= f_{\pi}^{-6}C^{\text{2box}}_{(12|34)} \left[\frac{(s_{12}I_2^D(k_{12}))^2}{2}\left(\frac{s_{14}-s_{13}}{(D-1)^2}+s_{12}\right)\right]
\\
&+f_{\pi}^{-6}C^{\text{2box}}_{(12|43)} \left[\frac{(s_{12}I_2^D(k_{12}))^2}{2}\left(\frac{s_{13}-s_{14}}{(D-1)^2}+s_{12}\right)\right]\,.
\end{aligned}
\end{equation}
Likewise, we can perform a similar reduction to the ostrich integrals -- but with a small twist. The double-bubble integral above is completely separable, which allows one to treat the tensor reduction of each loop integral as independent. This is not so for the ostrich-diagram integral. The ostrich-diagram carries an internal bubble, $I^D_2(q)$, with a loop dependent scale factor of $q=l_2+k_1$. Thus, we must first reduce the powers of $l_1$ and {then} perform a tensor reduction on all the remaining $l_2$ factors. Doing so yields the following result:
\begin{equation}\label{eq:NLSMostrich}
\begin{aligned}
\scaleIntBscalarsmall{1}{2}{3}{4} &= f_{\pi}^{-6}C^{\text{2box}}_{(12|34)} \frac{s_{12}}{3} \left[\frac{(D-1)(D-4)s_{14}+2(D-2)^2 s_{13}}{(D-1)(4-3D)}\right][I_3\circ I_2]^D(k_{12})
\\
&+f_{\pi}^{-6}C^{\text{2box}}_{(12|43)} \frac{s_{12}}{3} \left[\frac{(D-1)(D-4)s_{13}+2(D-2)^2 s_{14}}{(D-1)(4-3D)}\right][I_3\circ I_2]^D(k_{12})
\\
&+f_{\pi}^{-6}C^{\text{Xbox}}_{([12]|34)}\left[\frac{s_{12}^2}{3}\frac{D+1}{D-1}\right][I_3\circ I_2]^D(k_{12})\,.
\end{aligned}
\end{equation}
where we have defined the scalar ostrich-diagram integral, $[I_3\circ I_2]^D(k_{12})$, as a $D$-dimensional convolution of $I_2^D$ and $I_3^D$, provided below:
\begin{equation}
[I_3\circ I_2]^D(k_{12}) \equiv \int \frac{d^Dl_1}{(2\pi)^D} \frac{d^Dl_2}{(2\pi)^D} \frac{(l_2+k_1)^2}{l_1^2  (l_1+l_2+k_1)^2l_2^2 (l_2+k_{12})^2}\,.
\end{equation}
We have added the additional factor of $(l_2+k_1)^2$ in the definition of $[I_3\circ I_2]^D(k_{12})$ to simplify the final expression in \eqn{eq:NLSMostrich}. We will use this as our basis integral throughout the text. This two-loop integral can be evaluated analytically using the $D$-dimensional single scale bubble and triangle integral expressions in \eqn{eq:integrals}. Thus, we have completely reduced the full two-loop NLSM amplitude to a linear combination of scalar integrals, $I_2^D(k_{ij})$ and $[I_3\circ I_2]^D(k_{ij})$. Now we will proceed to the final step of evaluating the amplitude in particular dimensions.
\paragraph{\textbf{Exponentiation}} While there is plenty of physics to be extracted from the combination of \eqn{eq:NLSM2bub} and \eqn{eq:NLSMostrich}, we will focus on the behavior of the leading divergence for $D=2-2\epsilon$ for the target space model describing pions on $\mathbb{CP}^1\cong SU(2) /U(1)$. The tree amplitudes for this theory can be obtained from \eqn{eq:NLSMamp} by plugging in multi-trace color factors of the form,
\begin{equation}
\cHL_{(ijkl)} = \delta_{ik}\delta_{jl} -\delta_{il}\delta_{jk} \,.
\end{equation}
The delta function indices select out (anti)-holomorphic pion fields that live on the $\mathbb{CP}^1$ target space. Plugging in explicit values for the ``color" indices, we obtain the following tree-level amplitude:
\begin{equation}
\mathcal{A}^{\mathbb{CP}^1}_{\text{tree}}(Z_1,\bar{Z}_2,Z_3,\bar{Z}_4) = f_\pi^{-2}s_{13}\,.
\end{equation}
Using this tree-level contact as the seed in our EMU construction, we can directly compute the one- and two-loop amplitudes from \eqn{eq:NLSM1loop} and \eqn{eq:NLSM2loop}. Furthermore, the leading divergences for the loop integrals in $D=2-2\epsilon$ dimensions are the following:
\begin{align}
I_2^{2-2\epsilon}(k_{ij})I_2^{2-2\epsilon}(k_{ij}) &= -\frac{1}{s_{ij }^2}\frac{1}{(2\pi \epsilon)^2} + \mathcal{O}(\epsilon^{-1}) \,,
\\
 [I_3\circ I_2]^{2-2\epsilon}{(k_{ij})} &= -\frac{3}{8s_{ij }}\frac{1}{(2\pi \epsilon)^2} + \mathcal{O}(\epsilon^{-1}) \,.
\end{align}
The analogous divergences for $D=4-2\epsilon$ can be extracted directly from the integral expressions in \eqn{eq:integrals}. Plugging these values in for the scattering process $\mathcal{A}(Z_1,\bar{Z}_2,Z_3,\bar{Z}_4) $, we can extract the full leading logarithmic divergence from the one-loop color dressed amplitude in \eqn{eq:NLSM1loop},
\begin{equation}\label{eq:1loopCP1}
\mathcal{A}^{\mathbb{CP}^1}_{\text{1-loop}} = - \left[\frac{if_{\pi}^{-2} }{4\pi \epsilon}\right]f_\pi^{-2}s_{13}+\mathcal{O}(\epsilon^0)\,,
\end{equation}
and likewise for the two-loop result upon evaluating the integrals in $D=2-2\epsilon$,
\begin{equation}
\mathcal{A}^{\mathbb{CP}^1}_{\text{2-loop}} = \frac{1}{2} \left[\frac{if_{\pi}^{-2} }{4\pi \epsilon}\right]^2f_\pi^{-2}s_{13} +\mathcal{O}(\epsilon^{-1})\,.
\end{equation}
We write the two-loop amplitude in this suggestive form to emphasize that the logarithmic divergence present at one-loop in \eqn{eq:1loopCP1} appears to exponentiate! More concretely, we have demonstrated through explicit calculation that through two-loop order, the leading divergence of the full color dressed amplitude goes as follows:
\begin{eBox}
\begin{equation}\label{eq:expNLSM}
\mathcal{A}_{2-2\epsilon}^{\mathbb{CP}^1}\bigg|_{\text{div.}} = \mathcal{A}^{\text{tree}}\left( 1 + \frac{\mathcal{A}^{\text{1-loop}}}{ \mathcal{A}^{\text{tree}}} + \frac{1}{2}\left[\frac{\mathcal{A}^{\text{1-loop}}}{ \mathcal{A}^{\text{tree}}} \right]^2 + \cdots \right)\,.
\end{equation}
\end{eBox}
This iterative form suggests that the IR divergence of the $\mathbb{CP}^1$ model can be neatly extracted from the perturbative series as an overall factor, $e^\Omega$, where $\Omega$ is the divergent piece of the one-loop amplitude. Similar exponentiation of the IR has a long history in the context of gravity amplitudes \cite{Weinberg:1965nx}, for which IR divergences appear due to soft graviton emissions that mediate long range interactions. Considering the established relationship \cite{deWit:1983xhu,deWit:1992cr,deWit:1992wf, Chiodaroli:2014xia} between coset manifolds (target spaces) and supergravity theories, there's a possibility that the exponentiation of the $\mathbb{CP}^1$ model in \eqn{eq:expNLSM} is related to the analogous behavior of supergravity amplitudes \cite{Naculich:2008ew,White:2011yy,DiVecchia:2019myk,DiVecchia:2019kta,Heissenberg:2021tzo}.

In addition to the exponential structure of \eqn{eq:expNLSM} that we have computed for the $\mathbb{CP}^1$ model, there is potentially a parallel story for the planar limit of the chiral NLSM amplitudes. Indeed, if one takes $D\rightarrow 2$ in \eqn{eq:NLSM2bub}, we find that the leading divergence of the double bubble in $D=2-2\epsilon$ is precisely one half the square of the one-loop divergence in \eqn{eq:1loopPionD2}. However, while the cross-box term appearing in \eqn{eq:NLSMostrich} is subleading when $N_c\rightarrow 0$, there are still relevant contributions from the double-boxes appearing in the ostrich integral. In principle one could add additional operators \cite{Elvang:2018dco} or particle states \cite{Cachazo:2016njl} to \eqn{eq:NLSMLag} that cancel the additional double-box contributions from the ostrich integral, while preserving the soft behavior needed for our EMU construction. Doing so would extend the exponential behavior to the subleading logarithms in the planar limit, similar to what was found in \cite{Anastasiou:2003kj,Bern:2005iz} for planar $\mathcal{N}=4$ sYM in $D=4-2\epsilon$. The iterated structure of planar $\mathcal{N}=4$ has been linked to the integrability of the theory \cite{Sterman:2002qn,Minahan:2002ve,Bena:2003wd,Beisert:2003jj,Beisert:2003jb,Beisert:2003yb,Dolan:2003uh,Arutyunov:2003rg,Ryzhov:2004nz,Frolov:2005iq}. We leave identifying such an extended theory consistent with exponentiation as a compelling direction of future study.

Now that we have walked through our procedure for computing two-loop even point EFT amplitudes with NLSM as an exemplar, we are prepared to proceed to the main results of the text. While the construction of DBIVA integrands is more involved than constructing the simple expressions of NLSM, the general procedure is exactly the same. The only differences being additional gauge independent state sums, and the generally more complex tensor reduction. 

\subsection{DBIVA via EMU}\label{sec:DBIU}
In the remainder of this section, we will construct four-photon matrix elements in DBIVA theories through two-loop order. 
While we will consider a number of different internal matter states inside the loops, the interactions for which are not captured by the DBIVA Lagrangian \eqn{dbivaLag}, we will require that all the external states are vectors. This will allow us to map our loop-level results to the basis of gauge invariant tensors described in \sect{sec:basisT}. Furthermore, in the last section, we will investigate how the loop level contributions turn on new operators in the EFT written in \eqn{biLag}. 
 
As we noted in the introduction, DBIVA tree-level amplitudes, which appear in the field theory limit of the abelianized open superstring, can be realized as a {double copy} between NLSM and sYM \cite{Cachazo:2014xea}, 
\begin{equation}
\mathcal{M}^{\text{DBIVA}} = \mathcal{A}^{\text{NLSM}} \otimes \mathcal{A}^{\text{sYM}}\,.
\end{equation}
To carry out this construction at the (multi-)loop level, we would simply replace the color factors of NLSM with color-dual loop-level numerators, and then integrate the result. In the case of $\mathcal{N}=4$ sYM, color-dual representations are known through four-loop \cite{BCJLoop, Bern:2012uf}. Indeed, in \sect{sec:DBIvDC} we will use use double-copy to verify the $\mathcal{N}=4$ DBIVA results we construct here via unitarity. If we had access to color-dual representations for NLSM at two-loop, we additionally could plug those into the sYM integrands constructed from simplified methods of supersymmetric sums \cite{SuperSum}.

Just as in the previous section where we computed NLSM amplitudes, the recursively one-loop behavior will allow us to only consider the four-point contacts when constructing the physical parts of the two-loop integrand. We will start by reproducing the known one-loop results from a completely $D$-dimensional framework, and then move onto our novel two-loop results. 

\subsubsection{One-loop DBIVA}\label{sec:1loopDBIU}
Analogous to our procedure in the previous section where we defined $\mathcal{A}^{\text{NLSM}}_{(1|23|4)}$, first we will define a set of four-point operators for each distinct set of particle interaction. The $D$-dimensional contacts needed for DBIVA amplitudes at one-loop are as follows:
\begin{align}
\mathcal{M}(1_\gamma,2_\gamma,3_\gamma,4_\gamma) &= 2\text{tr}(F_1F_2F_3F_4) - \frac{1}{2}\text{tr}(F_1F_2)\text{tr}(F_3F_4)+\text{cyc}(1,2,3) \equiv t_8 F^4\,,
\\
\mathcal{M}(1_\lambda,2_\gamma,3_\gamma,4_{\overline{\lambda}}) &= s_{13}\bar{u}_1(\slashed{\varepsilon}_2\slashed{k}_{12}\slashed{\varepsilon}_3) \bar{v}_4 + s_{12}\bar{u}_1(\slashed{\varepsilon}_3\slashed{k}_{13}\slashed{\varepsilon}_2) \bar{v}_4\,,
\\
\mathcal{M}(1_X,2_\gamma,3_\gamma,4_{\overline{X}}) &= 2 (k_1 F_2 F_3 k_1) +2 (k_4 F_3 F_2 k_4)\,,
\end{align}
where we have introduced notation $(k_a F_b F_c k_a) \equiv k_a^\mu F^{\mu\nu}_bF^{\nu\rho}_c k_a^\rho$ for the mixed scalar-vector amplitude. The particle content is labeled by $\gamma$ for the BI photons, $\lambda$ for the VA fermions, and $X$ for the Dirac scalars. With these tree-level amplitudes in hand, we define the following set of matrix-elements needed for integrand construction:
\begin{align}
\mathcal{M}^{\gamma \gamma \gamma \gamma }_{(1|23|4)} &= \mathcal{M}(1_\gamma,2_\gamma,3_\gamma,4_\gamma)   \,,
\\
\mathcal{M}^{\lambda \gamma \gamma \bar{\lambda} }_{(1|23|4)} &= \mathcal{M}(1_\lambda,2_\gamma,3_\gamma,4_{\overline{\lambda}})\,,
\\
\mathcal{M}^{X \gamma \gamma \bar{X} }_{(1|23|4)} &= \mathcal{M}(1_X,2_\gamma,3_\gamma,4_{\overline{X}}) \,.
\end{align}
While these are a subset of the four-point operators we will use in the two-loop construction, they are sufficient for all the one-loop external-photon amplitudes. To construct the integrands, we will use the $D$-dimensional state sums for vectors and fermions, which we provide below:
\begin{align}
\sum_{\text{states}}\varepsilon^\mu_{(l)}\varepsilon^\nu_{(-l)} &= \eta^{\mu\nu} - \frac{l^\mu q^\nu+l^\nu q^\mu}{l\cdot q},
\\
\sum_{\text{states}}u_{(l)}\bar{v}_{(-l)} &= \frac{1}{2}(1\pm \Gamma_5)\Gamma_\mu l^\mu \,,
\end{align}
where $q^2=0$ is a null-reference momentum. Here, $\Gamma_\mu$ are $D$-dimensional gamma matrices endowed with all the normal Clifford algebraic relations. Since we want to keep this as $D$-dimensional as possible, we define $\Gamma_5$ as the symbol representing the $(D+1)$-{th} gamma matrix that anti-commutes with all other $\Gamma_\mu$. Using this definition of $\Gamma_5$, we could also define a chiral projection state operator $P_{\pm}=\frac{1}{2}(1\pm \Gamma_5)$ in a dimension agnostic form. Chiral operators become relevant for our external photon amplitudes when computing internal fermion contributions at two-loop. To see why, note that since we are only computing external photon amplitudes, the matrix elements must be {parity even}. This means that if there is a single chiral trace, the $\Gamma_5$ contribution must integrate to zero:
\begin{equation}
2\text{tr}_{\pm}[\cdots] \equiv \text{tr}[(1\pm \Gamma_5)\cdots]  \xrightarrow{\int d\Pi_{\text{loop}}} \text{tr}[\cdots]\,.
\end{equation}
This property significantly simplifies the two-loop reduction in the presence of {single} chiral trace. However, at two-loop we can also get multi-trace contributions. In this case, the parity odd contribution (sourced by odd powers of $\Gamma_5$) must vanish after integration as follows:
\begin{equation}
4\text{tr}_{\pm}[\cdots] \text{tr}_{\pm}[\cdots]  \xrightarrow{\int d\Pi_{\text{loop}}} \text{tr}[\cdots]\text{tr}[\cdots]+\text{tr}[\Gamma_5\cdots]\text{tr}[\Gamma_5\cdots]\,.
\end{equation}
The second term is also parity even, and will contribute to $D$-dimensional Gramm determinants in the presence of two internal fermion loops. While in principle this term is relevant for two-loop diagrams with internal fermions, due to the simplicity of $\mathcal{N}=4$ DBIVA state sums, we won't need to account for it in our analysis at two-loop. 

Before constructing the integrands, below we provide our conventions for the $D$-dimensional spinors and gamma matrices. We will normalize the gamma matrices as follows:
\begin{equation}
\text{Tr}(\Gamma_\mu\Gamma^\mu) = 2^{D/2-1}D\,.
\end{equation}
Furthermore, since we will eventually be evaluating our matrix elements in $D=4$ after integration, we assume that the $D$-dimensional generalization of the Majorana condition holds throughout the calculation. That is, the $\bar{u}$ and $v$ spinors obey the following relationship:
\begin{equation}
\bar{u} = v^T \mathcal{C} \qquad v = - \mathcal{C} \bar{u}^T\,,
\end{equation}
where the $D$-dimensional charge conjugation operator can be defined in terms of the gamma matrices, $\Gamma_\mu$, as follows:
\begin{equation}
\Gamma_\mu = - \mathcal{C} ^{-1}\Gamma_\mu ^T\, \mathcal{C}\,.
\end{equation}
From this, the spinor strings obey another in addition to the normal Clifford algebra identities \cite{Chiodaroli2013upa}:
\begin{equation}
\bar{u}(\Gamma_{\mu_1} \cdots \Gamma_{\mu _n}) v =(-1)^n \bar{u} (\Gamma_{\mu_n} \cdots \Gamma_{\mu _1} )v\,.
\end{equation}
This relationship essentially just imposes Fermi statistics on the identical Majorana fermions \cite{Carrasco:2023vjg}. Now we are prepared to constructing the integrand with the state sums and operators defined above. 
\paragraph{\textbf{Construction}} Just as before, the first step in our procedure is to construct the integrand with all the internal loop dependence present. Taking the operators defined above, and applying the appropriate state sums, we obtain:

\begin{align}\label{eq:1loopIntegrands}
\text{Cut}\left(\!\scaleIntAvector{1}{2}{3}{4}\!\right) &= \sum_{ \text{states}}\mathcal{M}^{\gamma \gamma \gamma \gamma }_{(q_1|12|q_2)}\mathcal{M}^{\gamma \gamma \gamma \gamma }_{(\bar{q}_2|34|\bar{q}_1)} \equiv \text{Cut}(\mathcal{I}^{\text{1-loop}}_{N_\gamma} )\,,
\\
\text{Cut}\left(\!\scaleIntAfermion{1}{2}{3}{4}\!\right)  &= \sum_{ \text{states}}\mathcal{M}^{\lambda \gamma \gamma \bar{\lambda} }_{(q_1|12|q_2)} \mathcal{M}^{\lambda \gamma \gamma \bar{\lambda} }_{(\bar{q}_2|34|\bar{q}_1)}\equiv \text{Cut}(\mathcal{I}^{\text{1-loop}}_{N_\lambda} )\,,
\\
\text{Cut}\left(\!\scaleIntAScalar{1}{2}{3}{4}\!\right)  &= \sum_{ \text{states}}\mathcal{M}^{X \gamma \gamma \bar{X} }_{(q_1|12|q_2)}\mathcal{M}^{X \gamma \gamma \bar{X} }_{(\bar{q}_2|34|\bar{q}_1)}\equiv \text{Cut}(\mathcal{I}^{\text{1-loop}}_{N_X} )\,,
\end{align}
where exposed internal particles are taken to be on-shell. Above we have defined the internal momenta as $q_1 = l$ and $q_2 = -(l_1+k_{12})$. The vector and fermion contributions are rather complicated -- however, to get a sense of what pops out of this $D$-dimensional construction, we provide the scalar cut below as an example:
\begin{equation}
\text{Cut}(\mathcal{I}^{\text{1-loop}})= 16(q_1 F_1F_2 q_1)(q_1 F_3F_4 q_1)\,.
\end{equation}
Constructing the full amplitude is then just a matter of summing over all cut contributions, each weighted by symmetry factors $S_\alpha$ and the number, $N_\alpha$, of $\alpha$-type particles:
\begin{equation}\label{oneLoopCuts}
\mathcal{M}^{\text{1-loop}} = \sum_{\alpha } \frac{N_\alpha}{S_{\alpha}}\int \frac{d^D l}{(2\pi)^D} \frac{\text{Cut}(\mathcal{I}^{\text{1-loop}}_{N_\alpha} )}{l^2(l+k_{12})^2} +\text{cyc}(2,3,4)\,.
\end{equation}
Since the scalars are complex, and the fermions are oriented, they come dressed with symmetry factors of $S_X = - S_\lambda = 1$. The minus sign is due to the presence of a single fermion loop. Furthermore, since the photons are indistinguishable, they carry an internal symmetry factor of $S_\gamma = 2$. Now we are prepared to carry out the integral reduction and evaluate each contribution to the one-loop amplitude above. 
\paragraph{\textbf{Reduction}} While the full $D$-dimensional integral at one-loop is rather complicated for the vectors and fermions, the tensor reduced integrals are actually quite simple -- and its worth showing the result of applying \eqn{eq:bubRed} to the integrands above in \eqn{eq:1loopIntegrands}. After reducing the tensor integrals that appear in \eqn{oneLoopCuts}, we obtain

\begin{equation}
\begin{aligned}
\scaleIntAvector{1}{2}{3}{4} &=\left[\begin{aligned}
&\frac{(D^3- 24 D^2+ 68 D +72 )}{4}f_{12}f_{34}
\\
&+(2D^2-3D-8 )(f_{1234} + f_{1243}) 
\\
&+6(D+1)(f_{1324}-f_{13}f_{24}-f_{14}f_{23})
\end{aligned}\right]\frac{s_{12}^2 I_2^D(k_{12})}{D^2-1}\,,
\\\\
\scaleIntAfermion{1}{2}{3}{4} &= \left[\begin{aligned}
&\,\,\frac{(D-4)(D+1)}{8}(f_{1324}-f_{13}f_{24}-f_{14}f_{23})
\\
&-\frac{( D+8)}{8}f_{12}f_{34}+\frac{(D-1)}{4}(f_{1234} + f_{1243}) 
\end{aligned}\right]\frac{s_{12}^2 I_2^D(k_{12})}{D^2-1}2^{D/2-1}\,,
\\\\
\scaleIntAScalar{1}{2}{3}{4} &= \left[ \frac{D(D-6)}{4}f_{12}f_{34}+(f_{1234} + f_{1243}) \right]\frac{s_{12}^2 I_2^D(k_{12})}{D^2-1}\,.
\end{aligned}
\end{equation}
In terms of these integrals, the full amplitude at one-loop can be expressed as a sum over all distinct permutations of external legs, with each graph weighted by their associated symmetry factors:\footnote{Equivalently, the scalars could be real, in which case they would carry a symmetry factor of $S_{\text{Re}[X]} =\frac{1}{2}$. Complex scalars are then recovered by two real scalars multiplying the symmetry factor, $2S_{\text{Re}[X]} = S_X=\frac{2}{2}$}
\begin{equation}
\mathcal{M}^{\text{1-loop}} = \left[\frac{N_\gamma}{2}\!\!\scaleIntAvector{}{}{}{}-N_\lambda\!\!\scaleIntAfermion{}{}{}{} +N_X\!\!\scaleIntAScalar{}{}{}{} \!\!\right] +\text{cyc}(2,3,4)\,.
\end{equation}
Now we can proceed to evaluating the integrals in the dimension of interest. 
\paragraph{\textbf{Integration}} While the procedure that we have employed thus far allows us to extract all of the logarithmic dependence of the loop integrals, for the remainder of this paper we are going for focus our attention to the {leading order} divergences in $\epsilon$. The purpose of doing so is to identify the class of photon effective operators that we must add to the Born-Infeld action in order to cancel off anomalous rational terms in the $S$-matrix. We will leave future analysis of the novel DBIVA amplitudes in this paper as a direction of future study.

In order to capture all the tensor structures that survive after integration, including the evanescent contributions, we will map our results to the basis of tensor structures identified in \sect{sec:basisT}:
\begin{equation}
\mathcal{M}^{\text{photon-EFT}} = \sum_{x,y} a_{(x,y)}^{F^2F^2}\mathcal{T}^{F^2F^2}_{(x,y)}+a_{(x,y)}^{F^4}\mathcal{T}^{F^4}_{(x,y)}+a_{(x,y)}^{F^3}\mathcal{T}^{F^3}_{(x,y)}\,.
\end{equation}
We remind the reader that the photon structures given above are defined in \eqn{eq:basisTensors}. At one-loop mass-dimension, there are only four tensor structures that contribute:
\begin{equation}
\mathcal{M}^{\text{photon-EFT}}_{\text{1-loop}} = a_{(2,0)}^{F^2F^2}\mathcal{T}^{F^2F^2}_{(2,0)}+a_{(2,0)}^{F^4}\mathcal{T}^{F^4}_{(2,0)}+a_{(0,1)}^{F^2F^2}\mathcal{T}^{F^2F^2}_{(0,1)}+a_{(0,1)}^{F^4}\mathcal{T}^{F^4}_{(0,1)}\,.
\end{equation}
Plugging in $D=4-2\epsilon$ to the above evaluated one-loop integrals, the EFT expansion above forms a basis for all the algebraic (non-transcendental) parts of the one-loop integral. Explicitly, the Wilson coefficients above take on the follow values when we expand the integral in $D=4-2\epsilon$ dimensions,
\begin{align}
a_{(2,0)}^{F^2F^2}&=\frac{i}{(4\pi)^2}\frac{1}{\epsilon}\left[\frac{N_\gamma }{75}(-60+61\epsilon)+\frac{N_X }{225}(30+47\epsilon)-\frac{N_\lambda }{75}(15+\epsilon)\right]+\mathcal{O}(\epsilon)
\\
a_{(0,1)}^{F^2F^2}&=\frac{i}{(4\pi)^2}\frac{1}{\epsilon}\left[\frac{N_\gamma }{3}(6+4\epsilon)-\frac{N_\lambda}{3}\epsilon\right]+\mathcal{O}(\epsilon)
\\
a_{(2,0)}^{F^4}&=\frac{i}{(4\pi)^2}\frac{1}{\epsilon}\left[\frac{N_\gamma }{75}(105+17\epsilon)+\frac{N_X }{225}(15+16\epsilon)+\frac{N_\lambda }{150}(15-19\epsilon)\right]+\mathcal{O}(\epsilon)
\\
a_{(0,1)}^{F^4}&=\frac{i}{(4\pi)^2}\frac{1}{\epsilon}\left[\frac{N_\gamma }{25}(-20+22\epsilon)-\frac{N_X }{225}(30+32\epsilon)-\frac{N_\lambda }{25}(5+2\epsilon)\right]+\mathcal{O}(\epsilon)
\end{align}
We remind the reader that while there are four spanning $D$-dimensional four-photon structures, they have a non-trivial null space in four-dimensions where only three structures span the non-vanishing space. Since all of the above Wilson coefficients come dressed with a leading $1/\epsilon$ divergence for pure photon amplitudes, this indicates the emergence of a divergent {evanescent operator} that will contribute at two-loop order. We will see the consequences of this when calculating the anomalous matrix elements for pure BI at the next loop order. 
\paragraph{\textbf{Projection}} While the expressions above capture the full behavior of the photon amplitudes, they obscure the 4D physics captured by spinor helicity variables. Rather than plugging in explicit values for $D$ in the integral, it can be more informative to first project down onto a 4D basis of states, leaving the internal dimensional dependence untouched. Indeed, below we can see the effect of plugging in all-plus helicity configurations into the evaluated one-loop integrals above:
\begin{equation}
\begin{aligned}
\frac{1}{2}\scaleIntAvector{+}{+}{+}{+} &=\left[\frac{(D-4)(D-2)}{16(D^2-1)}s_{12}^2[12]^2[34]^2I^D_2(k_{12})\right]\frac{(D-2)}{2}\,,
\\\\
\scaleIntAfermion{+}{+}{+}{+}  &=\left[\frac{(D-4)(D-2)}{16(D^2-1)}s_{12}^2[12]^2[34]^2I^D_2(k_{12})\right]\frac{2^{D/2-1}}{2}\,,
\\\\
\frac{1}{2}\scaleIntAScalar{+}{+}{+}{+}  &= \left[\frac{(D-4)(D-2)}{16(D^2-1)}s_{12}^2[12]^2[34]^2I^D_2(k_{12})\right]\frac{2}{2}\,.
\end{aligned}
\end{equation}
Here we can see two characteristic properties of the anomaly cancellation that takes place when we introduce scalars and fermions into our spectrum. 
\begin{itemize}
\item First, all integrals carry a factor of $(D-4)$. This reflects the property that in $D=4$, all the tree-level amplitudes that contribute to the cut must vanish outside the MHV sector. Thus, any contribution to the all plus matrix element must be suppressed by a factor of $\epsilon = (4-D)/2$ in order to push the logarithms to $\mathcal{O}(\epsilon)$. 
\item Second, the common bracketed factor in all three integrals is weighted differently depending on whether the internal particle is a boson or a fermion. Bosonic contributions are weighted by the number of particles in the loop (in this case, 2 real scalars or $(D-2)$ helicity states), whereas the fermion weight is dependent on the dimension of the gamma matrix representation, where $\text{tr}[\Gamma_\mu \Gamma^\mu ]=2^{D/2-1} D $. In $D=4$, all of these contribute equal magnitude to the full amplitude. 
\end{itemize} 
Putting this all together, we obtain the following non-vanishing algebraic parts of the one-loop matrix elements:
\begin{align}
\mathcal{M}^{\text{DBIVA}} _{(--++)}&= \frac{1}{\epsilon}\frac{i}{(4\pi)^2}\left[\frac{N_\gamma}{2}s_{12}^2+\left(\frac{N_\gamma}{5}+\frac{N_\lambda}{20}+\frac{N_X}{30}\right)(s_{13}^2+s_{14}^2)\right][12]^2\langle34\rangle^2+\mathcal{O}(\epsilon^0)
\\
\mathcal{M}^{\text{DBIVA}}_{(-+++)}&= 0
\\
\mathcal{M}^{\text{DBIVA}}_{(++++)}&= -\frac{i}{(4\pi)^2}\frac{1}{60}\left(N_\gamma+N_X-N_\lambda\right) (s_{12}^4+s_{13}^4+s_{14}^4)\frac{[12][34]}{\langle12\rangle\langle 34\rangle}+\mathcal{O}(\epsilon^1) \label{eq:generalDBI1loop}
\end{align}
The $(-+++)$ matrix element is identically zero due to $D$-dimensional four-point kinematics --- the only available helicity structure is $\langle 1|2|3]^2[24]^2$ which must be weighted by a mass-dimension 2 permutation invariant. The only such permutation invariant is $s+t+u =0$, which vanishes regardless of the integration dimension. As a check, above we have reproduced the results found in \cite{Elvang:2020kuj}, which used four-dimensional spinor helicity and dimension-shifting relations when performing the integration. This serves as a nice verification of our $D$-dimensional methods. Now we will proceed with the two-loop calculation. 
\subsubsection{Two-loop Born-Infeld}\label{sec:2loopBIU}
For all of the two-loop calculations, we will drop the header labelling which step in the process we are presenting. While we will omit these markers, our procedure is still the same: \textbf{construction}, \textbf{reduction}, \textbf{integration} and then \textbf{projection}. Due to the formidably large expressions that result from doing this calculation completely covariantly, most of our amplitudes will be presented {after} projecting down to 4D helicity states. 

To begin, we will compute the pure Born-Infeld two-loop amplitude. Like pions, there are only two diagrams that contribute at this loop order. Including symmetry factors, the full Born-Infeld amplitude at two-loop can be expressed as follows:
\begin{equation}
\mathcal{M}^{\text{BI}}_{\text{2-loop}}=\frac{1}{2}\,\,\scaleIntBtune{photon}{photon}{photon}{photon}{}{}{}{}\!\!+ \frac{1}{4}\scaleIntCtune{photon}{}{photon}{} \!\!+\text{perms}\,.
\end{equation}
As in the previous section, exposed legs implicitly sum over internal states. The grey blobs above represent $D$-dimensional $t_8F^4$ operator insertions from the four-point BI tree amplitudes, which we labelled as $\mathcal{M}^{\gamma\gamma\gamma\gamma}_{(1234)}$. We give the internal loop momenta the same internal labels as we did the pions in \eqn{eq:NLSM2Bub} and \eqn{eq:NLSMostrich}, giving us the following integrals to evaluate for two-loop pure BI, 
\begin{align}
\scaleIntCtune{photon}{}{photon}{}&= \sum_{\text{states}}\int \frac{d^D l_{1}d^D l_{2}}{(2\pi)^{2D}}\frac{\mathcal{M}^{\gamma\gamma\gamma\gamma}_{(p_412p_3)}\mathcal{M}^{\gamma\gamma\gamma\gamma}_{(\bar{p}_1\bar{p}_4\bar{p}_3\bar{p}_2)}\mathcal{M}^{\gamma\gamma\gamma\gamma}_{(p_234p_1)}}{l_1^2(l_1+k_{12})^2l_2^2(l_2+k_{12})^2}\,,
\\
\scaleIntBtune{photon}{photon}{photon}{photon}{}{}{}{}\,\,\,\,&=\sum_{\text{states}}\int \frac{d^D l_{1}d^D l_{2}}{(2\pi)^{2D}}\frac{\mathcal{M}^{\gamma\gamma\gamma\gamma}_{(2q_4\bar{q}_3\bar{q}_1)}\mathcal{M}^{\gamma\gamma\gamma\gamma}_{(1\bar{q}_4q_3\bar{q}_2)}\mathcal{M}^{\gamma\gamma\gamma\gamma}_{(q_234q_1)}}{l_1^2(l_1+l_2+k_1)^2l_2^2(l_2+k_{12})^2}\,.
\end{align}
where the internal momenta obey the same convetion, $\bar{p}_i= -p_i$ and $\bar{q}_i = -q_i$. Performing the tensor reduction on the internal loop momenta and projecting to 4D helicity states yields divergent quantities for all helicity configurations. In the interest of projecting to a $D$-dimensional basis of operators, and analyzing the $U(1)$ anomaly present at two-loop, we will just focus on the leading divergences for each helicity configuration. 
\paragraph{Leading $(++++)$ divergence} As we saw above, the one-loop matrix element has a rational all-plus contribution. This manifestly breaks the $U(1)$ duality invariance present as tree-level. Moreover, this will contribute to non-vanishing 4D cut of the form:
\begin{equation}
\text{Cut}\left[\mathcal{M}^{\text{BI,2-loop}}_{(++++)} \right]^{D=4} = \mathcal{M}^{\text{BI,1-loop}}_{(++++)} \times \mathcal{M}^{\text{BI,tree}}_{(--++)}\,.
\end{equation}
This cut should source a logarithmic discontinuity, and thus, the leading contribution for two-loop all-plus matrix element should diverge as $1/\epsilon$. Indeed this is what we find:
\begin{eBox}
\begin{equation}
\label{eq:allPlus2}
\mathcal{M}^{\text{BI,2-loop}}_{(++++)} =\frac{1}{\epsilon} \frac{29}{600} \frac{1}{(4\pi)^4}(s_{12}^6+s_{13}^6+s_{14}^6)\frac{[12][34]}{\langle 12\rangle \langle 34\rangle }+\mathcal{O}(\epsilon^0)\,.
\end{equation}
\end{eBox}
In principle, this divergence could be cancelled by the addition of a counterterm at $\mathcal{O}(\alpha'^4)$ inserted into a one-loop matrix element with $t_8F^4$ at $\mathcal{O}(\alpha'^{\,2})$. We will explore the effect of such anomaly cancelling counterterms in the next section, and we will see that this alone is not sufficient to cancel the divergence above. As we noted previously, there are additional 4D operators that appear at one-loop that will vanish when plugging in $(++++)$ physical states. These too could secretly contribute to the divergence expressed above. We can see this more clearly for the $(-+++)$ result below. 
\paragraph{Leading $(-+++)$ divergence} The same cut construction above suggests a different story for $(-+++)$ at two-loop. Indeed, the following cut should vanish when taken on-shell in $D=4$:
\begin{equation}
\text{Cut}\left[\mathcal{M}^{\text{BI,2-loop}}_{(-+++)} \right]^{D=4} = \mathcal{M}^{\text{BI,1-loop}}_{(-+++)} \times \mathcal{M}^{\text{BI,tree}}_{(--++)}\,,
\end{equation}
However, we found that $\mathcal{M}^{\text{BI,1-loop}}_{(-+++)} =0$ due to $D$-dimensional four-point kinematics. Despite this, we find similarly to the all-plus helicity configuration above in \eqn{eq:allPlus2}, the one-minus matrix element {also} carries a leading order $1/\epsilon$ divergence:
\begin{eBox}
\begin{equation}
\mathcal{M}^{\text{BI,2-loop}}_{(-+++)} =-\frac{1}{\epsilon} \frac{1}{75} \frac{1}{(4\pi)^4}(s_{12}^3+s_{13}^3+s_{14}^3) \langle 1|2|3]^2[24]^2 +\mathcal{O}(\epsilon^0)\,.
\label{eq:oneMinus2loop}
\end{equation}
\end{eBox}
At first glance, this appears to be a violation of the Optical Theorem. However, this divergence is sourced by one-loop evanescent operators that vanish in $D=4$, but which carries a $1/\epsilon$ divergence in $D=4-2\epsilon$. In the \sect{sec:Anomalies}, we will demonstrate this in more detail, and show how higher derivative four-photon operators can be used to cancel the $U(1)$ anomaly. In doing so, we can construct a $D$-dimensional quantum effective action that satisfies duality invariance in $D=4$ through two-loop order in perturbation theory.  
\paragraph{Leading $(--++)$ divergence} 
Finally, we express below the leading divergence for the aligned-helicity matrix elements. After reducing to the two-loop scalar integral basis, and projecting along 4D helicity states, we obtain the following expression at leading order in the $\epsilon$-expansion:
\begin{eBox}
\begin{equation}
\mathcal{M}^{\text{BI,2-loop}}_{(--++)} = -\frac{1}{\epsilon^2}\frac{\langle12\rangle^2[34]^2}{(4\pi)^4}\left[\frac{19}{60}s_{12}^4+\frac{17}{150}(s_{13}^4+s_{14}^4)\right]+\mathcal{O}(\epsilon^{-1})\,.
\end{equation}
\end{eBox}
Similar to the one-loop result, we can see that $s_{12}^4 \langle12\rangle^2[34]^2$ and $(s_{13}^4+s_{14}^4)\langle12\rangle^2[34]^2$ helicity sectors are asymmetrically weighted in pure photon amplitudes. In the next section, we will show that these two operators carry equal weight when introducing additional states consistent with maximal supersymmetry. 

Moreover, adding additional scalar and fermion states consistent with supersymmetry is the simplest way to cancel the $U(1)$ anomaly computed above outside of the aligned helicity sectors. When summing over superfield contributions at loop level, the $U(1)$ symmetry at tree-level is promoted to a $U(1)_R$ symmetry, and thus is protected pertubatively by supersymmetric Ward Identities. We demonstrate this explicitly in the next section by computing the two-loop four-photon matrix element in $\mathcal{N}=4$ DBIVA via our $D$-dimensional unitarity methods. 

\subsubsection{Two-loop $\mathcal{N}=4$ DBIVA}\label{sec:2loopN4U}
Much of the complication in performing a two-loop calculation in pure BI theory is the proliferation of high power of loop momenta left over after the state sum. These factors of loop momenta not only conspire with external momenta, but also mingle with external polarizations. There are a number of integral reduction algorithms \cite{Anastasiou:2004vj,vonManteuffel:2012np,Smirnov:2014hma,vonManteuffel:2014ixa,Smirnov:2019qkx,Smirnov:2020quc,Usovitsch:2020jrk,Maierhofer:2018gpa} that are very effective for reducing factors of $(k_i \cdot l)$, since they can trivially be expressed as a linear combinations of inverse propagators,
\begin{equation}
(k_i \cdot l) = \frac{1}{2}\left[(l+k_i)^2-l^2\right]\,.
\end{equation}
However, the factors of $(\varepsilon_i\cdot l_j)$ can be more tedious, as they do not permit an inverse propagator expansion. This is part of the motivation for applying Passarino-Veltman to the recursively one-loop integrals present in the two-loop Born-Infeld amplitude.  

Luckily, the state-sewing for maximal supersymmetry is dramatically simpler than less than maximal supersymmetry \cite{BDKUniarityReview,BRY}. This is most easily seen by considering the covariant operators we defined for one-loop DBIVA amplitudes. When one applies the conditions for maximal supersymmetry, stated below,
\begin{align}
D=10 \qquad N_\lambda &=1 \qquad N_X=0\,,
\\
D=6 \qquad N_\lambda &=2 \qquad N_X=1\,,
\\
D=4 \qquad N_\lambda &=4 \qquad N_X=3\,,
\\
D=3 \qquad N_\lambda &=8 \qquad N_X=8\,,
\end{align}
then the state sum is {completely independent of loop momenta} and precisely reproduces the supersymmetric matrix element $s_{12}^2 (t_8F^4)$ when cut along the $s_{12}$-channel.  This statement holds covariantly in general dimension. Concretely, we find that for theories with maximal supersymmetry:
\begin{equation}
\sum_{\text{states}}(t_8F^4)^{(\text{max})}_{(12l_1l_2)}(t_8F^4)^{(\text{max})}_{(\bar{l}_1\bar{l}_234)} = s_{12}^2 (t_8F^4)^{(\text{max})}_{(1234)}\,.
\end{equation}
This can similarly be used in the integrand construction $\mathcal{N}=4$ super-Yang-Mills, which has been computed to six-loop order \cite{Carrasco:2021otn}. In the case of maximal supersymmetry in $D=4$, we can write the  $(t_8F^4)^{(\text{max})}_{(1234)}$ operator as a supersymmetric delta function with the all-plus permutation invariant introduced previously:
\begin{equation}
\mathcal{M}^{\mathcal{N}=4\text{DBIVA}}_{(1234)} = \delta^{(8)}{(Q)} \frac{[12][34]}{\langle 12\rangle \langle 34\rangle} \equiv (t_8F^4)^{(\text{max})}_{(1234)}\Big|_{D=4}
\end{equation}
where the delta function is a Grassmann valued polynomial of spinor-helicity variables:
\begin{equation}
\delta^{(8)}{(Q)} = \prod_{a=1}^4\sum_{i\neq j} \langle ij \rangle \eta_i^a\eta_j^a\,.
\end{equation}
By applying this supersymmetric state sum, we find that the integrands for two-loop $\mathcal{N}=4$ DBIVA is trivially easy to construct - even simpler than NLSM integrands. Below we represent an internal on-shell superfield with a green line, and obtain the following integrand for the double-bubble:
\begin{equation}\label{eq:N42bub}
\text{Cut}\left(\scaleIntCtune{}{hgreen}{}{hgreen}\right)=\sum_{\text{states}}  (t_8F^4)^{(\text{max})}_{(p_4 12p_3)} (t_8F^4)^{(\text{max})}_{(\bar{p}_1\bar{p}_4\bar{p}_3\bar{p}_2)} (t_8F^4)^{(\text{max})}_{(p_234p_1)}= s_{12}^4 (t_8 F^4)
\end{equation}
and similarly so for the ostrich-diagram integral contribution:
\begin{equation}\label{eq:N4ostrich}
\text{Cut}\left(\scaleIntBtune{}{}{}{}{hgreen}{hgreen}{hgreen}{hgreen}\,\,\right)=\sum_{\text{states}}  (t_8F^4)^{(\text{max})}_{(2q_4\bar{q}_3\bar{q}_1)} (t_8F^4)^{(\text{max})}_{(1\bar{q}_4q_3\bar{q}_2)} (t_8F^4)^{(\text{max})}_{(q_234q_1)}=s_{12}^2 \tau^{(1)}_{\bar{2}}\tau^{(2)}_{\bar{1}}(t_8 F^4)
\end{equation}
We should point out an additional verification can occur for any SUSY amount of DBIVA.  We can recycle multiloop sYM integrands to verify our cuts.  In general this is because double-copy works at tree-level, and we can always write the Jacobi-satisfying Yang-Mills cut numerators in terms of color-ordered sYM amplitudes, $A^{\text{sYM}}$, offering us a KLT between cut-integrands.  When we cut to four-point amplitudes, as we do for the cuts above, this is particularly simple as only one order needs contribute for each four-point tree, yielding:
\begin{align}
\text{Cut}({\text{DBIVA}})&= \sum_\text{states} \prod_i \mathcal{A}^{\text{DBIVA}}(p^{(i)}_1,p^{(i)}_2,p^{(i)}_3,p^{(i)}_4)\,\\
&= \sum_\text{states} \prod_i  \left( s^{(i)}_{12} s^{(i)}_{23}  A^{\text{sYM}}(p^{(i)}_1,p^{(i)}_2,p^{(i)}_3,p^{(i)}_4)\right)\,\\
&= \left[\prod_j  \left(s^{(j)}_{12} s^{(j)}_{23}\right)  \right] \sum_\text{states}   \prod_i  A^{\text{sYM}}(p^{(i)}_1,p^{(i)}_2,p^{(i)}_3,p^{(i)}_4)\,\\
&= \left[\prod_j  \left(s^{(j)}_{12} s^{(j)}_{23}\right) \right] \times \text{Cut}({\text{sYM}})\,.
\end{align}
Here  the products run over the number of trees contributing to the cut,  $s^{(i)}_{mn} = ( p^{(i)}_m + p^{(i)}_n )^2$, and $p^{(i)}_m$ is the $m$-th label of the $i$-th tree contributing to the cut. The second line follows from the double-copy structure of four-point amplitudes in DBIVA, where:
\begin{align}
\mathcal{A}_4^{\text{DBIVA}} = s_{12} \, s_{23} \, A^{\text{sYM}}(1,2,3,4)\,.
\end{align}
Carrying out such sYM recycled cut verifications of \eqn{eq:N42bub} and \eqn{eq:N4ostrich} in the case of maximally supersymmetric Yang-Mills in four-dimensions is straightforward -- only one (non-vanishing) diagram contributes to each cut. 

Since there are no loop momenta interfering with the tensor structures of the external photons, the result of integration will just be proportional to a $(t_8F^4)$ tensor, which is manifestly $U(1)$ duality invariant. Thus, as promised, adding in $\mathcal{N}=4$ states cancels the $U(1)$ anomaly present in the pure Born-Infeld $S$-matrix. After applying the tensor reduction to the ostrich-diagram, we obtain the following expression for the two diagrams:
\begin{align}
\scaleIntCtune{}{hgreen}{}{hgreen} &= s_{12}^4 (I_2^{D}(k_{12}))^2(t_8 F^4)  \,,
\\
\scaleIntBtune{}{}{}{}{hgreen}{hgreen}{hgreen}{hgreen} \,\,\,\,&= \frac{2}{3}s_{12}^3[I_3\circ I_2]^{D}(k_{12})(t_8 F^4)\,.
\end{align}
This is easily evaluated in $D=4-2\epsilon$, from which we find the following expression for the leading order divergence of the four-photon two-loop amplitude in $\mathcal{N}=4$ DBIVA theory:
\begin{eBox}
\begin{equation}
\mathcal{M}^{\mathcal{N}=4\text{ DBIVA}}_{\text{2-loop}} = -\frac{1}{12\epsilon^2}\frac{1}{(4\pi)^4}(s_{12}^4+s_{13}^4+s_{14}^4)(t_8 F^4) + \mathcal{O}(\epsilon^{-1})\,.
\end{equation}
\end{eBox}
With this result in hand, we will now demonstrate that the same calculation performed via generalized unitarity can be reproduced using loop-level double copy construction. 
\subsection{DBIVA via double copy}\label{sec:DBIvDC}
The calculation above was completely agnostic to (but verified by) the known tree-level relationship of DBIVA amplitudes as a double copy of NLSM and super Yang-Mills. As we will now show, the amplitude can be equivalently produced using the two-loop color-dual numerators of $\mathcal{N}=4$ sYM, and the NLSM integrands constructed earlier in the section. This observation provides further evidence for the consistency of double-copy construction at multi-loop order in perturbation theory, and serves as an existence proof for color-dual NLSM numerators at two-loop. 

At loop-level, the double-copy construction amounts to replacing the color factors in a cubic-graph representation of the integrand with a set of color-dual numerators. That is, starting with an $L$-loop NLSM integrand of the form:
\begin{equation}
\mathcal{A}^{\text{NLSM}}_{n} = \int \prod_{i=1}^L \frac{d^D l_i}{(2\pi)^d} \sum_{g\in \Gamma^{(3)}} \frac{1}{S_g}\frac{C_g N^{\text{NLSM}}_g}{D_g}\,,
\end{equation}
we can construct $\mathcal{N}=4$ DBIVA by replacing the color factors, $C_g$, with the kinematic numerators, $N^{\mathcal{N}=4}_g$, of sYM amplitudes:
\begin{equation}
C_g \rightarrow N^{\mathcal{N}=4}_g \quad \Rightarrow \quad \mathcal{M}^{\text{DBIVA}}_{n} = \int \prod_{i=1}^L \frac{d^D l_i}{(2\pi)^d} \sum_{g\in \Gamma^{(3)}} \frac{1}{S_g}\frac{N^{\mathcal{N}=4}_g N^{\text{NLSM}}_g}{D_g}\,,
\end{equation}
where the kinematic numerators depend on particular choice of generalized gauge. In order for this construction to work, the kinematic numerators on at least one side of the double copy must satisfy all the same algebraic relations as the color factors. Since generalized unitarity allows us to construct the integrands on-shell, this construction conjecturally holds at loop-level as long as the gauge-theory is tree-level color-dual. 

While there are currently no color-dual NLSM numerators identified in the literature beyond one-loop, there are color-dual representation available for $\mathcal{N}=4$ sYM through four-loop four-point \cite{GravityFour}. Below are the one- and two-loop basis numerators relevant for our construction: 
\begin{equation}
N^{\text{box}}_{\mathcal{N}=4} =\simpleBox=  (t_8F^4)^{\text{(max)}}_{(1234)}\,,
\end{equation}
and similarly so at two-loop four-point for the double-box and cross-box, which take on identical expressions:
\begin{align}
N^{(12|34)}_{\mathcal{N}=4}  \equiv \dBox =s_{12} (t_8F^4)^{\text{(max)}}_{(1234)}\,,
\\
N^{([12]|34)}_{\mathcal{N}=4} \equiv \xBox =s_{12} (t_8F^4)^{\text{(max)}}_{(1234)}\,.
\end{align}
These two-loop numerators can be constructed via the rung-rule \cite{BRY} and the no-triangle hypothesis \cite{BernNoTriangle}. We can see that both the one- and two-loop kinematic numerators are completely independent of loop momenta. This allows us to take our NLSM amplitudes from earlier in the section, and simply replace the color factors {post integration}. We will show the results of this procedure for both one-loop and two-loop DBIVA in the proceeding sections.
\subsubsection{One-loop $\mathcal{N}=4$ DBIVA} \label{sec:DBIvDC1loop}
At one-loop, performing the double copy is a simple task. As we noted above, since the box numerator for $\mathcal{N}=4$ sYM has no kinematic dependence on internal loop momenta, we can pull it out of the integration process. Thus, making the following replacement 
\begin{equation}
C^{\text{box}}_{(1234)} \rightarrow N^{\text{box}}_{(1234)}\,,
\end{equation}
on the integrated NLSM four-point one-loop amplitude. This yields the following expression in $D=4-2\epsilon$ for the four-point one-loop $\mathcal{N}=4$ DBIVA matrix element:
\begin{equation}
\mathcal{M}^{\mathcal{N}=4}_{\text{DBIVA}} = \frac{i}{(4\pi)^2}\left[\frac{s_{12}^2}{2}\left(\frac{1}{\epsilon} +\text{ln}(-s_{12})\right)+\text{cyc}(2,3,4)\right](t_8F^4)^{\text{(max)}}_{(1234)} \,.
\end{equation}
This is in agreement with the leading divergence of $\mathcal{N}=4$ DBIVA previously found in \cite{Elvang:2020kuj}, along with our result in \eqn{eq:generalDBI1loop} when taking $N_\gamma =1$, $N_\lambda=4$ and $N_X=3$, in concordance with $D=4$ maximal supersymmetry 

\subsubsection{Two-loop $\mathcal{N}=4$ DBIVA} \label{sec:DBIvDC2loop}
At two-loop it sufficient to show that replacing the color factors reproduces the integrals we found above with the unitarity construction. Starting with the NLSM integrands in \eqn{eq:NLSM2bub} and \eqn{eq:NLSMostrich}, we make the same replacement of color factors post integration as was done at one-loop.  We carry this out first for the double-bubble, and find that the full $D$-dimensional NLSM integral simplifies dramatically:
\begin{equation}
\begin{aligned}
\scaleIntCsmall \bigg|^{C_g \rightarrow N^{\mathcal{N}=4}_g}  \!\!&= N^{(12|34)}_{\mathcal{N}=4}\left[\frac{(s_{12}I_2^D(k_{12}))^2}{2}\left(\frac{s_{14}-s_{13}}{(D-1)^2}+s_{12}\right)\right]
\\
&+N^{(12|43)}_{\mathcal{N}=4}\left[\frac{(s_{12}I_2^D(k_{12}))^2}{2}\left(\frac{s_{13}-s_{14}}{(D-1)^2}+s_{12}\right)\right]
\\
&=s_{12}^4 (I_2^{D}(k_{12}))^2(t_8 F^4)^{\text{(max)}}_{(1234)}   \equiv \scaleIntCtune{}{hgreen}{}{hgreen} \,.
\end{aligned}
\end{equation}
The cancellation between dimension dependent factors is even more startling for the ostrich-diagram integral:
\begin{equation}
\begin{aligned}
\scaleIntBscalarsmall{1}{2}{3}{4}\bigg|^{C_g \rightarrow N^{\mathcal{N}=4}_g} \!\!&= N^{(12|34)}_{\mathcal{N}=4}\frac{s_{12}}{3} \left[\frac{(D-1)(D-4)s_{14}+2(D-2)^2 s_{13}}{(D-1)(4-3D)}\right][I_3\circ I_2]^D(k_{12})
\\
&+N^{(12|34)}_{\mathcal{N}=4}\frac{s_{12}}{3} \left[\frac{(D-1)(D-4)s_{13}+2(D-2)^2 s_{14}}{(D-1)(4-3D)}\right][I_3\circ I_2]^D(k_{12})
\\
&+N^{([12]|34)}_{\mathcal{N}=4}\left[\frac{s_{12}^2}{3}\frac{D+1}{D-1}\right][I_3\circ I_2]^D(k_{12})
\\
&=\frac{2}{3}s_{12}^3[I_3\circ I_2]^{D}(k_{12})(t_8 F^4)^{\text{(max)}}_{(1234)} \equiv \scaleIntBtune{}{}{}{}{hgreen}{hgreen}{hgreen}{hgreen} \,.
\end{aligned}
\end{equation}
This alone is sufficient to demonstrate the validity of the double-copy, as these integrated quantities are exactly the same as those produced via generalized unitarity of $\mathcal{N}=4$ DBIVA at two-loop in \eqn{eq:N42bub} and \eqn{eq:N4ostrich}. We emphasize that based on our analysis, the double-copy at two loop clearly holds in {any} dimension, as all the $D$-dependent prefactors drop out when replacing the color factors with $\mathcal{N}=4$ numerators. Considering the consistency of these two construction, this also serves as strong evidence for the existence of color-dual representation for two-loop pion integrands. We leave identifying such valid representation as an enticing future direction worth investigation. 

\section{Effective Actions} \label{sec:Actions}
In this section we demonstrate how our basis of higher-derivative four-photon operators in \eqn{eq:basisTensors} can be used to construct quantum effective actions that capture loop-level effects. We will distinguish between position space operators that appear in the Lagrangian, $\mathcal{O}$, and their corresponding matrix elements, $\mathcal{T}$, which are on-shell quantities in momentum space:
\begin{equation}\label{eq:OpsVsTens}
\mathcal{T} \equiv \langle \text{out}| \mathcal{O}|\text{in}\rangle\,.
\end{equation}
Similar to the matrix elements, the operators can be expressed in both $D$-dimensional and 4D representations. For example, we can define $\mathcal{O}^{F^2F^2}$ operators as follows:
\begin{align}
\mathcal{O}^{F^2F^2}_{(2,0)} &\sim (D_\mu F_{\rho\sigma}D^\mu F^{\rho\sigma})^2\,,
\\
\mathcal{O}^{F^2F^2}_{(0,1)} &\sim (D_\mu F_{\rho\sigma}D^\nu F^{\rho\sigma})(D_\nu F_{\alpha\beta}D^\mu F^{\alpha\beta})\,,
\end{align}
where the spacetimes indices run from $1,2,...,D$, and $F_{\mu\nu} = \partial_\mu A_\nu - \partial_\nu A_\mu$ are operator valued abelian field strengths. These operator valued expressions can be normalized so that for an all outgoing four-photon scattering process, $|\text{out}\rangle = |k_1,k_2,k_3,k_4\rangle$, they are related to the on-shell tensor basis elements defined previously in \eqn{eq:basisTensors} as follows:
\begin{align}
\mathcal{T}^{F^2F^2}_{(2,0)} =\langle \text{out}\big|\mathcal{O}^{F^2F^2}_{(2,0)} \big|\text{in}\rangle\,,
\\
\mathcal{T}^{F^2F^2}_{(0,1)} =\langle \text{out} \big|\mathcal{O}^{F^2F^2}_{(0,1)}\big |\text{in}\rangle\,.
\end{align}
Likewise we can construct field theory operators for the split helicity operators using the spinor form of the field strengths:
\begin{equation}
F_\pm = (F^{\mu\nu} \pm i \tilde{F}^{\mu\nu})\sigma^{\mu\nu}_\pm\,,
\end{equation}
where the dual field strength are defined as $\tilde{F}^{\mu\nu}= \frac{1}{2}\epsilon^{\mu\nu\rho\sigma}F^{\mu\nu}$, and the spin matrices are expressed as $\sigma^{\mu\nu}_\pm = \frac{1}{4} \sigma^{[\mu}_\pm \sigma^{\nu]}_\mp$ in terms of the Pauli matrices as follows:
\begin{equation}
\sigma_\pm^\mu = (\mathbb{1},\pm \vec{\sigma})^\mu\,.
\end{equation}
In the above expressions, the spacetime indices now run from $1$ to $4$. Using these manifestly 4D operators, we can similarly define operators that select out particular helicity states and derivative structures:
\begin{align}
\mathcal{O}^{\text{4D,1}}_{(--++)} &\sim (D_\mu F_- D_\nu F_-)(D^\mu F_+ D^\nu F_+)\,,
\\
\mathcal{O}^{\text{4D,2}}_{(--++)} &\sim (D_\mu F_- D^\mu F_-)(D_\nu F_+ D^\nu F_+)\,,
\end{align}
where the trace is over the spinor indices of the Pauli matrices. As defined, there operators will produce the 4D helicity structures defined in \eqn{eq:4D1Tens} and \eqn{eq:4D2Tens} when contracted between the outgoing helicity state, $|\text{out}\rangle = |k_1^-,k_2^-,k_3^+,k_4^+\rangle$, and the incoming vacuum state:
\begin{align}
\mathcal{T}^{\text{4D,1}}_{(--++)} &=\langle \text{out}\big| \mathcal{O}^{\text{4D,1}}_{(--++)}  \big|\text{in}\rangle = (s_{13}^2 +s_{14}^2)\langle 12\rangle^2[34]^2\,,
\\
\mathcal{T}^{\text{4D,2}}_{(--++)} &= \langle \text{out}\big|\mathcal{O}^{\text{4D,2}}_{(--++)}  \big|\text{in}\rangle =s_{12}^2 \langle 12\rangle^2[34]^2\,.
\end{align}
In the duality invariant quantum actions we construct in this section, we will reserve $\mathcal{O}$ for effective actions, and $\mathcal{T}$ for on-shell matrix elements used in the EMU construction. However, since the field theory operators are redundant up to field redefinition and equations of motion, we'll find it convenient to work in terms of the on-shell matrix elements, and then implicitly define the operators in terms of these on-shell expressions. 

Below we first study anomaly cancellation for the multi-loop photon amplitudes computed above. After this, we proceed by interpreting these higher derivative operators as double copies between NLSM and higher derivative Yang-Mills amplitudes with off-shell higher-spin modes. As was demonstrated in previous work by the authors \cite{Carrasco:2022jxn}, the full set of $D$-dimensional four-photon operators can be constructed via adjoint double-copy, but at the cost of introducing off-shell higher spin modes in the single-copy vector theory. However, these higher-spin modes can be absorbed consistently in a double-copy framework by introducing symmetric algebraic structures. We discuss future applications of this construction in \sec{sec:Discussion}.

\subsection{Anomaly cancellation}\label{sec:Anomalies}
Now we begin with our study of higher-derivative extensions to BI theory. Our goal is to identify the higher derivative four-photon operators that are needed to cancel the $U(1)$ anomalous matrix elements computed in the previous section. We will start with the one-loop corrections, captured by tree level insertions at $\mathcal{O}(\alpha'^4)$, and then proceed with the two loop corrections, which combine both $\mathcal{O}(\alpha'^4)$ operator insertions at one-loop and $\mathcal{O}(\alpha'^6)$ operator insertions at tree-level. In doing so, we demonstrate that cancelling the $U(1)$ anomaly through two loop order can be achieved with local finite counterterms if and only if we introduce an evanescent operator at $\mathcal{O}(\alpha'^4)$ to the Born-Infeld action.

\subsubsection{One-loop}\label{sec:Anomalies1loop}
In \sect{sec:DBIU} we computed the one-loop matrix element for a general DBIVA theory. Plugging in the values $N_{\gamma}=1$ and $N_\lambda = N_X = 0$ we obtain the following anomalous all-plus matrix element for pure Born-Infeld theory:
\begin{equation}
\mathcal{M}^{\text{BI,1-loop}}_{(++++)} = -\frac{i}{(4\pi)^2}\frac{1}{60}(s_{12}^4+s_{13}^4+s_{14}^4)\frac{[12][34]}{\langle12\rangle\langle 34\rangle}+\mathcal{O}(\epsilon)\,.
\end{equation}
In Ref. \cite{Elvang:2020kuj},  Elvang, Hadjiantonis, Jones, and Paranjape  identify a candidate 4D counterterm that cancels this anomalous matrix element, whose prediction we have called $\mathcal{T}^{\text{4D}}_{(++++)}$, thereby restoring duality invariance through one-loop. As noted in the previous section, the one-loop matrix element can be mapped to our $D$-dimensional operator basis. In general, all available 4D tensor structures at $\mathcal{O}(\alpha'^4)$ map onto our $D$-dimensional basis. One particular map we provide below:
\begin{align}
\mathcal{T}^{\text{4D}}_{(++++)} &=  a_{({\text{ev.}})}\mathcal{T}^{\text{ev.}}+2\mathcal{T}^{4+}\,,
\\
\mathcal{T}^{\text{4D,1}}_{(--++)} &= a_{({\text{ev.}})}\mathcal{T}^{\text{ev.}}+2\mathcal{T}_{(2,0)}^{F^4}- 4\mathcal{T}_{(0,1)}^{F^2F^2}\,,
\\ 
\mathcal{T}^{4\text{D,}2}_{(--++)} &= a_{({\text{ev.}})}\mathcal{T}^{\text{ev.}}+2\mathcal{T}_{(2,0)}^{F^4}+ 4\mathcal{T}_{(0,1)}^{F^2F^2}\,,
\end{align}
where we have defined the following $D$-dimensional operator that projects down to the all-plus configuration,
\begin{equation}
\mathcal{T}^{4+}= 2 \mathcal{T}_{(2,0)}^{F^2F^2}-\mathcal{T}_{(2,0)}^{F^4}-2\mathcal{T}_{(0,1)}^{F^4}\,,
\end{equation}
and all the 4D operators have the freedom to add the previously defined evanescent operator, $\mathcal{T}^{\text{ev.}}$, 
\begin{equation}
\mathcal{T}^{\text{ev.}} = \mathcal{T}_{(2,0)}^{F^2F^2}-\mathcal{T}_{(0,1)}^{F^2F^2} + \mathcal{T}_{(0,1)}^{F^4}\,.
\end{equation}
Thus, we can construct the new effective photon Lagrangian, $\mathcal{L}^{\text{BI}+\text{CT}}$, with the addition of the all-plus counter-terms to our Born-Infeld Lagrangian:
\begin{equation}
\label{eq:1loopU1Action}
\mathcal{L}^{\text{BI}+\text{CT}} = \mathcal{L}^{\text{BI}} + \frac{\alpha'^4}{(4\pi)^2}\frac{1}{30}\left(\mathcal{O}^{4+}+ a_{({\text{ev.}})}\mathcal{O}^{\text{ev.}}\right)\,,
\end{equation}
where we have used \eqn{eq:OpsVsTens} to implicitly define the operators appearing in the quantum effective action above. While there are a number of perturbatively equivalent construction of these operators, below we provide a couple expressions that resemble the on-shell basis elements:
\begin{equation}
\begin{split} 
\mathcal{O}^{4+} &\sim 2 (D_\mu F_{\alpha \beta} D^\mu  F^{\alpha \beta})^2 - \eta^{\mu(\nu}\eta^{\rho\sigma)}(D_\mu F_{\alpha \beta} D_\nu  F^{\gamma\delta} D_\rho F_{\beta \gamma } D_\sigma F^{\delta \alpha})\,,
\\\\
\mathcal{O}^{\text{ev.}} &\sim
(D_\mu F_{\alpha \beta} D^\mu  F^{\alpha \beta})^2 - (D_\mu F_{\alpha \beta} D_\nu  F^{\alpha \beta})(D^\mu F_{\alpha \beta} D^\nu  F^{\alpha \beta})
\\
&\quad +(D_\mu F_{\alpha \beta} D^\mu  F^{\gamma\delta} D_\nu F_{\beta \gamma } D^\nu F^{\delta \alpha})\,.
\end{split}
\end{equation}
Computing the one-loop amplitudes from the Lagrangian of \eqn{eq:1loopU1Action} yields the following matrix elements at $\mathcal{O}(\alpha'^4)$:
\begin{align}
\mathcal{M}^{\text{BI+CT,1-loop}}_{(--++)}\big|_{\alpha'^4} &= \mathcal{M}^{\text{BI,1-loop}}_{(--++)}  \,,
\\
 \mathcal{M}^{\text{BI+CT,1-loop}}_{(-+++)}\big|_{\alpha'^4} &= 0\,,
 \\
 \mathcal{M}^{\text{BI+CT,1-loop}}_{(++++)}\big|_{\alpha'^4}  &= \mathcal{O}(\epsilon)\,.
\end{align}
Thus, \eqn{eq:1loopU1Action} constitutes a duality invariant photon theory through one-loop order. With this, we can identify what additional operators will be needed to cancel the anomaly through two-loop. 
\subsubsection{Two-loop}\label{sec:Anomalies2loop}
The first step in identifying the requisite operators needed to cancel the two-loop anomaly at $\mathcal{O}(\alpha'^6)$ is to perform another one-loop calculation at this mass dimension, which includes the counterterms of $\mathcal{L}^{\text{BI}+\text{CT}}$ defined above. The one-loop amplitude is constructed as follows:
\begin{equation}
\mathcal{M}^{\text{BI+CT}}_{\text{1-loop}}\big|_{\alpha'^6} = \frac{1}{2}+ \frac{\alpha'^4}{(4\pi)^2}\frac{1}{30}\left[\scaleIntAvectorODD{}{}{}{}{$\mathcal{T}^{4+}$}{$t_8F^4$}+a_{\text{ev.}}\scaleIntAvectorODD{}{}{}{}{$\mathcal{T}^{\text{ev.}}$}{$t_8F^4$}\right]+\text{perms}\,.
\end{equation}
Both of these operator insertions can be evaluated using the same $D$-dimensional procedure used throughout the text. The all-plus counterterm yields the following contributions to $(++++)$ and $(-+++)$ helicity configurations:
\begin{align}
\scaleIntAvectorODD{+}{+}{+}{+}{$\mathcal{T}^{4+}$}{$t_8F^4$} &= \left[\frac{7}{10}+\frac{79}{300}\epsilon\right] s_{12}^4 [12]^2[34]^2 I_2^{4-2\epsilon}(k_{12})+\mathcal{O}(\epsilon)
\\
\scaleIntAvectorODD{$-$}{+}{+}{$+$}{$\mathcal{T}^{4+}$}{$t_8F^4$} &= \mathcal{O}(\epsilon)\qquad \qquad \quad \scaleIntAvectorODD{$+$}{+}{+}{$-$}{$\mathcal{T}^{4+}$}{$t_8F^4$} = 0
\end{align}
We note that there is a distinction between the first and second $(-+++)$ expressions. The first expression is dressed with $(D-4)^2$, which pushes the leading contribution to $\mathcal{O}(\epsilon)$. Whereas the second term is identically zero because the 4D helicity structure carries an overall factor of $(s+t+u)=0$. In addition, since there is a non-vanishing 4D residue for the all-plus integrand, the integral has a leading order divergence in $\epsilon$. 

Below we find it instructive to show the $D$-dependence of the evanescent operator insertion, which yields the following matrix element contributions:
\begin{align}
\scaleIntAvectorODD{+}{+}{+}{+}{$\mathcal{T}^{\text{ev.}}$}{$t_8F^4$} &= -\left[\frac{(D-4)(D^2-7D-4)}{32(D^2-1)}\right] s_{12}^4 [12]^2[34]^2 I_2^{D}(k_{12})
\\
\scaleIntAvectorODD{$+$}{+}{+}{$-$}{$\mathcal{T}^{\text{ev.}}$}{$t_8F^4$} &= -\left[\frac{(D-4)(D+2)}{8(D^2-1)} \right]s_{12}^3 \langle4|3|2]^2[13]^2 I_2^{D}(k_{12})
\\
\scaleIntAvectorODD{$-$}{+}{+}{$+$}{$\mathcal{T}^{\text{ev.}}$}{$t_8F^4$} &=0
\end{align}
This will produce $\mathcal{O}(\epsilon^0)$ matrix elements in the $(-+++)$ helicity sector. 
Thus, in order to cancel the divergent part of the two-loop $(-+++)$ anomaly computed in \eqn{eq:oneMinus2loop}, we must weight the evanescent operator by a numerical factor that {diverges} in $D=4$. 
Given the particular numerical value computed in the previous section, we find the evanescent Wilson coefficient must take the following $D$-dependent value:
\begin{eBox}
\begin{equation}\label{eq:BI+CTwithOev}
\mathcal{L}^{\text{BI}+\text{CT}} = \mathcal{L}^{\text{BI}} + \frac{\alpha'^4}{(4\pi)^2}\frac{1}{30}\left[\mathcal{O}^{4+}- \frac{8}{(D-4)}\mathcal{O}^{\text{ev.}}\right]+\mathcal{O}(\alpha'^6)\,,
\end{equation}
\end{eBox}
where the $(D-4)$ in the denominator cancels the factor in the numerator above. In order to further absorb the remaining rational terms, we must introduce an additional set of tree-level operators at $\mathcal{O}(\alpha'^6)$. At this order in mass-dimension, there are seven distinct operators:
\begin{equation}\label{eq:Oa6counterTerms}
\{\mathcal{O}^{F^2F^2}_{(4,0)},\mathcal{O}^{F^2F^2}_{(2,1)},\mathcal{O}^{F^2F^2}_{(0,2)},\mathcal{O}^{F^4}_{(4,0)},\mathcal{O}^{F^4}_{(2,1)},\mathcal{O}^{F^4}_{(0,2)},\mathcal{O}^{F^3}_{(1,0)}\}\,.
\end{equation}
By adding these operators to the effective Lagrangian above, we have verified that there is sufficient freedom to absorb the remaining rational terms present at two-loop. 
Of these available operators, only the $F^3$ tensor structure is non-vanishing when projected along the $(-+++)$ helicity configuration. 
Furthermore, just as at $\mathcal{O}(\alpha'^4)$, there is a single evanescent matrix element which we define below:
\begin{equation} 
\mathcal{T}^{\text{ev.}}_{\alpha'^6} =\mathcal{T}^{F^2F^2}_{(4,0)}-2\mathcal{T}^{F^2F^2}_{(2,1)}+\mathcal{T}^{F^2F^2}_{(0,2)}+\mathcal{T}^{F^4}_{(2,1)}-\mathcal{T}^{F^4}_{(0,2)}\,.
\end{equation}
Thus, of the seven available $D$-dimensional operators, they are projected to only six distinct 4D tensor structures. 
We will describe the counting of 4D versus general dimension photon operators in generality at all orders in $\alpha'$ in more depth at the end of this section.
\subsection{Double copy construction}\label{sec:ActionDC}
As we have stated in the text, it is well known that DBIVA theory can be constructed at tree-level as an adjoint double copy between NLSM and sYM amplitudes. There is now a large body of literature studying double-copy construction of higher derivative gauge theory counterterms \cite{Carrasco:2019yyn,Carrasco:2021ptp,Chi:2021mio,Bonnefoy:2021qgu,Carrasco:2022lbm,Carrasco:2022sck,Pavao:2022kog,Chen:2022shl,Chen:2023dcx,Brown:2023srz}, like those used above to cancel $U(1)$ anomalous matrix elements in pure Born-Infeld theory. Indeed, recent work by the authors demonstrated that all four-photon operators can be constructed consistently via the double-copy \cite{Carrasco:2022jxn}. Here we briefly describe the single-copy gauge theory that when double copied with NLSM produces the higher derivative operators of the previous section. 
\subsubsection{Symmetric-structure double-copy}\label{sec:symDC}
To realize the double-copy construction that produces the counterterms above, Ref. \cite{Carrasco:2022jxn} first decomposed NLSM pions amplitude into symmetric structure constants using the $U(N)$ color identity
\begin{equation}
f^{abe}f^{ecd} = d^{ade}d^{ebc}- d^{ace}d^{ebd}\,,
\end{equation}
where the symmetric structure constant is defined as follows
\begin{equation}
d^{abc} = \text{tr}[T^a\{T^b,T^c\}]\,.
\end{equation}
By applying this color algebra identity to the four-point NLSM amplitudes of \eqn{eq:NLSMamp}, one finds that pions can similarly be expressed as a {symmetric-structure} double copy:
\begin{equation}
\mathcal{M}^{\text{NLSM}}_4 =  \sum_{g\in \Gamma^3} \frac{c_g^{\text{dd}}n^{\text{dd,}\pi}_g}{d_g}= d^{abe}d^{ecd}\,s+d^{ade}d^{ebd}\,t+d^{ace}d^{ebd}\,u \,,
\end{equation}
where $c_s^{\text{dd}} \equiv d^{abe}d^{ecd}$ and the NLSM symmetric $s$-channel numerator is $n^{\text{dd,}\pi}_g=s^2$. By identifying a set of gauge theory numerators that obey the same algebraic relations as the color factors, one can construct consistent double copy theories. 

For example, consider the two-loop divergence for the $(++++)$ anomalous matrix element in pure BI theory. The 4D helicity structure can be captured by a symmetric structure double-copy between NLSM pion numerators and a local gauge theory contact at $\mathcal{O}(\alpha'^5)$. The $s$-channel numerators for this symmetric-structure double copy are as follows
\begin{equation}\label{eq:symNum2loop}
n_s^{\text{dd,}\pi} = s^2 \qquad n_s^{\text{dd,HD}} = s^5 \frac{[12][34]}{\langle 12\rangle \langle 34\rangle}\,.
\end{equation}
Double-copying these kinematic numerators yields a matrix element of the form:
\begin{equation} \label{eq:2loopCTsymDC}
\mathcal{M}^{\text{BI}+\text{HD}} = \sum_{g\in \Gamma^3} \frac{n_g^{\text{dd,HD}}n^{\text{dd,}\pi}_g}{d_g} = (s^6+t^6+u^6) \frac{[12][34]}{\langle 12\rangle \langle 34\rangle}\,.
\end{equation}
While this construction lacks any algebraic relations between the four-point kinematic-factors, similar symmetric numerators were found at six-point for NLSM, which obey non-trivial algebraic relations \cite{Carrasco:2022jxn}. In addition to the photon counterterms constructed above, symmetric double copy is likewise a natural description of gravitational counterterms. Rather than composing the local vector numerators, $n^{\text{dd,HD}}$, with symmetric NLSM numerators, $n^{\text{dd,}\pi}_g$, they can also be squared -- yielding a gravitational counterterm:
\begin{equation} \label{eq:2loopCTsymDC}
\mathcal{M}^{\text{GR}+\text{HD}} = \sum_{g\in \Gamma^3} \frac{(n^{\text{dd,HD}}_g)^2}{d_g} = (s^9+t^9+u^9)  \frac{[12][34]}{\langle 12\rangle \langle 34\rangle}\,.
\end{equation}
In general, any four-point symmetric vector numerator can constructed from two $D$-dimensional gauge theory building blocks \cite{Carrasco:2022jxn},
\begin{equation}
n_s^{\text{dd,vec,1}} = f_{12}f_{34}\,, \qquad n_s^{\text{dd,vec,2}} = f_{1324}\,,
\end{equation}
by composing with the symmetric scalar numerators, $n_s^{\text{dd,1}} = s$ and  $n_s^{\text{dd,2}} = tu$. With these building blocks, symmetric double copy construction captures the exceptional four graviton amplitude of Ref.~\cite{Boels:2016xhc}, which only considered linear combinations of gauge theory amplitudes rather than gauge theory numerators. We note that a similar construction was recently identified in Ref.~\cite{Bourjaily:2023tcc} beyond four-point, which demonstrated that gravity amplitudes permit double copy construction in terms of gauge-invariant numerators that do not obey any algebraic relations between themselves, much like the symmetric numerators above in \eqn{eq:symNum2loop}. We see further exploring higher multiplicity examples of symmetric double copy as an important direction of future study.

We will now show that the matrix element above in \eqn{eq:2loopCTsymDC} needed to cancel the two-loop anomaly can be constructed from an equivalent adjoint double, at the cost of introducing a spin-5 off-shell mode in the single copy gauge theory. This is a special property of symmetric double copy between NLSM numerators and symmetric vector building blocks. 

\subsubsection{Higher-spin $\otimes$ Adler Zero}\label{sec:HspinDC}
Guided by the dual description of NLSM pion amplitudes as symmetric {and} adjoint double copies, one can easily cast the symmetric-structure numerators back to adjoint kinematics. Due to the duality between color and kinematic factors, we construct partner adjoint numerators using the following color relation:
\begin{equation}
c_s^{\text{ff}} = c_t^{\text{dd}}-c_u^{\text{dd}} \quad \Leftrightarrow \quad n_s^{\text{ff}} = n_t^{\text{dd}}-n_u^{\text{dd}}\,. 
\end{equation}
Applying this identity to the symmetric vector numerator needed to reproduce the two-loop all-plus counterterm yields the following adjoint color-dual $s$-channel numerator:
\begin{equation}\label{eq:2loopNumerator}
n^{\text{HD},(2)}_s = (t^{5}-u^{5})\frac{[12][34]}{\langle 12\rangle \langle 34\rangle}\,.
\end{equation}
Following the argument of \cite{Pavao:2022kog}, this degree five kinematic numerator indicates that there is a spin-5 mode on top of the $s$-channel pole. However, when double-copied with NLSM the residue is suppressed by the Adler zero satisfying four-point contact of pion amplitudes. Thus, this adjoint color-dual numerator serves as a consistent single-copy theory when composed with color-dual NLSM numerators. 

Guided by the structure of the anomalous BI matrix elements computed through two-loop in the text, a possible guess for the single-copy HD vector numerators needed to cancel the leading $L$-loop divergence might go as
\begin{equation}
n^{\text{HD},(L)}_s \stackrel{?}{=} (t^{2L+1}-u^{2L+1})\frac{[12][34]}{\langle 12\rangle \langle 34\rangle}\,.
\end{equation}
This all-loop guess mirrors the structure of the one-loop adjoint numerator identified in \cite{Carrasco:2022jxn}, and the two-loop counterterm numerator expressed above in \eqn{eq:2loopNumerator}, in that at each loop order we would require the addition of a higher odd-integer-spin mode. 

The physical picture one should have for this class of photon effective operators, is that symmetric-structure double-copy and adjoint double-copy with higher spin modes are one in the same. In exchange for constructing color-dual adjoint numerators needed adjoint double-copy, or equivalently the KLT kernel, one must admit the addition of higher spin modes. However, as long as these higher spin modes are composed via adjoint double-copy, $\stackrel{\text{adj.}}{\otimes}$, with contact numerators, they map to the same local vector numerators one would achieve with the symmetric double-copy kernel, $\stackrel{\text{sym.}}{\otimes}$,
\begin{equation}
\Hspin \stackrel{\text{adj.}}{\otimes} \pionCon\quad \Leftrightarrow \quad \vectorCon \stackrel{\text{sym.}}{\otimes} \pionCon\,.
\end{equation}
It would be fascinating to determine if the resulting adjoint higher-spin theory that is sourced by the anomolous matrix elements in pure BI theory is in any way a consistent physical theory. One natural guess would be some sort of tensionless limit of string theory, where the tower of massless higher spin modes are needed for the theory to be unitary. Along these lines, there have been many recent studies into building higher-spin gauge theories constructively from general principles of locality and unitarity \cite{Caron-Huot:2016icg, Chiodaroli:2021eug,Cangemi:2022abk,Cangemi:2022bew,Geiser:2022exp,Cheung:2022mkw}. There is also a possibility that these higher-derivative operators could be required for double-copy consistency at higher-multiplicity. Such behavior was demonstrated for $\text{YM}+F^3$ gauge theory \cite{Carrasco:2022lbm}, where double-copy consistency demanded a tower of four-point contacts that conjecturally resums to $DF^2+\text{YM}$ theory of Johansson and Nohle \cite{Johansson:2017srf,Johansson:2018ues}. More recently, similar structure was likewise identified for higher-derivative corrections to biadjoint scalar EFT \cite{Chen:2022shl,Chen:2023dcx}. We see better understanding the double-copy consistency of higher-spin gauge theory as an exciting direction of future study. 
\subsection{Evanescent operator counting}\label{sec:EOpsCounting}
Before concluding, in this section we compute the counting of evanescent operators at higher orders in mass dimension needed for $U(1)$ duality satisfying quantum effective actions. We will do so using Hilbert series, which are an effective method for counting the number of independent operators at a particular order in mass dimension \cite{Henning:2015daa,Lehman:2015via} and have been used in abundance in the SMEFT literature \cite{Fonseca:2019yya,Hays:2018zze,Alioli:2022fng}. Hilbert series account for Lagrangian-level redundancy by counting the number of distinct on-shell quantities. For example, the following two operators would only be counted once since they are equivalent up to a boundary term,
\begin{equation}
(\partial_\rho F_{\mu\nu})(\partial^\rho F^{\mu\nu}) \sim \frac{1}{2}F_{\mu\nu}\partial_\rho \partial^\rho F^{\mu\nu}\,,
\end{equation}
due to the Bianchi identity, $D_{(\mu} F_{\nu \rho)} = 0$. As such, we will use the on-shell matrix elements, $\mathcal{T}$, for the computations in this section. Below we construct the Hilbert series for general dimension photon operators, $\mathcal{H}^{\text{gen.}D}$, and similarly so for $D=4$ operators, $\mathcal{H}^{D=4}$, for which we are restricted to a subset of 4D helicity operators. 

To construct the Hilbert series for photon operators, we will define the coefficient at order-$n$ of a polynomial in $\alpha$ to be the number of operators that appear at $\mathcal{O}(\alpha'^{n+2})$ in our dimensionful coupling. It turns out that there are only two integer sequences that control the operator counting, which are common feature when enumerating operator bases for 2-to-2 scattering events \cite{Damgaard:2019lfh,Haddad:2020que,Bern:2020uwk,Balkin:2021dko,Liu:2023jbq,Haddad:2023ylx}. The first sequence is produced by Hilbert series that counts four-point {permutation invariants}, $\sigma_3^x \sigma_2^y$, which we call $\mathcal{H}^{(ijkl)}$, 
\begin{equation}
[\mathcal{H}^{(ijkl)}] = 1,0,1,1,1,1,2,1,2,2,2,2,3,2,3,3,3,...
\end{equation}
and the second sequence counts {symmetric invariants}, $s_{ij}^x(s_{ik}s_{jk})^y$, which we denote as $\mathcal{H}^{(ij)(kl)}$,
\begin{equation}
[\mathcal{H}^{(ij)(kl)}] = 1,1,2,2,3,3,4,4,5,5,6,6,7,7,8,8,...
\end{equation}
Above, we have defined the bracket $[\mathcal{H}]$ such that it maps Hilbert series with integer coefficients into a sequence of numbers at successive orders in $\alpha$. Both of these Hilbert series are simple rational functions of $\alpha$, which we state below:
\begin{align}
\mathcal{H}^{(ij)(kl)} &= \frac{1}{(\alpha-1)^2(\alpha+1)} = \alpha^0+\alpha^2+\alpha^3+\alpha^4+\alpha^5+2\alpha^6+\alpha^7+2\alpha^8+\cdots
\\
\mathcal{H}^{(ijkl)} &= \frac{1}{(\alpha-1)^2(\alpha+1)(\alpha^2+\alpha+1)} = \alpha^0+\alpha^1+2\alpha^2+2\alpha^3+3\alpha^4+3\alpha^5+\cdots
\end{align}
Using these, we can infer the operator counting for both general dimension and the $D=4$ operators. The general dimension operator basis that we have used throughout scales as follows:
\begin{equation}
\mathcal{T}^{F^2F^2}_{(x,y)} \sim s_{ij}^x(s_{ik}s_{jk})^y \qquad \mathcal{T}^{F^4}_{(x,y)} \sim s_{ij}^x(s_{ik}s_{jk})^y \qquad \mathcal{T}^{F^3}_{(x,y)} \sim \sigma_3^x\sigma_2^y\,,
\end{equation}
where $\mathcal{T}^{F^4}_{(x,y)}$ begins at $\mathcal{O}(\alpha'^3)$ and both $\mathcal{T}^{F^2F^2}_{(x,y)}$ and $\mathcal{T}^{F^4}_{(x,y)} $ begin at $\mathcal{O}(\alpha'^{\,2})$. Thus, the Hilbert series $\mathcal{H}^{\text{gen.}D}$ can be defined as follows:
\begin{equation}
\mathcal{H}^{\text{gen.}D} = 2\mathcal{H}^{(ij)(kl)} + \alpha \mathcal{H}^{(ijkl)} \,.
\end{equation}
In contrast, the 4D helicity structures scale as,
\begin{equation}
\mathcal{T}^{(--++)}_{(x,y)} \sim s_{ij}^x(s_{ik}s_{jk})^y \qquad \mathcal{T}^{(-+++)}_{(x,y)} \sim \sigma_3^x\sigma_2^y\qquad \mathcal{T}^{(++++)}_{(x,y)} \sim \sigma_3^x\sigma_2^y\,.
\end{equation}
Similar to the $D$-dimensional operators above, $\mathcal{T}^{(--++)}_{(x,y)}$ starts at $\mathcal{O}(\alpha'^{\,2})$ and $\mathcal{T}^{(-+++)}_{(x,y)}$ begins the sequence at at $\mathcal{O}(\alpha'^3)$. However, the all plus behavior is slightly abnormal relative to the other counting sequences. Rather than pushing of the sequence to higher orders in $\alpha'$, it starts the sequence specified by $\mathcal{H}^{(ijkl)}$ at the {third entry} at $\mathcal{O}(\alpha'^{\,2})$. Thus, the $D=4$ Hilbert series can be defined as follows:
\begin{equation}
\mathcal{H}^{D=4} = \mathcal{H}^{(ij)(kl)} + \alpha \mathcal{H}^{(ijkl)} + (1+\alpha-\alpha^3)\mathcal{H}^{(ijkl)}\,.
\end{equation}
Putting this all together we obtain the following expression for the general dimension and 4D four-photon operator Hilbert series:
\begin{eBox}
\begin{equation}
\begin{aligned}
\mathcal{H}^{\text{gen.}D} &= \frac{(\alpha+2)(\alpha+1)+\alpha^2}{(\alpha
-1)^2(\alpha+1)(\alpha^2+\alpha+1)}\,,
\\
\mathcal{H}^{D=4} &= \frac{(\alpha+2)(\alpha+1)-\alpha^3}{(\alpha-1)^2(\alpha+1)(\alpha^2+\alpha+1)}\,.
\end{aligned}
\end{equation}
\end{eBox}
With this we can determine the number of four-photon evanescent operators that contribute at each successive order in $\alpha'$. The number of evanescent operators is determined by the difference in operator dimension between the general-$D$ Hilbert series and the $D=4$ Hilbert series, $\mathcal{H}^{\text{ev.}}=\mathcal{H}^{\text{gen.}D} - \mathcal{H}^{D=4}$. Thus we obtain the following Hilbert series for the number of evanescent four-photon operators at $\mathcal{O}(\alpha'^{n+2})$:
\begin{eBox}
\begin{equation}
\mathcal{H}^{\text{ev.}} = \frac{\alpha^2}{(\alpha-1)^2(\alpha^2+\alpha+1)}\,.
\end{equation}
\end{eBox}
It would be interesting to determine whether the three Hilbert series stated above owe their construction to some hidden geometric origin. Indeed all of the operator counting in the SMEFT literature can be traced to the geometry of the group theory representations that underly the Standard Model \cite{Henning:2015daa,Lehman:2015via}. We leave identifying these concealed mathematical structures as an exciting direction of future study. 
\section{Conclusions}\label{sec:Discussion}
In this manuscript, we have carried out a detailed analysis of even-point effective field theories through two-loop in the perturbative expansion. In \sect{sec:Review}, we provided a review of the generalized unitarity and integration methods employed throughout the text. Then in \sect{sec:EMU} we introduced and developed the on-shell constructive method of Even-point Multi-loop Unitarity (EMU), and computed the tensor reduction for triangle and bubble integrals in $D$-dimensions of arbitrary rank along with a spanning set of four-photon operators needed to capture higher-loop effects. Due to the simplicity of even-point multi-loop amplitudes of the nonlinear sigma model (NLSM) and Dirac-Born-Infeld-Volkov-Akulov (DBIVA) theories, these methods allowed us to compute fully integrated two-loop amplitudes for NLSM, pure Born-Infeld, and $\mathcal{N}=4$ DBIVA theory in \sect{sec:Loops}. Finally, in \sect{sec:Actions} we studied the quantum effective actions that capture the aforementioned loop effects, and studied the all-order counting of higher-derivative four-photon operators using Hilbert series. In doing so, we have identified a variety of rich physical structures that we summarize below: 

\paragraph{\textbf{Exponentiation}} In \eqn{eq:NLSM2bub} and \eqn{eq:NLSMostrich} we computed in general dimension, $D$, the two contributions to NLSM two-loop amplitudes. Plugging in explicit color structures, we found that the leading divergence of NLSM amplitudes on a $\mathbb{CP}^1$ target space exponentiate in $D=2-2\epsilon$, akin to the IR exponentiation of gravity amplitudes \cite{Weinberg:1965nx,Naculich:2008ew,White:2011yy,DiVecchia:2019myk,DiVecchia:2019kta,Heissenberg:2021tzo}. In addition, evaluating these diagrams in the planar limit, where $N_c \rightarrow \infty$, we found that the double-bubble integral iterates the leading divergence and subleading logarithms in the full one-loop amplitude. Thus, identifying corrections to the NLSM Lagrangian that absorb the ostrich diagram would imbue the scale-dependent logarithms with exponential structure at loop-level. This non-trivial property is found in theories that are conjectured to be integrable \cite{Shankar:1977cm,Zamolodchikov:1977nu, Komatsu:2019hgc}, like $\mathcal{N}=4$ super-Yang-Mills \cite{Anastasiou:2003kj,Bern:2005iz} in $D=4-2\epsilon$. In future work, we hope to study whether this iterative structure can be further applied beyond the $\mathbb{CP}^1$ model. 

\paragraph{\textbf{Anomalies}} Equipped with our $D$-dimensional integration methods, we performed a similar calculation at two-loop for pure Born-Infeld in \sect{sec:2loopBIU}. In doing so, we demonstrated that the previously identified one-loop counterterm \cite{Elvang:2020kuj} is not sufficient to cancel the two-loop anomaly. In fact, the $(-+++)$ anomaly of \eqn{eq:oneMinus2loop}, which was absent at one-loop, diverges at two-loop order due to the presence of a one-loop evanescent operator in the $D$-dimensional formulation of Born-Infeld theory.~One resolution to this anomaly comes in the form of introducing $\mathcal{N}=4$ DBIVA superfields in the two-loop state-sum. This protects the $S$-matrix from anomalies by promoting the classically conserved $U(1)$ duality to a supersymmetric $R$-symmetry. The evaluated integrals that contribute to the $\mathcal{N}=4$ DBIVA two-loop amplitudes are provided in \eqn{eq:N42bub} and \eqn{eq:N4ostrich}.  

As mentioned at the close of \sec{subsec:EPEFTReview}, higher-multiplicity tree-level abelianized open-superstring amplitudes, in four-dimensions, violates $U(1)$ duality at higher-derivative corrections beyond the leading DBIVA predictions. It would be fascinating if these higher derivative violations of $U(1)$ duality in the OSS spectrum are precisely those needed to cancel anomalous behavior at the quantum level. To test this would require computing a six-point one-loop amplitude in supersymmetric DBIVA. Such a calculation could in principle be performed with the double copy using the color-dual integrands recently constructed in Ref.~\cite{Edison:2022jln}. We see this as a natural future direction and application of the methods we have developed here.

\paragraph{\textbf{Evanescence}}Another resolution to the two-loop anomaly comes in the form of higher-derivative pure-photon counterterms. As we demonstrated in \sect{sec:Anomalies}, one must introduce a divergent evanescent operator at one-loop in order to absorb the two-loop anomaly. This is similar to the anomalies of pure Einstein-Hilbert gravity \cite{Bern:2015xsa,Bern:2017puu}, which vanish at one-loop since the $R^2$ Gauss-Bonet term is evanescent in $D=4$, but which diverge at two-loop order \cite{Goroff:1985sz,Goroff:1985th,vandeVen:1991gw}. We have constructed the higher derivative Lagrangian through $\mathcal{O}(\alpha'^6)$ in \eqn{eq:BI+CTwithOev} that cancels the divergent part of the anomaly, along with a spanning set of counterterms in \eqn{eq:Oa6counterTerms} needed to absorb the finite part left over at two-loop. Given the complexity of gravity calculations at high loop order, further studies of multi-loop Born-Infeld amplitudes could serve as an accessible laboratory for studying evanescent effects beyond one-loop in double-copy constructible theories. To this end, we have used Hilbert series to count the number of four-photon evanescent operators to higher-order derivative corrections in \sect{sec:EOpsCounting} to aid in future studies. 

\paragraph{} In addition to these themes woven throughout the text, we have, \textit{en passant}, identified novel double-copy structures at two-loop. In \sect{sec:DBIvDC} we found that two-loop $\mathcal{N}=4$ DBIVA amplitudes can be constructed via the double copy of color-dual $\mathcal{N}=4$ sYM integrands with the generalized unitarity cuts of NLSM. This construction was $D$-dimensionally identical to the result obtained via generalized unitarity and maximal supersymmetric state sums. While not a proof, this provides strong evidence for the compatibility of NLSM with color-kinematics duality beyond one-loop. This result also serves as the first non-gravitational double-copy beyond one-loop, further supporting the consistency of color-kinematics duality at loop-level, which as of today remains a conjecture.

Moreover, recent work by the authors has demonstrated that color-kinematics duality can be used as a bootstrap principle to constrain higher derivative operators. This has been shown both for gauged NLSM amplitudes \cite{Carrasco:2022sck} and $\text{YM}+F^3$ theory \cite{Carrasco:2022lbm}, the later of which is particularly relevant for anomaly cancellation, and possibly UV completion, of both $\mathcal{N}=4$ supergravity and the $R^3$ modification to Einstein-Hilbert gravity \cite{Carrasco:2022lbm}. Indeed, similar structure has been recently identified in color-dual scalar theories \cite{Chen:2022shl,Chen:2023dcx,Brown:2023srz}. This observation suggests a new paradigm that elevates color-kinematics duality from a mathematical correspondence capable of encoding IR symmetries, to a principle that probes signatures of UV physics captured by higher-derivative corrections. 
Guided by this new paradigm, a natural next step is to determine whether the anomaly cancelling counterterms of \eqn{eq:allPlus2} and \eqn{eq:oneMinus2loop} source additional higher-loop counterterms constrained by double-copy consistency, in the spirit of \cite{Carrasco:2022lbm}. 
We see this as an exciting future direction in further understanding the loop-level constraints imposed by the duality between color and kinematics.

\paragraph{Acknowledgments} The authors would like to thank Rafael Aoude, Alex Edison, Kezhu Guo, Kays Haddad, Ian Low, James Mangan, Frank Petriello, Paolo Pichini, Nia Robles, Radu Roiban, Aslan Seifi, and Suna Zekio\u{g}lu for insightful conversations, related collaboration, and encouragement throughout the completion of this work. That authors additionally would like to thank James Mangan for incredibly thoughtful comments on earlier versions of the draft. The completion of this manuscript benefited from the hospitality of NORDITA during the workshop ``Amplifying Gravity
at All Scales". This work was supported by the DOE under contract DE-SC0015910 and by the Alfred P. Sloan Foundation. N.H.P. acknowledges the Northwestern University Amplitudes and Insight group, the Department of Physics and Astronomy, and Weinberg College for their generous support. 

\bibliographystyle{JHEP}
\bibliography{Refs_emu}

\providecommand{\href}[2]{#2}\begingroup\raggedright\begin{thebibliography}{100}

\bibitem{UnitarityMethod}
Z.~Bern, L.J.~Dixon, D.C.~Dunbar and D.A.~Kosower, \emph{{One loop $n$-point
  gauge theory amplitudes, unitarity and collinear limits}},
  \href{https://doi.org/10.1016/0550-3213(94)90179-1}{\emph{Nucl. Phys.}
  {\bfseries B425} (1994) 217}
  [\href{https://arxiv.org/abs/hep-ph/9403226}{{\ttfamily hep-ph/9403226}}].

\bibitem{Fusing}
Z.~Bern, L.J.~Dixon, D.C.~Dunbar and D.A.~Kosower, \emph{{Fusing gauge theory
  tree amplitudes into loop amplitudes}},
  \href{https://doi.org/10.1016/0550-3213(94)00488-Z}{\emph{Nucl. Phys.}
  {\bfseries B435} (1995) 59}
  [\href{https://arxiv.org/abs/hep-ph/9409265}{{\ttfamily hep-ph/9409265}}].

\bibitem{BDKUniarityReview}
Z.~Bern, L.J.~Dixon and D.A.~Kosower, \emph{{Progress in one loop QCD
  computations}},
  \href{https://doi.org/10.1146/annurev.nucl.46.1.109}{\emph{Ann. Rev. Nucl.
  Part. Sci.} {\bfseries 46} (1996) 109}
  [\href{https://arxiv.org/abs/hep-ph/9602280}{{\ttfamily hep-ph/9602280}}].

\bibitem{Forde:2007mi}
D.~Forde, \emph{{Direct extraction of one-loop integral coefficients}},
  \href{https://doi.org/10.1103/PhysRevD.75.125019}{\emph{Phys. Rev. D}
  {\bfseries 75} (2007) 125019}
  [\href{https://arxiv.org/abs/0704.1835}{{\ttfamily 0704.1835}}].

\bibitem{BCJ}
Z.~Bern, J.J.M.~Carrasco and H.~Johansson, \emph{{New relations for
  gauge-theory amplitudes}},
  \href{https://doi.org/10.1103/PhysRevD.78.085011}{\emph{Phys. Rev.}
  {\bfseries D78} (2008) 085011}
  [\href{https://arxiv.org/abs/0805.3993}{{\ttfamily 0805.3993}}].

\bibitem{BCJLoop}
Z.~Bern, J.J.M.~Carrasco and H.~Johansson, \emph{{Perturbative quantum gravity
  as a double copy of gauge theory}},
  \href{https://doi.org/10.1103/PhysRevLett.105.061602}{\emph{Phys. Rev. Lett.}
  {\bfseries 105} (2010) 061602}
  [\href{https://arxiv.org/abs/1004.0476}{{\ttfamily 1004.0476}}].

\bibitem{Adams:2006sv}
A.~Adams, N.~Arkani-Hamed, S.~Dubovsky, A.~Nicolis and R.~Rattazzi,
  \emph{{Causality, analyticity and an IR obstruction to UV completion}},
  \href{https://doi.org/10.1088/1126-6708/2006/10/014}{\emph{JHEP} {\bfseries
  10} (2006) 014} [\href{https://arxiv.org/abs/hep-th/0602178}{{\ttfamily
  hep-th/0602178}}].

\bibitem{Cheung:2014dqa}
C.~Cheung, K.~Kampf, J.~Novotny and J.~Trnka, \emph{{Effective Field Theories
  from Soft Limits of Scattering Amplitudes}},
  \href{https://doi.org/10.1103/PhysRevLett.114.221602}{\emph{Phys. Rev. Lett.}
  {\bfseries 114} (2015) 221602}
  [\href{https://arxiv.org/abs/1412.4095}{{\ttfamily 1412.4095}}].

\bibitem{Cheung:2015ota}
C.~Cheung, K.~Kampf, J.~Novotny, C.-H.~Shen and J.~Trnka, \emph{{On-Shell
  Recursion Relations for Effective Field Theories}},
  \href{https://doi.org/10.1103/PhysRevLett.116.041601}{\emph{Phys. Rev. Lett.}
  {\bfseries 116} (2016) 041601}
  [\href{https://arxiv.org/abs/1509.03309}{{\ttfamily 1509.03309}}].

\bibitem{Cheung:2016drk}
C.~Cheung, K.~Kampf, J.~Novotny, C.-H.~Shen and J.~Trnka, \emph{{A Periodic
  Table of Effective Field Theories}},
  \href{https://doi.org/10.1007/JHEP02(2017)020}{\emph{JHEP} {\bfseries 02}
  (2017) 020} [\href{https://arxiv.org/abs/1611.03137}{{\ttfamily
  1611.03137}}].

\bibitem{Cheung:2018oki}
C.~Cheung, K.~Kampf, J.~Novotny, C.-H.~Shen, J.~Trnka and C.~Wen, \emph{{Vector
  Effective Field Theories from Soft Limits}},
  \href{https://doi.org/10.1103/PhysRevLett.120.261602}{\emph{Phys. Rev. Lett.}
  {\bfseries 120} (2018) 261602}
  [\href{https://arxiv.org/abs/1801.01496}{{\ttfamily 1801.01496}}].

\bibitem{Low:2019ynd}
I.~Low and Z.~Yin, \emph{{Soft Bootstrap and Effective Field Theories}},
  \href{https://doi.org/10.1007/JHEP11(2019)078}{\emph{JHEP} {\bfseries 11}
  (2019) 078} [\href{https://arxiv.org/abs/1904.12859}{{\ttfamily
  1904.12859}}].

\bibitem{Carrasco:2019yyn}
J.J.M.~Carrasco, L.~Rodina, Z.~Yin and S.~Zekioglu, \emph{{Simple encoding of
  higher derivative gauge and gravity counterterms}},
  \href{https://doi.org/10.1103/PhysRevLett.125.251602}{\emph{Phys. Rev. Lett.}
  {\bfseries 125} (2020) 251602}
  [\href{https://arxiv.org/abs/1910.12850}{{\ttfamily 1910.12850}}].

\bibitem{Arkani-Hamed:2020blm}
N.~Arkani-Hamed, T.-C.~Huang and Y.-t.~Huang, \emph{{The EFT-Hedron}},
  \href{https://doi.org/10.1007/JHEP05(2021)259}{\emph{JHEP} {\bfseries 05}
  (2021) 259} [\href{https://arxiv.org/abs/2012.15849}{{\ttfamily
  2012.15849}}].

\bibitem{Carrasco:2021ptp}
J.J.M.~Carrasco, L.~Rodina and S.~Zekioglu, \emph{{Composing effective
  prediction at five points}},
  \href{https://doi.org/10.1007/JHEP06(2021)169}{\emph{JHEP} {\bfseries 06}
  (2021) 169} [\href{https://arxiv.org/abs/2104.08370}{{\ttfamily
  2104.08370}}].

\bibitem{Chi:2021mio}
H.-H.~Chi, H.~Elvang, A.~Herderschee, C.R.T.~Jones and S.~Paranjape,
  \emph{{Generalizations of the double-copy: the KLT bootstrap}},
  \href{https://doi.org/10.1007/JHEP03(2022)077}{\emph{JHEP} {\bfseries 03}
  (2022) 077} [\href{https://arxiv.org/abs/2106.12600}{{\ttfamily
  2106.12600}}].

\bibitem{Bonnefoy:2021qgu}
Q.~Bonnefoy, G.~Durieux, C.~Grojean, C.S.~Machado and J.~Roosmale~Nepveu,
  \emph{{The seeds of EFT double copy}},
  \href{https://doi.org/10.1007/JHEP05(2022)042}{\emph{JHEP} {\bfseries 05}
  (2022) 042} [\href{https://arxiv.org/abs/2112.11453}{{\ttfamily
  2112.11453}}].

\bibitem{Carrasco:2022lbm}
J.J.M.~Carrasco, M.~Lewandowski and N.H.~Pavao, \emph{{The color-dual fates of
  $F^3$, $R^3$, and $\mathcal{N}=4$ supergravity}},
  \href{https://arxiv.org/abs/2203.03592}{{\ttfamily 2203.03592}}.

\bibitem{Carrasco:2022sck}
J.J.M.~Carrasco, M.~Lewandowski and N.H.~Pavao, \emph{{Double-copy towards
  supergravity inflation with \ensuremath{\alpha}-attractor models}},
  \href{https://doi.org/10.1007/JHEP02(2023)015}{\emph{JHEP} {\bfseries 02}
  (2023) 015} [\href{https://arxiv.org/abs/2211.04441}{{\ttfamily
  2211.04441}}].

\bibitem{Green:2022slj}
D.~Green, Y.~Huang and C.-H.~Shen, \emph{{Inflationary Adler conditions}},
  \href{https://doi.org/10.1103/PhysRevD.107.043534}{\emph{Phys. Rev. D}
  {\bfseries 107} (2023) 043534}
  [\href{https://arxiv.org/abs/2208.14544}{{\ttfamily 2208.14544}}].

\bibitem{Pavao:2022kog}
N.H.~Pavao, \emph{{Effective observables for electromagnetic duality from novel
  amplitude decomposition}},
  \href{https://doi.org/10.1103/PhysRevD.107.065020}{\emph{Phys. Rev. D}
  {\bfseries 107} (2023) 065020}
  [\href{https://arxiv.org/abs/2210.12800}{{\ttfamily 2210.12800}}].

\bibitem{Chen:2022shl}
A.S.-K.~Chen, H.~Elvang and A.~Herderschee, \emph{{Emergence of String
  Monodromy in Effective Field Theory}},
  \href{https://arxiv.org/abs/2212.13998}{{\ttfamily 2212.13998}}.

\bibitem{Chen:2023dcx}
A.S.-K.~Chen, H.~Elvang and A.~Herderschee, \emph{{Bootstrapping the String KLT
  Kernel}},  \href{https://arxiv.org/abs/2302.04895}{{\ttfamily 2302.04895}}.

\bibitem{Brown:2023srz}
T.V.~Brown, K.~Kampf, U.~Oktem, S.~Paranjape and J.~Trnka, \emph{{Scalar BCJ
  Bootstrap}},  \href{https://arxiv.org/abs/2305.05688}{{\ttfamily
  2305.05688}}.

\bibitem{Li:2023wdm}
Y.~Li, D.~Roest and T.~ter Veldhuis, \emph{{Hybrid Goldstone Modes from the
  Double Copy Bootstrap}},  \href{https://arxiv.org/abs/2307.13418}{{\ttfamily
  2307.13418}}.

\bibitem{BCJreview}
Z.~Bern, J.J.~Carrasco, M.~Chiodaroli, H.~Johansson and R.~Roiban, \emph{{The
  Duality Between Color and Kinematics and its Applications}},
  \href{https://arxiv.org/abs/1909.01358}{{\ttfamily 1909.01358}}.

\bibitem{Bern:2022wqg}
Z.~Bern, J.J.~Carrasco, M.~Chiodaroli, H.~Johansson and R.~Roiban, \emph{{The
  SAGEX Review on Scattering Amplitudes, Chapter 2: An Invitation to
  Color-Kinematics Duality and the Double Copy}},
  \href{https://arxiv.org/abs/2203.13013}{{\ttfamily 2203.13013}}.

\bibitem{Adamo:2022dcm}
T.~Adamo, J.J.M.~Carrasco, M.~Carrillo-Gonz\'alez, M.~Chiodaroli, H.~Elvang,
  H.~Johansson et~al., \emph{{Snowmass White Paper: the Double Copy and its
  Applications}},  in \emph{{2022 Snowmass Summer Study}}, 4, 2022
  [\href{https://arxiv.org/abs/2204.06547}{{\ttfamily 2204.06547}}].

\bibitem{Bern:2012uf}
Z.~Bern, J.J.M.~Carrasco, L.J.~Dixon, H.~Johansson and R.~Roiban,
  \emph{{Simplifying Multiloop Integrands and Ultraviolet Divergences of Gauge
  Theory and Gravity Amplitudes}},
  \href{https://doi.org/10.1103/PhysRevD.85.105014}{\emph{Phys. Rev. D}
  {\bfseries 85} (2012) 105014}
  [\href{https://arxiv.org/abs/1201.5366}{{\ttfamily 1201.5366}}].

\bibitem{Neq44np}
Z.~Bern, J.J.M.~Carrasco, L.J.~Dixon, H.~Johansson and R.~Roiban, \emph{{The
  complete four-loop four-point amplitude in ${\cal N}=4$ super-Yang-Mills
  theory}}, \href{https://doi.org/10.1103/PhysRevD.82.125040}{\emph{Phys. Rev.}
  {\bfseries D82} (2010) 125040}
  [\href{https://arxiv.org/abs/1008.3327}{{\ttfamily 1008.3327}}].

\bibitem{GravityFour}
Z.~Bern, J.J.~Carrasco, L.J.~Dixon, H.~Johansson and R.~Roiban, \emph{{The
  ultraviolet behavior of ${\cal N}=8$ supergravity at four loops}},
  \href{https://doi.org/10.1103/PhysRevLett.103.081301}{\emph{Phys. Rev. Lett.}
  {\bfseries 103} (2009) 081301}
  [\href{https://arxiv.org/abs/0905.2326}{{\ttfamily 0905.2326}}].

\bibitem{Bjerrum-Bohr:2013iza}
N.E.J.~Bjerrum-Bohr, T.~Dennen, R.~Monteiro and D.~O'Connell, \emph{{Integrand
  Oxidation and one-loop colour-dual numerators in ${\cal N}=4$ gauge theory}},
  \href{https://doi.org/10.1007/JHEP07(2013)092}{\emph{JHEP} {\bfseries 07}
  (2013) 092} [\href{https://arxiv.org/abs/1303.2913}{{\ttfamily 1303.2913}}].

\bibitem{Edison:2020uzf}
A.~Edison, S.~He, O.~Schlotterer and F.~Teng, \emph{{One-loop Correlators and
  BCJ Numerators from Forward Limits}},
  \href{https://doi.org/10.1007/JHEP09(2020)079}{\emph{JHEP} {\bfseries 09}
  (2020) 079} [\href{https://arxiv.org/abs/2005.03639}{{\ttfamily
  2005.03639}}].

\bibitem{Edison:2022jln}
A.~Edison, S.~He, H.~Johansson, O.~Schlotterer, F.~Teng and Y.~Zhang,
  \emph{{Perfecting one-loop BCJ numerators in SYM and supergravity}},
  \href{https://doi.org/10.1007/JHEP02(2023)164}{\emph{JHEP} {\bfseries 02}
  (2023) 164} [\href{https://arxiv.org/abs/2211.00638}{{\ttfamily
  2211.00638}}].

\bibitem{Cheung2016prv}
C.~Cheung and C.-H.~Shen, \emph{{Symmetry for flavor-kinematics duality from an
  action}}, \href{https://doi.org/10.1103/PhysRevLett.118.121601}{\emph{Phys.
  Rev. Lett.} {\bfseries 118} (2017) 121601}
  [\href{https://arxiv.org/abs/1612.00868}{{\ttfamily 1612.00868}}].

\bibitem{He:2017spx}
S.~He, O.~Schlotterer and Y.~Zhang, \emph{{New BCJ representations for one-loop
  amplitudes in gauge theories and gravity}},
  \href{https://doi.org/10.1016/j.nuclphysb.2018.03.003}{\emph{Nucl. Phys. B}
  {\bfseries 930} (2018) 328}
  [\href{https://arxiv.org/abs/1706.00640}{{\ttfamily 1706.00640}}].

\bibitem{OneTwoLoopPureYMBCJ}
Z.~Bern, S.~Davies, T.~Dennen, Y.-t.~Huang and J.~Nohle,
  \emph{{Color-kinematics duality for pure Yang-Mills and gravity at one and
  two Loops}}, \href{https://doi.org/10.1103/PhysRevD.92.045041}{\emph{Phys.
  Rev.} {\bfseries D92} (2015) 045041}
  [\href{https://arxiv.org/abs/1303.6605}{{\ttfamily 1303.6605}}].

\bibitem{Mogull:2015adi}
G.~Mogull and D.~O'Connell, \emph{{Overcoming Obstacles to Colour-Kinematics
  Duality at Two Loops}},
  \href{https://doi.org/10.1007/JHEP12(2015)135}{\emph{JHEP} {\bfseries 12}
  (2015) 135} [\href{https://arxiv.org/abs/1511.06652}{{\ttfamily
  1511.06652}}].

\bibitem{Bern:2015ooa}
Z.~Bern, S.~Davies and J.~Nohle, \emph{{Double-Copy Constructions and Unitarity
  Cuts}}, \href{https://doi.org/10.1103/PhysRevD.93.105015}{\emph{Phys. Rev. D}
  {\bfseries 93} (2016) 105015}
  [\href{https://arxiv.org/abs/1510.03448}{{\ttfamily 1510.03448}}].

\bibitem{Geyer:2019hnn}
Y.~Geyer, R.~Monteiro and R.~Stark-Much\~ao, \emph{{Two-Loop Scattering
  Amplitudes: Double-Forward Limit and Colour-Kinematics Duality}},
  \href{https://doi.org/10.1007/JHEP12(2019)049}{\emph{JHEP} {\bfseries 12}
  (2019) 049} [\href{https://arxiv.org/abs/1908.05221}{{\ttfamily
  1908.05221}}].

\bibitem{Johansson:2017bfl}
H.~Johansson, G.~K\"alin and G.~Mogull, \emph{{Two-loop supersymmetric QCD and
  half-maximal supergravity amplitudes}},
  \href{https://doi.org/10.1007/JHEP09(2017)019}{\emph{JHEP} {\bfseries 09}
  (2017) 019} [\href{https://arxiv.org/abs/1706.09381}{{\ttfamily
  1706.09381}}].

\bibitem{Chen:2019ywi}
G.~Chen, H.~Johansson, F.~Teng and T.~Wang, \emph{{On the kinematic algebra for
  BCJ numerators beyond the MHV sector}},
  \href{https://doi.org/10.1007/JHEP11(2019)055}{\emph{JHEP} {\bfseries 11}
  (2019) 055} [\href{https://arxiv.org/abs/1906.10683}{{\ttfamily
  1906.10683}}].

\bibitem{Chen:2021chy}
G.~Chen, H.~Johansson, F.~Teng and T.~Wang, \emph{{Next-to-MHV Yang-Mills
  kinematic algebra}},
  \href{https://doi.org/10.1007/JHEP10(2021)042}{\emph{JHEP} {\bfseries 10}
  (2021) 042} [\href{https://arxiv.org/abs/2104.12726}{{\ttfamily
  2104.12726}}].

\bibitem{Brandhuber:2021bsf}
A.~Brandhuber, G.~Chen, H.~Johansson, G.~Travaglini and C.~Wen,
  \emph{{Kinematic Hopf Algebra for Bern-Carrasco-Johansson Numerators in
  Heavy-Mass Effective Field Theory and Yang-Mills Theory}},
  \href{https://doi.org/10.1103/PhysRevLett.128.121601}{\emph{Phys. Rev. Lett.}
  {\bfseries 128} (2022) 121601}
  [\href{https://arxiv.org/abs/2111.15649}{{\ttfamily 2111.15649}}].

\bibitem{Cheung:2021zvb}
C.~Cheung and J.~Mangan, \emph{{Covariant color-kinematics duality}},
  \href{https://doi.org/10.1007/JHEP11(2021)069}{\emph{JHEP} {\bfseries 11}
  (2021) 069} [\href{https://arxiv.org/abs/2108.02276}{{\ttfamily
  2108.02276}}].

\bibitem{Ben-Shahar:2021zww}
M.~Ben-Shahar and H.~Johansson, \emph{{Off-shell color-kinematics duality for
  Chern-Simons}}, \href{https://doi.org/10.1007/JHEP08(2022)035}{\emph{JHEP}
  {\bfseries 08} (2022) 035}
  [\href{https://arxiv.org/abs/2112.11452}{{\ttfamily 2112.11452}}].

\bibitem{Cheung:2022mix}
C.~Cheung, J.~Mangan, J.~Parra-Martinez and N.~Shah, \emph{{Non-perturbative
  Double Copy in Flatland}},
  \href{https://doi.org/10.1103/PhysRevLett.129.221602}{\emph{Phys. Rev. Lett.}
  {\bfseries 129} (2022) 221602}
  [\href{https://arxiv.org/abs/2204.07130}{{\ttfamily 2204.07130}}].

\bibitem{Ben-Shahar:2022ixa}
M.~Ben-Shahar, L.~Garozzo and H.~Johansson, \emph{{Lagrangians Manifesting
  Color-Kinematics Duality in the NMHV Sector of Yang-Mills}},
  \href{https://arxiv.org/abs/2301.00233}{{\ttfamily 2301.00233}}.

\bibitem{Cachazo:2014xea}
F.~Cachazo, S.~He and E.Y.~Yuan, \emph{{Scattering Equations and Matrices: From
  Einstein To Yang-Mills, DBI and NLSM}},
  \href{https://doi.org/10.1007/JHEP07(2015)149}{\emph{JHEP} {\bfseries 07}
  (2015) 149} [\href{https://arxiv.org/abs/1412.3479}{{\ttfamily 1412.3479}}].

\bibitem{Carrasco:2016ldy}
J.J.M.~Carrasco, C.R.~Mafra and O.~Schlotterer, \emph{{Abelian Z-theory: NLSM
  amplitudes and $\alpha$'-corrections from the open string}},
  \href{https://doi.org/10.1007/JHEP06(2017)093}{\emph{JHEP} {\bfseries 06}
  (2017) 093} [\href{https://arxiv.org/abs/1608.02569}{{\ttfamily
  1608.02569}}].

\bibitem{Carrasco:2022jxn}
J.J.M.~Carrasco and N.H.~Pavao, \emph{{Virtues of a symmetric-structure double
  copy}}, \href{https://doi.org/10.1103/PhysRevD.107.065005}{\emph{Phys. Rev.
  D} {\bfseries 107} (2023) 065005}
  [\href{https://arxiv.org/abs/2211.04431}{{\ttfamily 2211.04431}}].

\bibitem{Elvang:2020kuj}
H.~Elvang, M.~Hadjiantonis, C.R.T.~Jones and S.~Paranjape,
  \emph{{Electromagnetic Duality and D3-Brane Scattering Amplitudes Beyond
  Leading Order}}, \href{https://doi.org/10.1007/JHEP04(2021)173}{\emph{JHEP}
  {\bfseries 04} (2021) 173}
  [\href{https://arxiv.org/abs/2006.08928}{{\ttfamily 2006.08928}}].

\bibitem{Bern:2007xj}
Z.~Bern, J.J.~Carrasco, D.~Forde, H.~Ita and H.~Johansson, \emph{{Unexpected
  Cancellations in Gravity Theories}},
  \href{https://doi.org/10.1103/PhysRevD.77.025010}{\emph{Phys. Rev. D}
  {\bfseries 77} (2008) 025010}
  [\href{https://arxiv.org/abs/0707.1035}{{\ttfamily 0707.1035}}].

\bibitem{Carrasco:2013ypa}
J.J.M.~Carrasco, R.~Kallosh, R.~Roiban and A.A.~Tseytlin, \emph{{On the U(1)
  duality anomaly and the S-matrix of N=4 supergravity}},
  \href{https://doi.org/10.1007/JHEP07(2013)029}{\emph{JHEP} {\bfseries 07}
  (2013) 029} [\href{https://arxiv.org/abs/1303.6219}{{\ttfamily 1303.6219}}].

\bibitem{Craig:2019zkf}
N.~Craig, I.~Garcia~Garcia and G.D.~Kribs, \emph{{The UV fate of anomalous
  U(1)s and the Swampland}},
  \href{https://doi.org/10.1007/JHEP11(2020)063}{\emph{JHEP} {\bfseries 11}
  (2020) 063} [\href{https://arxiv.org/abs/1912.10054}{{\ttfamily
  1912.10054}}].

\bibitem{Monteiro:2022nqt}
R.~Monteiro, R.~Stark-Much\~ao and S.~Wikeley, \emph{{Anomaly and double copy
  in quantum self-dual Yang-Mills and gravity}},
  \href{https://arxiv.org/abs/2211.12407}{{\ttfamily 2211.12407}}.

\bibitem{Bern:2017tuc}
Z.~Bern, A.~Edison, D.~Kosower and J.~Parra-Martinez, \emph{{Curvature-squared
  multiplets, evanescent effects, and the U(1) anomaly in ${\cal N}=4$
  supergravity}}, \href{https://doi.org/10.1103/PhysRevD.96.066004}{\emph{Phys.
  Rev.} {\bfseries D96} (2017) 066004}
  [\href{https://arxiv.org/abs/1706.01486}{{\ttfamily 1706.01486}}].

\bibitem{Bern:2017rjw}
Z.~Bern, J.~Parra-Martinez and R.~Roiban, \emph{{Canceling the U(1) Anomaly in
  the $S$ Matrix of $N$=4 Supergravity}},
  \href{https://doi.org/10.1103/PhysRevLett.121.101604}{\emph{Phys. Rev. Lett.}
  {\bfseries 121} (2018) 101604}
  [\href{https://arxiv.org/abs/1712.03928}{{\ttfamily 1712.03928}}].

\bibitem{Bern:2019isl}
Z.~Bern, D.~Kosower and J.~Parra-Martinez, \emph{{Two-loop $n$-point anomalous
  amplitudes in ${\cal N}=4$ supergravity}},
  \href{https://arxiv.org/abs/1905.05151}{{\ttfamily 1905.05151}}.

\bibitem{Goroff:1985sz}
M.H.~Goroff and A.~Sagnotti, \emph{{QUANTUM GRAVITY AT TWO LOOPS}},
  \href{https://doi.org/10.1016/0370-2693(85)91470-4}{\emph{Phys. Lett. B}
  {\bfseries 160} (1985) 81}.

\bibitem{Goroff:1985th}
M.H.~Goroff and A.~Sagnotti, \emph{{The Ultraviolet Behavior of Einstein
  Gravity}}, \href{https://doi.org/10.1016/0550-3213(86)90193-8}{\emph{Nucl.
  Phys. B} {\bfseries 266} (1986) 709}.

\bibitem{vandeVen:1991gw}
A.E.M.~van~de Ven, \emph{{Two loop quantum gravity}},
  \href{https://doi.org/10.1016/0550-3213(92)90011-Y}{\emph{Nucl. Phys. B}
  {\bfseries 378} (1992) 309}.

\bibitem{Bern:2013uka}
Z.~Bern, S.~Davies, T.~Dennen, A.V.~Smirnov and V.A.~Smirnov,
  \emph{{Ultraviolet Properties of N=4 Supergravity at Four Loops}},
  \href{https://doi.org/10.1103/PhysRevLett.111.231302}{\emph{Phys. Rev. Lett.}
  {\bfseries 111} (2013) 231302}
  [\href{https://arxiv.org/abs/1309.2498}{{\ttfamily 1309.2498}}].

\bibitem{Alishahiha:2004eh}
M.~Alishahiha, E.~Silverstein and D.~Tong, \emph{{DBI in the sky}},
  \href{https://doi.org/10.1103/PhysRevD.70.123505}{\emph{Phys. Rev. D}
  {\bfseries 70} (2004) 123505}
  [\href{https://arxiv.org/abs/hep-th/0404084}{{\ttfamily hep-th/0404084}}].

\bibitem{Creminelli:2005hu}
P.~Creminelli, A.~Nicolis, L.~Senatore, M.~Tegmark and M.~Zaldarriaga,
  \emph{{Limits on non-gaussianities from wmap data}},
  \href{https://doi.org/10.1088/1475-7516/2006/05/004}{\emph{JCAP} {\bfseries
  05} (2006) 004} [\href{https://arxiv.org/abs/astro-ph/0509029}{{\ttfamily
  astro-ph/0509029}}].

\bibitem{Fergusson:2008ra}
J.R.~Fergusson and E.P.S.~Shellard, \emph{{The shape of primordial
  non-Gaussianity and the CMB bispectrum}},
  \href{https://doi.org/10.1103/PhysRevD.80.043510}{\emph{Phys. Rev. D}
  {\bfseries 80} (2009) 043510}
  [\href{https://arxiv.org/abs/0812.3413}{{\ttfamily 0812.3413}}].

\bibitem{Carrasco:2015pla}
J.J.M.~Carrasco, R.~Kallosh and A.~Linde, \emph{{$\alpha $-Attractors: Planck,
  LHC and Dark Energy}},
  \href{https://doi.org/10.1007/JHEP10(2015)147}{\emph{JHEP} {\bfseries 10}
  (2015) 147} [\href{https://arxiv.org/abs/1506.01708}{{\ttfamily
  1506.01708}}].

\bibitem{Carrasco:2015rva}
J.J.M.~Carrasco, R.~Kallosh and A.~Linde, \emph{{Cosmological Attractors and
  Initial Conditions for Inflation}},
  \href{https://doi.org/10.1103/PhysRevD.92.063519}{\emph{Phys. Rev. D}
  {\bfseries 92} (2015) 063519}
  [\href{https://arxiv.org/abs/1506.00936}{{\ttfamily 1506.00936}}].

\bibitem{Carrasco:2015uma}
J.J.M.~Carrasco, R.~Kallosh, A.~Linde and D.~Roest, \emph{{Hyperbolic geometry
  of cosmological attractors}},
  \href{https://doi.org/10.1103/PhysRevD.92.041301}{\emph{Phys. Rev. D}
  {\bfseries 92} (2015) 041301}
  [\href{https://arxiv.org/abs/1504.05557}{{\ttfamily 1504.05557}}].

\bibitem{BICEP:2021xfz}
{\scshape BICEP, Keck} collaboration, \emph{{Improved Constraints on Primordial
  Gravitational Waves using Planck, WMAP, and BICEP/Keck Observations through
  the 2018 Observing Season}},
  \href{https://doi.org/10.1103/PhysRevLett.127.151301}{\emph{Phys. Rev. Lett.}
  {\bfseries 127} (2021) 151301}
  [\href{https://arxiv.org/abs/2110.00483}{{\ttfamily 2110.00483}}].

\bibitem{Kallosh:2021mnu}
R.~Kallosh and A.~Linde, \emph{{BICEP/Keck and cosmological attractors}},
  \href{https://doi.org/10.1088/1475-7516/2021/12/008}{\emph{JCAP} {\bfseries
  12} (2021) 008} [\href{https://arxiv.org/abs/2110.10902}{{\ttfamily
  2110.10902}}].

\bibitem{Kleiss:1988ne}
R.~Kleiss and H.~Kuijf, \emph{{Multi - Gluon Cross-sections and Five Jet
  Production at Hadron Colliders}},
  \href{https://doi.org/10.1016/0550-3213(89)90574-9}{\emph{Nucl. Phys. B}
  {\bfseries 312} (1989) 616}.

\bibitem{Bern:1998sv}
Z.~Bern, L.J.~Dixon, M.~Perelstein and J.S.~Rozowsky, \emph{{Multileg one loop
  gravity amplitudes from gauge theory}},
  \href{https://doi.org/10.1016/S0550-3213(99)00029-2}{\emph{Nucl. Phys. B}
  {\bfseries 546} (1999) 423}
  [\href{https://arxiv.org/abs/hep-th/9811140}{{\ttfamily hep-th/9811140}}].

\bibitem{Vaman:2010ez}
D.~Vaman and Y.-P.~Yao, \emph{{Constraints and generalized gauge
  transformations on tree-level gluon and graviton amplitudes}},
  \href{https://doi.org/10.1007/JHEP11(2010)028}{\emph{JHEP} {\bfseries 11}
  (2010) 028} [\href{https://arxiv.org/abs/1007.3475}{{\ttfamily 1007.3475}}].

\bibitem{Bjerrum-Bohr:2010diw}
N.E.J.~Bjerrum-Bohr, P.H.~Damgaard, B.~Feng and T.~Sondergaard, \emph{{Gravity
  and Yang-Mills Amplitude Relations}},
  \href{https://doi.org/10.1103/PhysRevD.82.107702}{\emph{Phys. Rev. D}
  {\bfseries 82} (2010) 107702}
  [\href{https://arxiv.org/abs/1005.4367}{{\ttfamily 1005.4367}}].

\bibitem{Bjerrum-Bohr:2010kyi}
N.E.J.~Bjerrum-Bohr, P.H.~Damgaard, B.~Feng and T.~Sondergaard, \emph{{Proof of
  Gravity and Yang-Mills Amplitude Relations}},
  \href{https://doi.org/10.1007/JHEP09(2010)067}{\emph{JHEP} {\bfseries 09}
  (2010) 067} [\href{https://arxiv.org/abs/1007.3111}{{\ttfamily 1007.3111}}].

\bibitem{Bjerrum-Bohr:2010pnr}
N.E.J.~Bjerrum-Bohr, P.H.~Damgaard, T.~Sondergaard and P.~Vanhove, \emph{{The
  Momentum Kernel of Gauge and Gravity Theories}},
  \href{https://doi.org/10.1007/JHEP01(2011)001}{\emph{JHEP} {\bfseries 01}
  (2011) 001} [\href{https://arxiv.org/abs/1010.3933}{{\ttfamily 1010.3933}}].

\bibitem{Bjerrum-Bohr:2012kaa}
N.E.J.~Bjerrum-Bohr, P.H.~Damgaard, R.~Monteiro and D.~O'Connell,
  \emph{{Algebras for Amplitudes}},
  \href{https://doi.org/10.1007/JHEP06(2012)061}{\emph{JHEP} {\bfseries 06}
  (2012) 061} [\href{https://arxiv.org/abs/1203.0944}{{\ttfamily 1203.0944}}].

\bibitem{jjmcTASI2014}
J.J.M.~Carrasco, \emph{{Gauge and gravity amplitude relations}},  in
  \emph{{Proceedings, Theoretical Advanced Study Institute in Elementary
  Particle Physics: Journeys Through the Precision Frontier: Amplitudes for
  Colliders (TASI 2014): Boulder, Colorado, June 2-27, 2014}}, pp.~477--557,
  WSP, WSP, 2015, \href{https://doi.org/10.1142/9789814678766_0011}{DOI}
  [\href{https://arxiv.org/abs/1506.00974}{{\ttfamily 1506.00974}}].

\bibitem{Badger:2008cm}
S.D.~Badger, \emph{{Direct Extraction Of One Loop Rational Terms}},
  \href{https://doi.org/10.1088/1126-6708/2009/01/049}{\emph{JHEP} {\bfseries
  01} (2009) 049} [\href{https://arxiv.org/abs/0806.4600}{{\ttfamily
  0806.4600}}].

\bibitem{ElvangHuangReview}
H.~Elvang and Y.-t.~Huang, \emph{{Scattering amplitudes}},
  \href{https://arxiv.org/abs/1308.1697}{{\ttfamily 1308.1697}}.

\bibitem{Passarino:1978jh}
G.~Passarino and M.J.G.~Veltman, \emph{{One Loop Corrections for e+ e-
  Annihilation Into mu+ mu- in the Weinberg Model}},
  \href{https://doi.org/10.1016/0550-3213(79)90234-7}{\emph{Nucl. Phys. B}
  {\bfseries 160} (1979) 151}.

\bibitem{Bern:1993kr}
Z.~Bern, L.J.~Dixon and D.A.~Kosower, \emph{{Dimensionally regulated pentagon
  integrals}}, \href{https://doi.org/10.1016/0550-3213(94)90398-0}{\emph{Nucl.
  Phys. B} {\bfseries 412} (1994) 751}
  [\href{https://arxiv.org/abs/hep-ph/9306240}{{\ttfamily hep-ph/9306240}}].

\bibitem{Smirnov:2004ym}
V.A.~Smirnov, \emph{{Evaluating Feynman integrals}}, {\emph{Springer Tracts
  Mod. Phys.} {\bfseries 211} (2004) 1}.

\bibitem{Weinberg:1978kz}
S.~Weinberg, \emph{{Phenomenological Lagrangians}},
  \href{https://doi.org/10.1016/0378-4371(79)90223-1}{\emph{Physica A}
  {\bfseries 96} (1979) 327}.

\bibitem{Gasser:1983yg}
J.~Gasser and H.~Leutwyler, \emph{{Chiral Perturbation Theory to One Loop}},
  \href{https://doi.org/10.1016/0003-4916(84)90242-2}{\emph{Annals Phys.}
  {\bfseries 158} (1984) 142}.

\bibitem{Kampf:2006yf}
K.~Kampf, J.~Novotny and J.~Trnka, \emph{{On different lagrangian formalisms
  for vector resonances within chiral perturbation theory}},
  \href{https://doi.org/10.1140/epjc/s10052-006-0171-9}{\emph{Eur. Phys. J. C}
  {\bfseries 50} (2007) 385}
  [\href{https://arxiv.org/abs/hep-ph/0608051}{{\ttfamily hep-ph/0608051}}].

\bibitem{Kampf:2013vha}
K.~Kampf, J.~Novotny and J.~Trnka, \emph{{Tree-level Amplitudes in the
  Nonlinear Sigma Model}},
  \href{https://doi.org/10.1007/JHEP05(2013)032}{\emph{JHEP} {\bfseries 05}
  (2013) 032} [\href{https://arxiv.org/abs/1304.3048}{{\ttfamily 1304.3048}}].

\bibitem{Manohar:2008tc}
A.V.~Manohar and V.~Mateu, \emph{{Dispersion Relation Bounds for pi pi
  Scattering}}, \href{https://doi.org/10.1103/PhysRevD.77.094019}{\emph{Phys.
  Rev. D} {\bfseries 77} (2008) 094019}
  [\href{https://arxiv.org/abs/0801.3222}{{\ttfamily 0801.3222}}].

\bibitem{Bellazzini:2016xrt}
B.~Bellazzini, \emph{{Softness and amplitudes\textquoteright{} positivity for
  spinning particles}},
  \href{https://doi.org/10.1007/JHEP02(2017)034}{\emph{JHEP} {\bfseries 02}
  (2017) 034} [\href{https://arxiv.org/abs/1605.06111}{{\ttfamily
  1605.06111}}].

\bibitem{Guerrieri:2020bto}
A.L.~Guerrieri, J.~Penedones and P.~Vieira, \emph{{S-matrix bootstrap for
  effective field theories: massless pions}},
  \href{https://doi.org/10.1007/JHEP06(2021)088}{\emph{JHEP} {\bfseries 06}
  (2021) 088} [\href{https://arxiv.org/abs/2011.02802}{{\ttfamily
  2011.02802}}].

\bibitem{Bijnens:1995yn}
J.~Bijnens, G.~Colangelo, G.~Ecker, J.~Gasser and M.E.~Sainio, \emph{{Elastic
  pi pi scattering to two loops}},
  \href{https://doi.org/10.1016/0370-2693(96)00165-7}{\emph{Phys. Lett. B}
  {\bfseries 374} (1996) 210}
  [\href{https://arxiv.org/abs/hep-ph/9511397}{{\ttfamily hep-ph/9511397}}].

\bibitem{Bijnens:1997vq}
J.~Bijnens, G.~Colangelo, G.~Ecker, J.~Gasser and M.E.~Sainio, \emph{{Pion-pion
  scattering at low energy}},
  \href{https://doi.org/10.1016/S0550-3213(97)00621-4}{\emph{Nucl. Phys. B}
  {\bfseries 508} (1997) 263}
  [\href{https://arxiv.org/abs/hep-ph/9707291}{{\ttfamily hep-ph/9707291}}].

\bibitem{Girlanda:1997ed}
L.~Girlanda, M.~Knecht, B.~Moussallam and J.~Stern, \emph{{Comment on the
  prediction of two loop standard chiral perturbation theory for low-energy pi
  pi scattering}},
  \href{https://doi.org/10.1016/S0370-2693(97)00872-1}{\emph{Phys. Lett. B}
  {\bfseries 409} (1997) 461}
  [\href{https://arxiv.org/abs/hep-ph/9703448}{{\ttfamily hep-ph/9703448}}].

\bibitem{Low:2014nga}
I.~Low, \emph{{Adler\textquoteright{}s zero and effective Lagrangians for
  nonlinearly realized symmetry}},
  \href{https://doi.org/10.1103/PhysRevD.91.105017}{\emph{Phys. Rev. D}
  {\bfseries 91} (2015) 105017}
  [\href{https://arxiv.org/abs/1412.2145}{{\ttfamily 1412.2145}}].

\bibitem{Low:2017mlh}
I.~Low and Z.~Yin, \emph{{Ward Identity and Scattering Amplitudes for Nonlinear
  Sigma Models}},
  \href{https://doi.org/10.1103/PhysRevLett.120.061601}{\emph{Phys. Rev. Lett.}
  {\bfseries 120} (2018) 061601}
  [\href{https://arxiv.org/abs/1709.08639}{{\ttfamily 1709.08639}}].

\bibitem{Liu:2018vel}
D.~Liu, I.~Low and Z.~Yin, \emph{{Universal Imprints of a
  Pseudo-Nambu-Goldstone Higgs Boson}},
  \href{https://doi.org/10.1103/PhysRevLett.121.261802}{\emph{Phys. Rev. Lett.}
  {\bfseries 121} (2018) 261802}
  [\href{https://arxiv.org/abs/1805.00489}{{\ttfamily 1805.00489}}].

\bibitem{Low:2018acv}
I.~Low and Z.~Yin, \emph{{The Infrared Structure of Nambu-Goldstone Bosons}},
  \href{https://doi.org/10.1007/JHEP10(2018)078}{\emph{JHEP} {\bfseries 10}
  (2018) 078} [\href{https://arxiv.org/abs/1804.08629}{{\ttfamily
  1804.08629}}].

\bibitem{Born:1934gh}
M.~Born and L.~Infeld, \emph{{Foundations of the new field theory}},
  \href{https://doi.org/10.1098/rspa.1934.0059}{\emph{Proc. Roy. Soc. Lond. A}
  {\bfseries 144} (1934) 425}.

\bibitem{deRham:2010eu}
C.~de~Rham and A.J.~Tolley, \emph{{DBI and the Galileon reunited}},
  \href{https://doi.org/10.1088/1475-7516/2010/05/015}{\emph{JCAP} {\bfseries
  05} (2010) 015} [\href{https://arxiv.org/abs/1003.5917}{{\ttfamily
  1003.5917}}].

\bibitem{Volkov:1973ix}
D.V.~Volkov and V.P.~Akulov, \emph{{Is the Neutrino a Goldstone Particle?}},
  \href{https://doi.org/10.1016/0370-2693(73)90490-5}{\emph{Phys. Lett. B}
  {\bfseries 46} (1973) 109}.

\bibitem{Kallosh:1997aw}
R.~Kallosh, \emph{{Volkov-Akulov theory and D-branes}},
  \href{https://arxiv.org/abs/hep-th/9705118}{{\ttfamily hep-th/9705118}}.

\bibitem{Komargodski:2009rz}
Z.~Komargodski and N.~Seiberg, \emph{{From Linear SUSY to Constrained
  Superfields}},
  \href{https://doi.org/10.1088/1126-6708/2009/09/066}{\emph{JHEP} {\bfseries
  09} (2009) 066} [\href{https://arxiv.org/abs/0907.2441}{{\ttfamily
  0907.2441}}].

\bibitem{Kuzenko:2010ef}
S.M.~Kuzenko and S.J.~Tyler, \emph{{Relating the Komargodski-Seiberg and
  Akulov-Volkov actions: Exact nonlinear field redefinition}},
  \href{https://doi.org/10.1016/j.physletb.2011.03.020}{\emph{Phys. Lett. B}
  {\bfseries 698} (2011) 319}
  [\href{https://arxiv.org/abs/1009.3298}{{\ttfamily 1009.3298}}].

\bibitem{Rocek:1978nb}
M.~Rocek, \emph{{Linearizing the Volkov-Akulov Model}},
  \href{https://doi.org/10.1103/PhysRevLett.41.451}{\emph{Phys. Rev. Lett.}
  {\bfseries 41} (1978) 451}.

\bibitem{Casalbuoni:1988xh}
R.~Casalbuoni, S.~De~Curtis, D.~Dominici, F.~Feruglio and R.~Gatto,
  \emph{{Nonlinear Realization of Supersymmetry Algebra From Supersymmetric
  Constraint}}, \href{https://doi.org/10.1016/0370-2693(89)90788-0}{\emph{Phys.
  Lett. B} {\bfseries 220} (1989) 569}.

\bibitem{Ferrara:2014kva}
S.~Ferrara, R.~Kallosh and A.~Linde, \emph{{Cosmology with Nilpotent
  Superfields}}, \href{https://doi.org/10.1007/JHEP10(2014)143}{\emph{JHEP}
  {\bfseries 10} (2014) 143} [\href{https://arxiv.org/abs/1408.4096}{{\ttfamily
  1408.4096}}].

\bibitem{Kallosh:2013yoa}
R.~Kallosh, A.~Linde and D.~Roest, \emph{{Superconformal Inflationary
  $\alpha$-Attractors}},
  \href{https://doi.org/10.1007/JHEP11(2013)198}{\emph{JHEP} {\bfseries 11}
  (2013) 198} [\href{https://arxiv.org/abs/1311.0472}{{\ttfamily 1311.0472}}].

\bibitem{Tseytlin:1999dj}
A.A.~Tseytlin, \emph{{Born-Infeld action, supersymmetry and string theory}},
  \href{https://arxiv.org/abs/hep-th/9908105}{{\ttfamily hep-th/9908105}}.

\bibitem{Bergshoeff:2013pia}
E.~Bergshoeff, F.~Coomans, R.~Kallosh, C.S.~Shahbazi and A.~Van~Proeyen,
  \emph{{Dirac-Born-Infeld-Volkov-Akulov and deformation of supersymmetry}},
  \href{https://doi.org/10.1007/JHEP08(2013)100}{\emph{JHEP} {\bfseries 08}
  (2013) 100} [\href{https://arxiv.org/abs/1303.5662}{{\ttfamily 1303.5662}}].

\bibitem{Mafra:2011nv}
C.R.~Mafra, O.~Schlotterer and S.~Stieberger, \emph{{Complete N-Point
  Superstring Disk Amplitude I. Pure Spinor Computation}},
  \href{https://doi.org/10.1016/j.nuclphysb.2013.04.023}{\emph{Nucl. Phys. B}
  {\bfseries 873} (2013) 419}
  [\href{https://arxiv.org/abs/1106.2645}{{\ttfamily 1106.2645}}].

\bibitem{Mafra:2011nw}
C.R.~Mafra, O.~Schlotterer and S.~Stieberger, \emph{{Complete N-Point
  Superstring Disk Amplitude II. Amplitude and Hypergeometric Function
  Structure}},
  \href{https://doi.org/10.1016/j.nuclphysb.2013.04.022}{\emph{Nucl. Phys. B}
  {\bfseries 873} (2013) 461}
  [\href{https://arxiv.org/abs/1106.2646}{{\ttfamily 1106.2646}}].

\bibitem{Broedel:2013tta}
J.~Broedel, O.~Schlotterer and S.~Stieberger, \emph{{Polylogarithms, Multiple
  Zeta Values and Superstring Amplitudes}},
  \href{https://doi.org/10.1002/prop.201300019}{\emph{Fortsch. Phys.}
  {\bfseries 61} (2013) 812} [\href{https://arxiv.org/abs/1304.7267}{{\ttfamily
  1304.7267}}].

\bibitem{Carrasco:2016ygv}
J.J.M.~Carrasco, C.R.~Mafra and O.~Schlotterer, \emph{{Semi-abelian Z-theory:
  NLSM$+\phi^{3}$ from the open string}},
  \href{https://doi.org/10.1007/JHEP08(2017)135}{\emph{JHEP} {\bfseries 08}
  (2017) 135} [\href{https://arxiv.org/abs/1612.06446}{{\ttfamily
  1612.06446}}].

\bibitem{Mafra:2016mcc}
C.R.~Mafra and O.~Schlotterer, \emph{{Non-abelian $Z$-theory: Berends-Giele
  recursion for the $\alpha'$-expansion of disk integrals}},
  \href{https://doi.org/10.1007/JHEP01(2017)031}{\emph{JHEP} {\bfseries 01}
  (2017) 031} [\href{https://arxiv.org/abs/1609.07078}{{\ttfamily
  1609.07078}}].

\bibitem{KLT}
H.~Kawai, D.C.~Lewellen and S.H.H.~Tye, \emph{{A relation between tree
  amplitudes of closed and open strings}},
  \href{https://doi.org/10.1016/0550-3213(86)90362-7}{\emph{Nucl. Phys.}
  {\bfseries B269} (1986) 1}.

\bibitem{Green:1982sw}
M.B.~Green, J.H.~Schwarz and L.~Brink, \emph{{N=4 Yang-Mills and N=8
  Supergravity as Limits of String Theories}},
  \href{https://doi.org/10.1016/0550-3213(82)90336-4}{\emph{Nucl. Phys. B}
  {\bfseries 198} (1982) 474}.

\bibitem{Azevedo:2018dgo}
T.~Azevedo, M.~Chiodaroli, H.~Johansson and O.~Schlotterer, \emph{{Heterotic
  and bosonic string amplitudes via field theory}},
  \href{https://doi.org/10.1007/JHEP10(2018)012}{\emph{JHEP} {\bfseries 10}
  (2018) 012} [\href{https://arxiv.org/abs/1803.05452}{{\ttfamily
  1803.05452}}].

\bibitem{Bossard:2012xs}
G.~Bossard, P.S.~Howe and K.S.~Stelle, \emph{{Anomalies and divergences in N=4
  supergravity}},
  \href{https://doi.org/10.1016/j.physletb.2013.01.021}{\emph{Phys. Lett. B}
  {\bfseries 719} (2013) 424}
  [\href{https://arxiv.org/abs/1212.0841}{{\ttfamily 1212.0841}}].

\bibitem{Novotny:2018iph}
J.~Novotn\'y, \emph{{Self-duality, helicity conservation and normal ordering in
  nonlinear QED}},
  \href{https://doi.org/10.1103/PhysRevD.98.085015}{\emph{Phys. Rev. D}
  {\bfseries 98} (2018) 085015}
  [\href{https://arxiv.org/abs/1806.02167}{{\ttfamily 1806.02167}}].

\bibitem{Gibbons:1995ap}
G.W.~Gibbons and D.A.~Rasheed, \emph{{Sl(2,R) invariance of nonlinear
  electrodynamics coupled to an axion and a dilaton}},
  \href{https://doi.org/10.1016/0370-2693(95)01272-9}{\emph{Phys. Lett. B}
  {\bfseries 365} (1996) 46}
  [\href{https://arxiv.org/abs/hep-th/9509141}{{\ttfamily hep-th/9509141}}].

\bibitem{Babaei-Aghbolagh:2013hia}
H.~Babaei-Aghbolagh and M.R.~Garousi, \emph{{S-duality of tree-level S-matrix
  elements in D3-brane effective action}},
  \href{https://doi.org/10.1103/PhysRevD.88.026008}{\emph{Phys. Rev. D}
  {\bfseries 88} (2013) 026008}
  [\href{https://arxiv.org/abs/1304.2938}{{\ttfamily 1304.2938}}].

\bibitem{Schrodinger:1935oqa}
E.~Schr\"odinger, \emph{{Contributions to Born's new theory of the
  electromagnetic field}},
  \href{https://doi.org/10.1098/rspa.1935.0116}{\emph{Proc. Roy. Soc. Lond. A}
  {\bfseries 150} (1935) 465}.

\bibitem{Elvang:2019twd}
H.~Elvang, M.~Hadjiantonis, C.R.T.~Jones and S.~Paranjape,
  \emph{{All-Multiplicity One-Loop Amplitudes in Born-Infeld Electrodynamics
  from Generalized Unitarity}},
  \href{https://arxiv.org/abs/1906.05321}{{\ttfamily 1906.05321}}.

\bibitem{Heydeman:2017yww}
M.~Heydeman, J.H.~Schwarz and C.~Wen, \emph{{M5-Brane and D-Brane Scattering
  Amplitudes}}, \href{https://doi.org/10.1007/JHEP12(2017)003}{\emph{JHEP}
  {\bfseries 12} (2017) 003}
  [\href{https://arxiv.org/abs/1710.02170}{{\ttfamily 1710.02170}}].

\bibitem{Bern:1995db}
Z.~Bern and A.G.~Morgan, \emph{{Massive loop amplitudes from unitarity}},
  \href{https://doi.org/10.1016/0550-3213(96)00078-8}{\emph{Nucl. Phys. B}
  {\bfseries 467} (1996) 479}
  [\href{https://arxiv.org/abs/hep-ph/9511336}{{\ttfamily hep-ph/9511336}}].

\bibitem{Bern:1996ja}
Z.~Bern, L.J.~Dixon, D.C.~Dunbar and D.A.~Kosower, \emph{{One loop selfdual and
  ${\cal N}=4$ super-Yang-Mills}},
  \href{https://doi.org/10.1016/S0370-2693(96)01676-0}{\emph{Phys. Lett.}
  {\bfseries B394} (1997) 105}
  [\href{https://arxiv.org/abs/hep-th/9611127}{{\ttfamily hep-th/9611127}}].

\bibitem{collins_1984}
J.C.~Collins, \emph{Renormalization: An Introduction to Renormalization, the
  Renormalization Group and the Operator-Product Expansion}, Cambridge
  Monographs on Mathematical Physics, Cambridge University Press (1984),
  \href{https://doi.org/10.1017/CBO9780511622656}{10.1017/CBO9780511622656}.

\bibitem{Bern:2002zk}
Z.~Bern, A.~De~Freitas, L.J.~Dixon and H.L.~Wong, \emph{{Supersymmetric
  regularization, two loop QCD amplitudes and coupling shifts}},
  \href{https://doi.org/10.1103/PhysRevD.66.085002}{\emph{Phys. Rev. D}
  {\bfseries 66} (2002) 085002}
  [\href{https://arxiv.org/abs/hep-ph/0202271}{{\ttfamily hep-ph/0202271}}].

\bibitem{Bern:2007ct}
Z.~Bern, J.J.M.~Carrasco, H.~Johansson and D.A.~Kosower, \emph{{Maximally
  supersymmetric planar Yang-Mills amplitudes at five loops}},
  \href{https://doi.org/10.1103/PhysRevD.76.125020}{\emph{Phys. Rev. D}
  {\bfseries 76} (2007) 125020}
  [\href{https://arxiv.org/abs/0705.1864}{{\ttfamily 0705.1864}}].

\bibitem{Carrasco:2021bmu}
J.J.M.~Carrasco and I.A.~Vazquez-Holm, \emph{{Extracting Einstein from the
  loop-level double-copy}},
  \href{https://doi.org/10.1007/JHEP11(2021)088}{\emph{JHEP} {\bfseries 11}
  (2021) 088} [\href{https://arxiv.org/abs/2108.06798}{{\ttfamily
  2108.06798}}].

\bibitem{Anastasiou:2004vj}
C.~Anastasiou and A.~Lazopoulos, \emph{{Automatic integral reduction for higher
  order perturbative calculations}},
  \href{https://doi.org/10.1088/1126-6708/2004/07/046}{\emph{JHEP} {\bfseries
  07} (2004) 046} [\href{https://arxiv.org/abs/hep-ph/0404258}{{\ttfamily
  hep-ph/0404258}}].

\bibitem{vonManteuffel:2012np}
A.~von Manteuffel and C.~Studerus, \emph{{Reduze 2 - Distributed Feynman
  Integral Reduction}},  \href{https://arxiv.org/abs/1201.4330}{{\ttfamily
  1201.4330}}.

\bibitem{Smirnov:2014hma}
A.V.~Smirnov, \emph{{FIRE5: a C++ implementation of Feynman Integral
  REduction}}, \href{https://doi.org/10.1016/j.cpc.2014.11.024}{\emph{Comput.
  Phys. Commun.} {\bfseries 189} (2015) 182}
  [\href{https://arxiv.org/abs/1408.2372}{{\ttfamily 1408.2372}}].

\bibitem{vonManteuffel:2014ixa}
A.~von Manteuffel and R.M.~Schabinger, \emph{{A novel approach to integration
  by parts reduction}},
  \href{https://doi.org/10.1016/j.physletb.2015.03.029}{\emph{Phys. Lett. B}
  {\bfseries 744} (2015) 101}
  [\href{https://arxiv.org/abs/1406.4513}{{\ttfamily 1406.4513}}].

\bibitem{Smirnov:2019qkx}
A.V.~Smirnov and F.S.~Chuharev, \emph{{FIRE6: Feynman Integral REduction with
  Modular Arithmetic}},
  \href{https://doi.org/10.1016/j.cpc.2019.106877}{\emph{Comput. Phys. Commun.}
  {\bfseries 247} (2020) 106877}
  [\href{https://arxiv.org/abs/1901.07808}{{\ttfamily 1901.07808}}].

\bibitem{Smirnov:2020quc}
A.V.~Smirnov and V.A.~Smirnov, \emph{{How to choose master integrals}},
  \href{https://doi.org/10.1016/j.nuclphysb.2020.115213}{\emph{Nucl. Phys. B}
  {\bfseries 960} (2020) 115213}
  [\href{https://arxiv.org/abs/2002.08042}{{\ttfamily 2002.08042}}].

\bibitem{Usovitsch:2020jrk}
J.~Usovitsch, \emph{{Factorization of denominators in integration-by-parts
  reductions}},  \href{https://arxiv.org/abs/2002.08173}{{\ttfamily
  2002.08173}}.

\bibitem{Maierhofer:2018gpa}
P.~Maierh\"ofer and J.~Usovitsch, \emph{{Kira 1.2 Release Notes}},
  \href{https://arxiv.org/abs/1812.01491}{{\ttfamily 1812.01491}}.

\bibitem{Hartmann:2016pil}
C.~Hartmann, W.~Shepherd and M.~Trott, \emph{{The $Z$ decay width in the SMEFT:
  $y_t$ and $\lambda$ corrections at one loop}},
  \href{https://doi.org/10.1007/JHEP03(2017)060}{\emph{JHEP} {\bfseries 03}
  (2017) 060} [\href{https://arxiv.org/abs/1611.09879}{{\ttfamily
  1611.09879}}].

\bibitem{Chala:2021cgt}
M.~Chala, A.~D\'\i{}az-Carmona and G.~Guedes, \emph{{A Green\textquoteright{}s
  basis for the bosonic SMEFT to dimension 8}},
  \href{https://doi.org/10.1007/JHEP05(2022)138}{\emph{JHEP} {\bfseries 05}
  (2022) 138} [\href{https://arxiv.org/abs/2112.12724}{{\ttfamily
  2112.12724}}].

\bibitem{Aebischer:2022tvz}
J.~Aebischer, A.J.~Buras and J.~Kumar, \emph{{Simple rules for evanescent
  operators in one-loop basis transformations}},
  \href{https://doi.org/10.1103/PhysRevD.107.075007}{\emph{Phys. Rev. D}
  {\bfseries 107} (2023) 075007}
  [\href{https://arxiv.org/abs/2202.01225}{{\ttfamily 2202.01225}}].

\bibitem{Fuentes-Martin:2022vvu}
J.~Fuentes-Mart\'\i{}n, M.~K\"onig, J.~Pag\`es, A.E.~Thomsen and F.~Wilsch,
  \emph{{Evanescent operators in one-loop matching computations}},
  \href{https://doi.org/10.1007/JHEP02(2023)031}{\emph{JHEP} {\bfseries 02}
  (2023) 031} [\href{https://arxiv.org/abs/2211.09144}{{\ttfamily
  2211.09144}}].

\bibitem{Isidori:2023pyp}
G.~Isidori, F.~Wilsch and D.~Wyler, \emph{{The Standard Model effective field
  theory at work}},  \href{https://arxiv.org/abs/2303.16922}{{\ttfamily
  2303.16922}}.

\bibitem{Bern:2015xsa}
Z.~Bern, C.~Cheung, H.-H.~Chi, S.~Davies, L.~Dixon and J.~Nohle,
  \emph{{Evanescent Effects Can Alter Ultraviolet Divergences in Quantum
  Gravity without Physical Consequences}},
  \href{https://doi.org/10.1103/PhysRevLett.115.211301}{\emph{Phys. Rev. Lett.}
  {\bfseries 115} (2015) 211301}
  [\href{https://arxiv.org/abs/1507.06118}{{\ttfamily 1507.06118}}].

\bibitem{Bern:2017puu}
Z.~Bern, H.-H.~Chi, L.~Dixon and A.~Edison, \emph{{Two-Loop Renormalization of
  Quantum Gravity Simplified}},
  \href{https://doi.org/10.1103/PhysRevD.95.046013}{\emph{Phys. Rev. D}
  {\bfseries 95} (2017) 046013}
  [\href{https://arxiv.org/abs/1701.02422}{{\ttfamily 1701.02422}}].

\bibitem{Dugan:1990df}
M.J.~Dugan and B.~Grinstein, \emph{{On the vanishing of evanescent operators}},
  \href{https://doi.org/10.1016/0370-2693(91)90680-O}{\emph{Phys. Lett. B}
  {\bfseries 256} (1991) 239}.

\bibitem{Herrlich:1994kh}
S.~Herrlich and U.~Nierste, \emph{{Evanescent operators, scheme dependences and
  double insertions}},
  \href{https://doi.org/10.1016/0550-3213(95)00474-7}{\emph{Nucl. Phys. B}
  {\bfseries 455} (1995) 39}
  [\href{https://arxiv.org/abs/hep-ph/9412375}{{\ttfamily hep-ph/9412375}}].

\bibitem{Bell:2009nk}
G.~Bell, \emph{{NNLO vertex corrections in charmless hadronic B decays: Real
  part}}, \href{https://doi.org/10.1016/j.nuclphysb.2009.07.012}{\emph{Nucl.
  Phys. B} {\bfseries 822} (2009) 172}
  [\href{https://arxiv.org/abs/0902.1915}{{\ttfamily 0902.1915}}].

\bibitem{Becher:2004kk}
T.~Becher and R.J.~Hill, \emph{{Loop corrections to heavy-to-light form-factors
  and evanescent operators in SCET}},
  \href{https://doi.org/10.1088/1126-6708/2004/10/055}{\emph{JHEP} {\bfseries
  10} (2004) 055} [\href{https://arxiv.org/abs/hep-ph/0408344}{{\ttfamily
  hep-ph/0408344}}].

\bibitem{DiPietro:2017vsp}
L.~Di~Pietro and E.~Stamou, \emph{{Operator mixing in the
  $\boldsymbol{\epsilon}$-expansion: Scheme and evanescent-operator
  independence}}, \href{https://doi.org/10.1103/PhysRevD.97.065007}{\emph{Phys.
  Rev. D} {\bfseries 97} (2018) 065007}
  [\href{https://arxiv.org/abs/1708.03739}{{\ttfamily 1708.03739}}].

\bibitem{DixonMaltoni}
V.~Del~Duca, L.J.~Dixon and F.~Maltoni, \emph{{New color decompositions for
  gauge amplitudes at tree and loop level}},
  \href{https://doi.org/10.1016/S0550-3213(99)00809-3}{\emph{Nucl. Phys.}
  {\bfseries B571} (2000) 51}
  [\href{https://arxiv.org/abs/hep-ph/9910563}{{\ttfamily hep-ph/9910563}}].

\bibitem{Feng:2020opo}
B.~Feng and Y.~Zhang, \emph{{Note on the Labelled tree graphs}},
  \href{https://doi.org/10.1007/JHEP12(2020)096}{\emph{JHEP} {\bfseries 12}
  (2020) 096} [\href{https://arxiv.org/abs/2009.02394}{{\ttfamily
  2009.02394}}].

\bibitem{Alday:2023kfm}
L.F.~Alday, V.~Gon\c{c}alves, M.~Nocchi and X.~Zhou, \emph{{Six-Point AdS Gluon
  Amplitudes from Flat Space and Factorization}},
  \href{https://arxiv.org/abs/2307.06884}{{\ttfamily 2307.06884}}.

\bibitem{Weinberg:1965nx}
S.~Weinberg, \emph{{Infrared photons and gravitons}},
  \href{https://doi.org/10.1103/PhysRev.140.B516}{\emph{Phys. Rev.} {\bfseries
  140} (1965) B516}.

\bibitem{deWit:1983xhu}
B.~de~Wit, P.G.~Lauwers, R.~Philippe, S.Q.~Su and A.~Van~Proeyen, \emph{{Gauge
  and Matter Fields Coupled to N=2 Supergravity}},
  \href{https://doi.org/10.1016/0370-2693(84)90979-1}{\emph{Phys. Lett. B}
  {\bfseries 134} (1984) 37}.

\bibitem{deWit:1992cr}
B.~de~Wit and A.~Van~Proeyen, \emph{{Broken sigma model isometries in very
  special geometry}},
  \href{https://doi.org/10.1016/0370-2693(92)91485-R}{\emph{Phys. Lett. B}
  {\bfseries 293} (1992) 94}
  [\href{https://arxiv.org/abs/hep-th/9207091}{{\ttfamily hep-th/9207091}}].

\bibitem{deWit:1992wf}
B.~de~Wit, F.~Vanderseypen and A.~Van~Proeyen, \emph{{Symmetry structure of
  special geometries}},
  \href{https://doi.org/10.1016/0550-3213(93)90413-J}{\emph{Nucl. Phys. B}
  {\bfseries 400} (1993) 463}
  [\href{https://arxiv.org/abs/hep-th/9210068}{{\ttfamily hep-th/9210068}}].

\bibitem{Chiodaroli:2014xia}
M.~Chiodaroli, M.~G\"unaydin, H.~Johansson and R.~Roiban, \emph{{Scattering
  amplitudes in $ \mathcal{N}=2 $ Maxwell-Einstein and Yang-Mills/Einstein
  supergravity}}, \href{https://doi.org/10.1007/JHEP01(2015)081}{\emph{JHEP}
  {\bfseries 01} (2015) 081} [\href{https://arxiv.org/abs/1408.0764}{{\ttfamily
  1408.0764}}].

\bibitem{Naculich:2008ew}
S.G.~Naculich, H.~Nastase and H.J.~Schnitzer, \emph{{Two-loop graviton
  scattering relation and IR behavior in N=8 supergravity}},
  \href{https://doi.org/10.1016/j.nuclphysb.2008.07.001}{\emph{Nucl. Phys. B}
  {\bfseries 805} (2008) 40} [\href{https://arxiv.org/abs/0805.2347}{{\ttfamily
  0805.2347}}].

\bibitem{White:2011yy}
C.D.~White, \emph{{Factorization Properties of Soft Graviton Amplitudes}},
  \href{https://doi.org/10.1007/JHEP05(2011)060}{\emph{JHEP} {\bfseries 05}
  (2011) 060} [\href{https://arxiv.org/abs/1103.2981}{{\ttfamily 1103.2981}}].

\bibitem{DiVecchia:2019myk}
P.~Di~Vecchia, A.~Luna, S.G.~Naculich, R.~Russo, G.~Veneziano and C.D.~White,
  \emph{{A tale of two exponentiations in ${\cal N}=8$ supergravity}},
  \href{https://doi.org/10.1016/j.physletb.2019.134927}{\emph{Phys. Lett. B}
  {\bfseries 798} (2019) 134927}
  [\href{https://arxiv.org/abs/1908.05603}{{\ttfamily 1908.05603}}].

\bibitem{DiVecchia:2019kta}
P.~Di~Vecchia, S.G.~Naculich, R.~Russo, G.~Veneziano and C.D.~White, \emph{{A
  tale of two exponentiations in $ \mathcal{N} $ = 8 supergravity at subleading
  level}}, \href{https://doi.org/10.1007/JHEP03(2020)173}{\emph{JHEP}
  {\bfseries 03} (2020) 173}
  [\href{https://arxiv.org/abs/1911.11716}{{\ttfamily 1911.11716}}].

\bibitem{Heissenberg:2021tzo}
C.~Heissenberg, \emph{{Infrared divergences and the eikonal exponentiation}},
  \href{https://doi.org/10.1103/PhysRevD.104.046016}{\emph{Phys. Rev. D}
  {\bfseries 104} (2021) 046016}
  [\href{https://arxiv.org/abs/2105.04594}{{\ttfamily 2105.04594}}].

\bibitem{Elvang:2018dco}
H.~Elvang, M.~Hadjiantonis, C.R.T.~Jones and S.~Paranjape, \emph{{Soft
  Bootstrap and Supersymmetry}},
  \href{https://doi.org/10.1007/JHEP01(2019)195}{\emph{JHEP} {\bfseries 01}
  (2019) 195} [\href{https://arxiv.org/abs/1806.06079}{{\ttfamily
  1806.06079}}].

\bibitem{Cachazo:2016njl}
F.~Cachazo, P.~Cha and S.~Mizera, \emph{{Extensions of Theories from Soft
  Limits}}, \href{https://doi.org/10.1007/JHEP06(2016)170}{\emph{JHEP}
  {\bfseries 06} (2016) 170}
  [\href{https://arxiv.org/abs/1604.03893}{{\ttfamily 1604.03893}}].

\bibitem{Anastasiou:2003kj}
C.~Anastasiou, Z.~Bern, L.J.~Dixon and D.A.~Kosower, \emph{{Planar amplitudes
  in maximally supersymmetric Yang-Mills theory}},
  \href{https://doi.org/10.1103/PhysRevLett.91.251602}{\emph{Phys. Rev. Lett.}
  {\bfseries 91} (2003) 251602}
  [\href{https://arxiv.org/abs/hep-th/0309040}{{\ttfamily hep-th/0309040}}].

\bibitem{Bern:2005iz}
Z.~Bern, L.J.~Dixon and V.A.~Smirnov, \emph{{Iteration of planar amplitudes in
  maximally supersymmetric Yang-Mills theory at three loops and beyond}},
  \href{https://doi.org/10.1103/PhysRevD.72.085001}{\emph{Phys. Rev. D}
  {\bfseries 72} (2005) 085001}
  [\href{https://arxiv.org/abs/hep-th/0505205}{{\ttfamily hep-th/0505205}}].

\bibitem{Sterman:2002qn}
G.F.~Sterman and M.E.~Tejeda-Yeomans, \emph{{Multiloop amplitudes and
  resummation}},
  \href{https://doi.org/10.1016/S0370-2693(02)03100-3}{\emph{Phys. Lett. B}
  {\bfseries 552} (2003) 48}
  [\href{https://arxiv.org/abs/hep-ph/0210130}{{\ttfamily hep-ph/0210130}}].

\bibitem{Minahan:2002ve}
J.A.~Minahan and K.~Zarembo, \emph{{The Bethe ansatz for N=4 superYang-Mills}},
  \href{https://doi.org/10.1088/1126-6708/2003/03/013}{\emph{JHEP} {\bfseries
  03} (2003) 013} [\href{https://arxiv.org/abs/hep-th/0212208}{{\ttfamily
  hep-th/0212208}}].

\bibitem{Bena:2003wd}
I.~Bena, J.~Polchinski and R.~Roiban, \emph{{Hidden symmetries of the AdS(5) x
  S**5 superstring}},
  \href{https://doi.org/10.1103/PhysRevD.69.046002}{\emph{Phys. Rev. D}
  {\bfseries 69} (2004) 046002}
  [\href{https://arxiv.org/abs/hep-th/0305116}{{\ttfamily hep-th/0305116}}].

\bibitem{Beisert:2003jj}
N.~Beisert, \emph{{The complete one loop dilatation operator of N=4
  superYang-Mills theory}},
  \href{https://doi.org/10.1016/j.nuclphysb.2003.10.019}{\emph{Nucl. Phys. B}
  {\bfseries 676} (2004) 3}
  [\href{https://arxiv.org/abs/hep-th/0307015}{{\ttfamily hep-th/0307015}}].

\bibitem{Beisert:2003jb}
N.~Beisert, \emph{{Higher loops, integrability and the near BMN limit}},
  \href{https://doi.org/10.1088/1126-6708/2003/09/062}{\emph{JHEP} {\bfseries
  09} (2003) 062} [\href{https://arxiv.org/abs/hep-th/0308074}{{\ttfamily
  hep-th/0308074}}].

\bibitem{Beisert:2003yb}
N.~Beisert and M.~Staudacher, \emph{{The N=4 SYM integrable super spin chain}},
  \href{https://doi.org/10.1016/j.nuclphysb.2003.08.015}{\emph{Nucl. Phys. B}
  {\bfseries 670} (2003) 439}
  [\href{https://arxiv.org/abs/hep-th/0307042}{{\ttfamily hep-th/0307042}}].

\bibitem{Dolan:2003uh}
L.~Dolan, C.R.~Nappi and E.~Witten, \emph{{A Relation between approaches to
  integrability in superconformal Yang-Mills theory}},
  \href{https://doi.org/10.1088/1126-6708/2003/10/017}{\emph{JHEP} {\bfseries
  10} (2003) 017} [\href{https://arxiv.org/abs/hep-th/0308089}{{\ttfamily
  hep-th/0308089}}].

\bibitem{Arutyunov:2003rg}
G.~Arutyunov and M.~Staudacher, \emph{{Matching higher conserved charges for
  strings and spins}},
  \href{https://doi.org/10.1088/1126-6708/2004/03/004}{\emph{JHEP} {\bfseries
  03} (2004) 004} [\href{https://arxiv.org/abs/hep-th/0310182}{{\ttfamily
  hep-th/0310182}}].

\bibitem{Ryzhov:2004nz}
A.V.~Ryzhov and A.A.~Tseytlin, \emph{{Towards the exact dilatation operator of
  N=4 super Yang-Mills theory}},
  \href{https://doi.org/10.1016/j.nuclphysb.2004.07.037}{\emph{Nucl. Phys. B}
  {\bfseries 698} (2004) 132}
  [\href{https://arxiv.org/abs/hep-th/0404215}{{\ttfamily hep-th/0404215}}].

\bibitem{Frolov:2005iq}
S.A.~Frolov, R.~Roiban and A.A.~Tseytlin, \emph{{Gauge-string duality for
  (non)supersymmetric deformations of N=4 super Yang-Mills theory}},
  \href{https://doi.org/10.1016/j.nuclphysb.2005.10.004}{\emph{Nucl. Phys. B}
  {\bfseries 731} (2005) 1}
  [\href{https://arxiv.org/abs/hep-th/0507021}{{\ttfamily hep-th/0507021}}].

\bibitem{SuperSum}
Z.~Bern, J.J.M.~Carrasco, H.~Ita, H.~Johansson and R.~Roiban, \emph{{On the
  structure of supersymmetric sums in multi-loop unitarity cuts}},
  \href{https://doi.org/10.1103/PhysRevD.80.065029}{\emph{Phys. Rev.}
  {\bfseries D80} (2009) 065029}
  [\href{https://arxiv.org/abs/0903.5348}{{\ttfamily 0903.5348}}].

\bibitem{Chiodaroli2013upa}
M.~Chiodaroli, Q.~Jin and R.~Roiban, \emph{{Color/kinematics duality for
  general abelian orbifolds of ${\cal N}=4$ super-Yang-Mills theory}},
  \href{https://doi.org/10.1007/JHEP01(2014)152}{\emph{JHEP} {\bfseries 01}
  (2014) 152} [\href{https://arxiv.org/abs/1311.3600}{{\ttfamily 1311.3600}}].

\bibitem{Carrasco:2023vjg}
J.J.M.~Carrasco and A.~Seifi, \emph{{Loop-level double-copy for massive
  fermions in the fundamental}},
  \href{https://doi.org/10.1007/JHEP05(2023)217}{\emph{JHEP} {\bfseries 05}
  (2023) 217} [\href{https://arxiv.org/abs/2302.14861}{{\ttfamily
  2302.14861}}].

\bibitem{BRY}
Z.~Bern, J.S.~Rozowsky and B.~Yan, \emph{{Two loop four-gluon amplitudes in
  ${\cal N}=4$ superYang-Mills}},
  \href{https://doi.org/10.1016/S0370-2693(97)00413-9}{\emph{Phys. Lett.}
  {\bfseries B401} (1997) 273}
  [\href{https://arxiv.org/abs/hep-ph/9702424}{{\ttfamily hep-ph/9702424}}].

\bibitem{Carrasco:2021otn}
J.J.M.~Carrasco, A.~Edison and H.~Johansson, \emph{{Maximal Super-Yang-Mills at
  Six Loops via Novel Integrand Bootstrap}},
  \href{https://arxiv.org/abs/2112.05178}{{\ttfamily 2112.05178}}.

\bibitem{BernNoTriangle}
Z.~Bern, N.E.J.~Bjerrum-Bohr and D.C.~Dunbar, \emph{{Inherited twistor-space
  structure of gravity loop amplitudes}},
  \href{https://doi.org/10.1088/1126-6708/2005/05/056}{\emph{JHEP} {\bfseries
  05} (2005) 056} [\href{https://arxiv.org/abs/hep-th/0501137}{{\ttfamily
  hep-th/0501137}}].

\bibitem{Boels:2016xhc}
R.H.~Boels and R.~Medina, \emph{{Graviton and gluon scattering from first
  principles}},
  \href{https://doi.org/10.1103/PhysRevLett.118.061602}{\emph{Phys. Rev. Lett.}
  {\bfseries 118} (2017) 061602}
  [\href{https://arxiv.org/abs/1607.08246}{{\ttfamily 1607.08246}}].

\bibitem{Bourjaily:2023tcc}
J.L.~Bourjaily, N.~Kalyanapuram, K.~Patatoukos, M.~Plesser and Y.~Zhang,
  \emph{{Gauge-Invariant Double-Copies via Recursion}},
  \href{https://arxiv.org/abs/2307.02542}{{\ttfamily 2307.02542}}.

\bibitem{Caron-Huot:2016icg}
S.~Caron-Huot, Z.~Komargodski, A.~Sever and A.~Zhiboedov, \emph{{Strings from
  Massive Higher Spins: The Asymptotic Uniqueness of the Veneziano Amplitude}},
  \href{https://doi.org/10.1007/JHEP10(2017)026}{\emph{JHEP} {\bfseries 10}
  (2017) 026} [\href{https://arxiv.org/abs/1607.04253}{{\ttfamily
  1607.04253}}].

\bibitem{Chiodaroli:2021eug}
M.~Chiodaroli, H.~Johansson and P.~Pichini, \emph{{Compton black-hole
  scattering for s \ensuremath{\leq} 5/2}},
  \href{https://doi.org/10.1007/JHEP02(2022)156}{\emph{JHEP} {\bfseries 02}
  (2022) 156} [\href{https://arxiv.org/abs/2107.14779}{{\ttfamily
  2107.14779}}].

\bibitem{Cangemi:2022abk}
L.~Cangemi and P.~Pichini, \emph{{Classical Limit of Higher-Spin String
  Amplitudes}},  \href{https://arxiv.org/abs/2207.03947}{{\ttfamily
  2207.03947}}.

\bibitem{Cangemi:2022bew}
L.~Cangemi, M.~Chiodaroli, H.~Johansson, A.~Ochirov, P.~Pichini and
  E.~Skvortsov, \emph{{Kerr Black Holes Enjoy Massive Higher-Spin Gauge
  Symmetry}},  \href{https://arxiv.org/abs/2212.06120}{{\ttfamily 2212.06120}}.

\bibitem{Geiser:2022exp}
N.~Geiser and L.W.~Lindwasser, \emph{{Generalized Veneziano and Virasoro
  amplitudes}}, \href{https://doi.org/10.1007/JHEP04(2023)031}{\emph{JHEP}
  {\bfseries 04} (2023) 031}
  [\href{https://arxiv.org/abs/2210.14920}{{\ttfamily 2210.14920}}].

\bibitem{Cheung:2022mkw}
C.~Cheung and G.N.~Remmen, \emph{{Veneziano variations: how unique are string
  amplitudes?}}, \href{https://doi.org/10.1007/JHEP01(2023)122}{\emph{JHEP}
  {\bfseries 01} (2023) 122}
  [\href{https://arxiv.org/abs/2210.12163}{{\ttfamily 2210.12163}}].

\bibitem{Johansson:2017srf}
H.~Johansson and J.~Nohle, \emph{{Conformal Gravity from Gauge Theory}},
  \href{https://arxiv.org/abs/1707.02965}{{\ttfamily 1707.02965}}.

\bibitem{Johansson:2018ues}
H.~Johansson, G.~Mogull and F.~Teng, \emph{{Unraveling conformal gravity
  amplitudes}}, \href{https://doi.org/10.1007/JHEP09(2018)080}{\emph{JHEP}
  {\bfseries 09} (2018) 080}
  [\href{https://arxiv.org/abs/1806.05124}{{\ttfamily 1806.05124}}].

\bibitem{Henning:2015daa}
B.~Henning, X.~Lu, T.~Melia and H.~Murayama, \emph{{Hilbert series and operator
  bases with derivatives in effective field theories}},
  \href{https://doi.org/10.1007/s00220-015-2518-2}{\emph{Commun. Math. Phys.}
  {\bfseries 347} (2016) 363}
  [\href{https://arxiv.org/abs/1507.07240}{{\ttfamily 1507.07240}}].

\bibitem{Lehman:2015via}
L.~Lehman and A.~Martin, \emph{{Hilbert Series for Constructing Lagrangians:
  expanding the phenomenologist's toolbox}},
  \href{https://doi.org/10.1103/PhysRevD.91.105014}{\emph{Phys. Rev. D}
  {\bfseries 91} (2015) 105014}
  [\href{https://arxiv.org/abs/1503.07537}{{\ttfamily 1503.07537}}].

\bibitem{Fonseca:2019yya}
R.M.~Fonseca, \emph{{Enumerating the operators of an effective field theory}},
  \href{https://doi.org/10.1103/PhysRevD.101.035040}{\emph{Phys. Rev. D}
  {\bfseries 101} (2020) 035040}
  [\href{https://arxiv.org/abs/1907.12584}{{\ttfamily 1907.12584}}].

\bibitem{Hays:2018zze}
C.~Hays, A.~Martin, V.~Sanz and J.~Setford, \emph{{On the impact of
  dimension-eight SMEFT operators on Higgs measurements}},
  \href{https://doi.org/10.1007/JHEP02(2019)123}{\emph{JHEP} {\bfseries 02}
  (2019) 123} [\href{https://arxiv.org/abs/1808.00442}{{\ttfamily
  1808.00442}}].

\bibitem{Alioli:2022fng}
S.~Alioli et~al., \emph{{Theoretical developments in the SMEFT at dimension-8
  and beyond}},  in \emph{{Snowmass 2021}}, 3, 2022
  [\href{https://arxiv.org/abs/2203.06771}{{\ttfamily 2203.06771}}].

\bibitem{Damgaard:2019lfh}
P.H.~Damgaard, K.~Haddad and A.~Helset, \emph{{Heavy Black Hole Effective
  Theory}}, \href{https://doi.org/10.1007/JHEP11(2019)070}{\emph{JHEP}
  {\bfseries 11} (2019) 070}
  [\href{https://arxiv.org/abs/1908.10308}{{\ttfamily 1908.10308}}].

\bibitem{Haddad:2020que}
K.~Haddad and A.~Helset, \emph{{Tidal effects in quantum field theory}},
  \href{https://doi.org/10.1007/JHEP12(2020)024}{\emph{JHEP} {\bfseries 12}
  (2020) 024} [\href{https://arxiv.org/abs/2008.04920}{{\ttfamily
  2008.04920}}].

\bibitem{Bern:2020uwk}
Z.~Bern, J.~Parra-Martinez, R.~Roiban, E.~Sawyer and C.-H.~Shen, \emph{{Leading
  Nonlinear Tidal Effects and Scattering Amplitudes}},
  \href{https://doi.org/10.1007/JHEP05(2021)188}{\emph{JHEP} {\bfseries 05}
  (2021) 188} [\href{https://arxiv.org/abs/2010.08559}{{\ttfamily
  2010.08559}}].

\bibitem{Balkin:2021dko}
R.~Balkin, G.~Durieux, T.~Kitahara, Y.~Shadmi and Y.~Weiss, \emph{{On-shell
  Higgsing for EFTs}},
  \href{https://doi.org/10.1007/JHEP03(2022)129}{\emph{JHEP} {\bfseries 03}
  (2022) 129} [\href{https://arxiv.org/abs/2112.09688}{{\ttfamily
  2112.09688}}].

\bibitem{Liu:2023jbq}
H.~Liu, T.~Ma, Y.~Shadmi and M.~Waterbury, \emph{{An EFT hunter's guide to
  two-to-two scattering: HEFT and SMEFT on-shell amplitudes}},
  \href{https://arxiv.org/abs/2301.11349}{{\ttfamily 2301.11349}}.

\bibitem{Haddad:2023ylx}
K.~Haddad, \emph{{Recursion in the classical limit and the neutron-star Compton
  amplitude}},  \href{https://arxiv.org/abs/2303.02624}{{\ttfamily
  2303.02624}}.

\bibitem{Shankar:1977cm}
R.~Shankar and E.~Witten, \emph{{The S Matrix of the Supersymmetric Nonlinear
  Sigma Model}}, \href{https://doi.org/10.1103/PhysRevD.17.2134}{\emph{Phys.
  Rev. D} {\bfseries 17} (1978) 2134}.

\bibitem{Zamolodchikov:1977nu}
A.B.~Zamolodchikov and A.B.~Zamolodchikov, \emph{{Relativistic Factorized S
  Matrix in Two-Dimensions Having O(N) Isotopic Symmetry}},
  \href{https://doi.org/10.1016/0550-3213(78)90239-0}{\emph{JETP Lett.}
  {\bfseries 26} (1977) 457}.

\bibitem{Komatsu:2019hgc}
S.~Komatsu, R.~Mahajan and S.-H.~Shao, \emph{{An Index for Quantum
  Integrability}},
  \href{https://doi.org/10.21468/SciPostPhys.7.5.065}{\emph{SciPost Phys.}
  {\bfseries 7} (2019) 065} [\href{https://arxiv.org/abs/1907.07186}{{\ttfamily
  1907.07186}}].

\end{thebibliography}\endgroup
\end{document}